\documentclass[conference,compsoc]{IEEEtran}
\IEEEoverridecommandlockouts
\usepackage[dvipsnames, svgnames, x11names]{xcolor}
\usepackage{tikz}
\usepackage{amsmath}
\usepackage{filecontents}
\usepackage{amsfonts}
\usepackage{amssymb}
\usepackage{algorithmic}
\usepackage{graphicx}
\usepackage{textcomp}
\usepackage{xcolor}

\usepackage{mathrsfs}
\usepackage{amsthm}
\usepackage{epstopdf}
\usepackage{balance}
\usepackage{multirow}

\usepackage{epsfig,endnotes}
\usepackage{grffile}
\usepackage[font=bf]{caption}
\usepackage{color, url}
\usepackage{xspace} 
\usepackage[ruled,linesnumbered]{algorithm2e}

\usepackage[numbers,sort&compress,comma]{natbib}
\usepackage{epstopdf}
\usepackage{balance}
\usepackage{bm}
\usepackage{rotating}

\usepackage{enumitem}
\usepackage{makecell}
\usepackage{pdfx}
\usepackage{bm}

\usepackage{subcaption}

\usepackage{eso-pic}

\usepackage{fancyhdr}
\newcommand{\name}{\text{FCert}}

\newcommand\CR[1]{\textcolor{black}{#1}}

\newtheorem{theorem}{Theorem}

\newtheorem{example}{Example}

\renewcommand{\mathbf}[1]{\bm{#1}}
\newcommand\JY[1]{\textcolor{black}{#1}}

\newcommand{\lnorm}[1]{\ensuremath{\left\Vert#1\right\Vert}}

\newcommand{\argmax}{\operatornamewithlimits{argmax}}
\newcommand{\argmin}{\operatornamewithlimits{argmin}}

\newcommand{\myparatight}[1]{\smallskip\noindent{\bf {#1}:}~}

\allowdisplaybreaks

\begin{document}

\AddToShipoutPictureBG*{
  \AtPageUpperLeft{
    \setlength\unitlength{1in}
    \hspace*{\dimexpr0.5\paperwidth\relax}
    \makebox(0,-0.75)[c]{In IEEE Symposium on Security and Privacy, 2024.}
}}

\date{}

\title{\bf {\name}: Certifiably Robust Few-Shot Classification in the Era of Foundation Models}

\author{
 { Yanting Wang, Wei Zou, and Jinyuan Jia} \\
The Pennsylvania State University\\
\{yanting, weizou, jinyuan\}@psu.edu
}

\maketitle

\begin{abstract}
\emph{Few-shot classification} with foundation models (e.g., CLIP, DINOv2, PaLM-2) enables users to build an accurate classifier with a few labeled training samples (called \emph{support samples}) for a classification task. However, an attacker could perform data poisoning attacks by manipulating some support samples such that the classifier makes the attacker-desired, arbitrary prediction for a testing input. Empirical defenses cannot provide formal robustness guarantees, leading to a cat-and-mouse game between the attacker and defender. Existing certified defenses are designed for traditional supervised learning, resulting in sub-optimal performance when extended to few-shot classification. In our work, we propose {\name}, the \emph{first} certified defense against data poisoning attacks to few-shot classification. We show our {\name} provably predicts the same label for a testing input under arbitrary data poisoning attacks when the total number of poisoned support samples is bounded. We perform extensive experiments on benchmark few-shot classification datasets with foundation models released by OpenAI, Meta, and Google in both vision and text domains. Our experimental results show our {\name}: 1) maintains classification accuracy without attacks, 2) outperforms existing state-of-the-art certified defenses for data poisoning attacks, and 3) is efficient and general.

\end{abstract}


\section{Introduction}
Traditional supervised learning trains a model on a large amount of \emph{labeled} data to achieve good performance. However, collecting a large amount of labeled data could be time-consuming, expensive, or even impractical in many real-world applications. \emph{Few-shot classification} with \emph{ foundation models}~\cite{radford2021learning,bommasani2021opportunities,kirillov2023segment,oquab2023dinov2} aims to address this challenge. In particular, a resourceful foundation-model provider (e.g., OpenAI, Meta, and Google) collects a large amount of \emph{unlabeled} data to pre-train a foundation model, which is shared with users as a general \emph{feature extractor} to build downstream classifiers for various classification tasks.

Suppose a user has a classification task such as object classification. The user first collects a few training samples (called \emph{support samples}) for each class in the classification task. For instance, the user could \JY{download a few} support samples (e.g., images) \JY{for each class} from the Internet. Given a foundation model, the user could use a few-shot classification algorithm~\cite{snell2017prototypical,alain2016understanding,hu2022pushing,xu2023improving}  to build an accurate classifier with support samples. 
For instance, the user could use the foundation model to extract a feature vector for each support sample and then train a linear classifier (called \emph{linear probing}~\cite{alain2016understanding}) using the extracted feature vectors and labels of support samples.

\begin{figure}[!t]
    \centering
\includegraphics[width=1.0\linewidth]{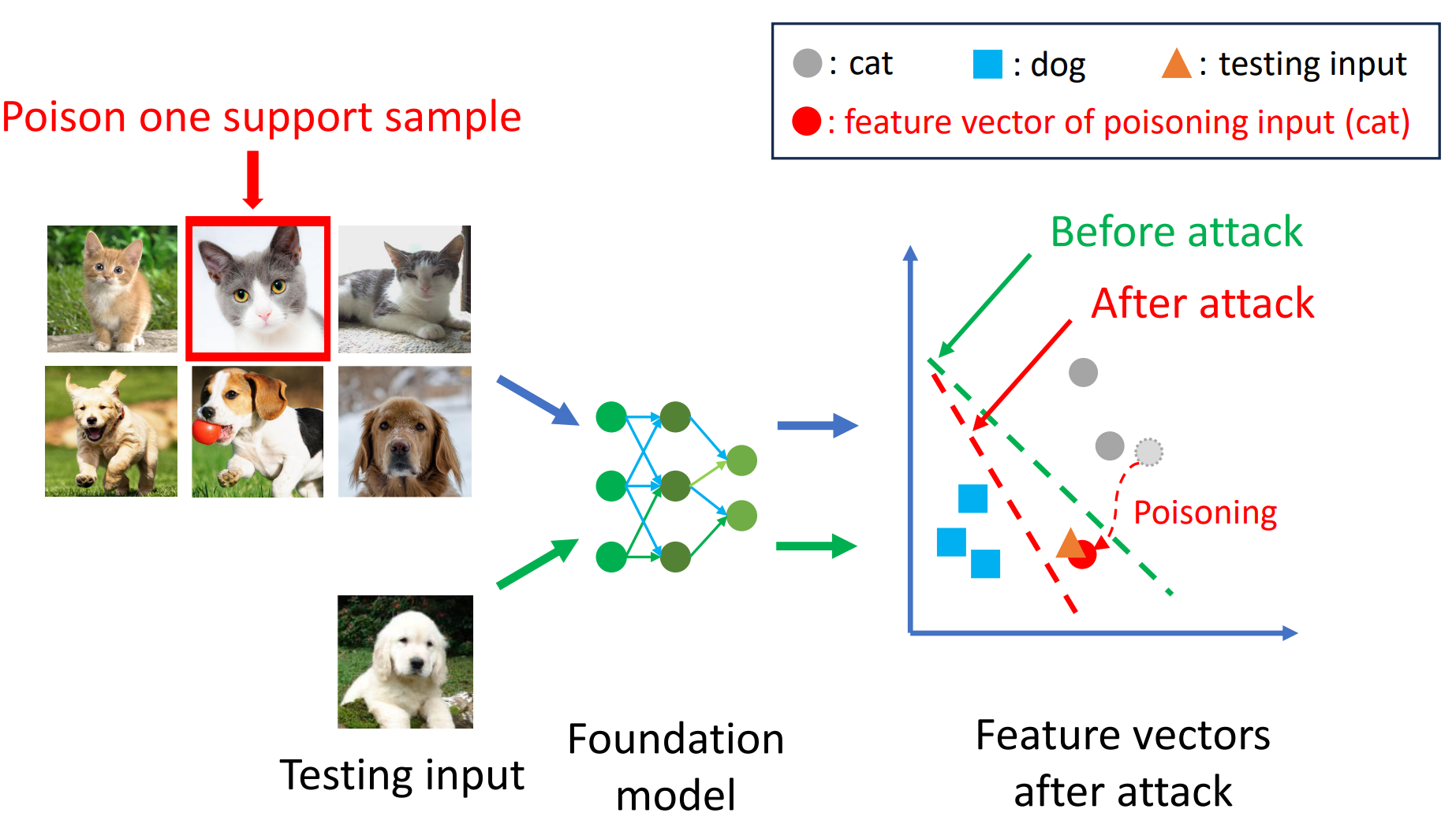}
    \caption{Illustration of few-shot classification with a foundation model using a linear classifier. An attacker could manipulate the classification boundary of the linear classifier by poisoning one support sample. The testing input is correctly classified as ``dog'' before attack and is misclassified as ``cat'' after attack.}\label{fig-linear}
    \label{fig-mr-net}
    \vspace{-1.8em}
\end{figure}

However, similar to traditional supervised learning, few-shot classification is also vulnerable to data poisoning attacks~\cite{munoz2017towards,shafahi2018poison,oldewage2023adversarial,xu2021yet,liu2023does}. Specifically, an attacker could poison support samples such that a classifier makes an attacker-desired, arbitrary prediction for a testing input (we defer the detailed discussion to Section~\ref{background}). 
For instance, Oldewage et al.~\cite{oldewage2023adversarial} showed that an attacker could add human imperceptible perturbations to support samples such that a few-shot classifier makes incorrect predictions. 
We also find that it is very challenging to manually check whether a support sample (e.g., an image) is poisoned or not when an attacker conducts clean-label attacks~\cite{shafahi2018poison} (i.e., the poisoned image is correctly labeled). Figure~\ref{clean-label-poisoning-attack} (in Appendix) shows an example of clean-label attack to few-shot classification. With the growing deployment of few-shot learning classifiers in security-critical applications such as biometric security systems~\cite{puch2019few}, medical diagnosis~\cite{ge2022few}, and autonomous driving~\cite{zhou2020few, cantarini2022few}, it is becoming increasingly crucial to develop defenses with formal security guarantees to defend against data poisoning attacks targeting few-shot learning systems. Figure~\ref{fig-mr-net} shows an example of data poisoning attacks to few-shot classification with a foundation model.

Existing defenses to data poisoning attacks can be divided into \emph{empirical defenses}~\cite{shan2021traceback,chen2021pois,zeng2023meta,peri2020deep,koh2022stronger,shokri2020bypassing,yang2022not,tran2018spectral,liu2022friendly,gao2019strip,chou2020sentinet,chen2018detecting,liu2018fine, qiu2021deepsweep,wang2019neural,qiao2019defending} and \emph{certified defenses}~\cite{jia2020intrinsic,levine2020deep,steinhardt2017certified,wang2022improved, ma2019data,rosenfeld2020certified,wang2020certifying,zhang2022bagflip,rezaei2023run,weber2023rab}. Empirical defenses cannot provide formal robustness guarantees, and thus they could be broken by strong, adaptive attacks~\cite{koh2022stronger,chen2023clean}. By contrast, certified defenses could provide a lower bound on their performance under \emph{arbitrary} data poisoning attacks, once the total number of poisoned samples is bounded. However, existing certified defenses are mainly designed for traditional supervised learning. Our results show they achieve sub-optimal robustness guarantees when extended to few-shot classification with foundation models. The key reason is that, by design, they didn't consider the high-quality feature vectors produced by a foundation model for support samples in few-shot classification. 

\myparatight{Our contribution} In our work, we propose {\name}, the \emph{first} certified defense against data poisoning attacks to few-shot classification. Our {\name} is based on two key observations. First, given a foundation model, the feature vector produced by the foundation model for a testing input would be close (or not close) to that of support samples whose labels are the same as (or different from)  the ground truth label of the testing input. Second, when fewer than half of the support samples from a class are poisoned, the feature vector distances between the majority of those support samples and the testing input remain unaffected. Based on those two observations, our {\name} calculates a robust distance between the test input and support samples of each class. In particular, our {\name} first calculates the feature vector distance between the test input and each support sample within a class, and then removes a certain number of largest or smallest feature vector distances and takes an average over the remaining feature vector distances as the robust distance. Finally, {\name} predicts the class with the smallest robust distance for a testing input. 

Our {\name} provably predicts the same label for a testing input under data poisoning attacks when the total number of poisoned samples is no larger than a threshold (called \emph{certified poisoning size}). To reach the goal, we derive a lower or upper bound for the robust distance under data poisoning attacks. In particular, we formulate the derivation of those lower or upper bounds as optimization problems and derive analytic solutions to solve them. By comparing those lower or upper bounds, we could compute the certified poisoning size. We also prove that our derived certified poisoning size is tight. In other words, without making extra assumptions, it is impossible to obtain a  certified poisoning size that is larger than ours. Our key idea in proving the tightness is to construct an empirical attack under which those lower or upper bounds could be reached. 

To evaluate the effectiveness of {\name}, we conduct experiments on three benchmark datasets: CUB200-2011~\cite{triantafillou2019meta}, CIFAR-FS~\cite{bertinetto2018meta}, and \emph{tiered}ImageNet~\cite{ren2019incremental} for few-shot classification. These experiments were carried out using two foundational models: CLIP~\cite{radford2021learning} from OpenAI and DINOv2~\cite{oquab2023dinov2} from Meta. We have the following three observations from our experimental results. First, without attacks, the classification accuracy of our {\name} is as accurate as existing state-of-the-art few-shot classification methods~\cite{snell2017prototypical, chen2019closer}. Second, our {\name} is more robust than existing state-of-the-art certified defenses~\cite{levine2020deep,jia2022certified,jia2020intrinsic} and few-shot classification methods~\cite{snell2017prototypical, chen2019closer}. Third, our {\name} achieves similar computation costs to those of existing state-of-the-art few-shot classification methods. Our three observations demonstrate that our {\name} is accurate, robust, and efficient. Additionally, we apply our {\name} to the natural language processing (NLP) domain. Our experimental results with PaLM-2 API~\cite{palm_api} (deployed by Google) and OpenAI API~\cite{openai_api} demonstrate the effectiveness of our {\name} in the text domain. Our major contributions are as follows:
\begin{itemize}
    \item We propose {\name}, the first certified defense against data poisoning attacks to few-shot classification. 
    \item We derive the certified robustness guarantees of {\name} and show its tightness.
    \item We perform a comprehensive evaluation for {\name}. Moreover, we compare {\name} with state-of-the-art few-shot classification methods and existing certified defenses against data poisoning attacks. 
\end{itemize}

\section{Background and Related Work}
\label{background}
\subsection{Background on Few-Shot Classification with Foundation Models} 
Suppose we have a foundation model $g$, where $g(\mathbf{x})$ represents the feature vector outputted by the foundation model $g$ for an input $\mathbf{x}$.
Given a foundation model $g$ and a few labeled training samples, \emph{few-shot classification} aims to build a classifier to predict the label of a testing input $\mathbf{x}_{test}$. Without loss of generality, we assume the total number of classes in the classification task is $C$. A sample is called a \emph{support sample} for a class $c$ if the ground truth label of the sample is $c$. We use $K$ to denote the number of support samples for each class $c$, where we denote by $\mathbf{x}_i^c$ the $i$th support sample for class $c$ and $c=1,2,\cdots, C$. We call the set of $C\cdot K$ support samples \emph{support set} (denoted by $\mathcal{D}$). A few-shot classification task with $C$ classes and $K$ support samples for each class is also called \emph{$C$-way-$K$-shot classification}. Next, we introduce two state-of-the-art methods for few-shot classification, namely ProtoNet~\cite{snell2017prototypical} and Linear Probing (LP)~\cite{chen2019closer}.

\myparatight{ProtoNet~\cite{snell2017prototypical}}ProtoNet first creates a prototype for each class based on its $K$ support samples and then uses those prototypes to build a classifier. In particular, the prototype for class $c$ is the mean of the feature vectors of its $K$ support samples. For simplicity, we use $\mathbf{e}_c$ to denote the prototype for the class $c$, where $c=1,2,\cdots, C$. Then, we have $\mathbf{e}_c =  \frac{1}{K}\sum_{i=1}^{K}g(\mathbf{x}_i^c)$, where $g(\mathbf{x}_i^c)$ is the feature vector produced by the foundation model $g$ for the support input $\mathbf{x}_i^c$. Given a testing input $\mathbf{x}_{test}$, the class whose prototype is closest to the feature vector of $\mathbf{x}_{test}$ is viewed as the final predicted label for $\mathbf{x}_{test}$. Formally, the label of the testing input $\mathbf{x}_{test}$ is predicted as $\hat{y}_{test} = \argmin_{c \in \{1,2,\cdots, C\}} Dist(g(\mathbf{x}_{test}), \mathbf{e}_c)$, where $Dist$ is a distance metric (e.g., $\ell_2$-distance).

\myparatight{Linear Probing (LP)~\cite{alain2016understanding, chen2019closer}} Linear Probing (LP) is widely used to build a classifier for few-shot classification. In particular, given support samples and a foundation model, LP uses the foundation model to compute a feature vector for each support sample. Then, LP trains a linear classifier based on the feature vectors and labels of support samples. Given a testing input, LP first uses the foundation model to compute a feature vector and then uses the linear classifier to predict a label for the feature vector, which is viewed as the final prediction for the testing input.

\myparatight{Limitations of existing state-of-the-art few-shot classification methods} The key limitation of existing state-of-the-art few-shot classification methods~\cite{snell2017prototypical,alain2016understanding, chen2019closer} is that they are not robust to data poisoning attacks as shown in our experimental results. Moreover, they cannot provide formal robustness guarantees against data poisoning attacks.
\subsection{Data Poisoning Attacks to Few-Shot Classification}

Few-shot classification is vulnerable to data poisoning attacks~\cite{shafahi2018poison,oldewage2023adversarial,xu2021yet}, where an attacker could poison the support samples in a support set. The classifier built upon the poisoned support set makes an attacker-desired prediction for a testing input. 
For instance, Shafahi et al.~\cite{shafahi2018poison} proposed a feature collision attack. Given a foundation model $g$ and a testing input $\mathbf{x}_{test}$, the feature collision attack adds a small perturbation to a clean support input whose ground truth label is an attacker-chosen label $l$ such that the foundation model outputs similar feature vectors for the poisoned support input and the testing input $\mathbf{x}_{test}$. The perturbation is crafted by solving the following optimization problem:
\begin{align}
\label{eqn-collision}
    \delta^*=\argmin_{\delta}\lnorm{g(\mathbf{x}^l+\delta), g(\mathbf{x}_{test})}_2 + \lambda \cdot \lnorm{\delta}_2,
\end{align}
where $\mathbf{x}^l$ is a clean support input with label $l$ and $\lambda$ is a hyper-parameter to balance the two terms. As a result, a classifier (e.g., trained by LP) is very likely to predict $\mathbf{x}_{test}$ as the attacker-chosen label $l$. In general, the feature collision attack could be very stealthy since it is a clean-label attack. (i.e., the poisoned support input is correctly labeled). 
Figure~\ref{fig-linear} and~\ref{clean-label-poisoning-attack} (in Appendix) visualizes this attack. 

\subsection{Existing Certified Defenses}

Defenses against data poisoning attacks could be divided into \emph{empirical defenses}~\cite{shan2021traceback,chen2021pois,zeng2023meta,peri2020deep,koh2022stronger,shokri2020bypassing,yang2022not,tran2018spectral,liu2022friendly,gao2019strip,chou2020sentinet} and \emph{certified defenses}~\cite{steinhardt2017certified,wang2020certifying,jia2020intrinsic,jia2022certified,cao2021provably,levine2020deep,zhang2021backdoor,wang2022improved,cao2022flcert,weber2023rab,li2023sok,jia2023pore,pei2023textguard}. Empirical defenses cannot provide formal robustness guarantees, and thus they could be broken by strong, adaptive attacks~\cite{koh2022stronger,chen2023clean}. Thus, we focus on certified defenses in this work. Next, we introduce existing state-of-the-art certified defenses, including Bagging~\cite{jia2020intrinsic}, DPA~\cite{peri2020deep}, and $k$-NN~\cite{jia2022certified}.

\myparatight{Bagging~\cite{jia2020intrinsic} and DPA~\cite{levine2020deep}}
Given a dataset that contains a set of samples, Bagging (or DPA) first creates many sub-datasets, each of which contains a subset of samples. Then, Bagging (or DPA) uses each sub-dataset to build a classifier. Given a testing input, Bagging (or DPA) uses each classifier to predict a label for the testing input and takes a majority vote over those predicted labels to make the final prediction for the testing input. Existing studies~\cite{jia2020intrinsic,levine2020deep} show the majority vote result of Bagging (or DPA) is unaffected when the total number of poisoned samples in the dataset is bounded. 
The key insight in deriving the certified robustness guarantees is that most of the sub-datasets are unaffected by the poisoned samples when the total number of poisoned samples is small. The difference between Bagging and DPA is that they use different ways to create sub-datasets, e.g., Bagging creates each sub-dataset by subsampling a number of samples from the dataset uniformly at random with replacement; DPA uses a hash function to divide a dataset into multiple disjoint sub-datasets.

\myparatight{$k$-NN}Given a dataset that contains a set of samples and a testing input, $k$-NN finds the $k$ samples in the dataset that have the smallest distance (e.g., $\ell_2$-distance) to the testing input and then takes a majority vote over the labels of those $k$ samples as the final predicted label of the testing input. Jia et al.~\cite{jia2022certified} showed that $k$-NN could provide certified robustness guarantees against data poisoning attacks due to its intrinsic majority vote mechanism, i.e., most of $k$ samples are unaffected when the total number of poisoned samples is small and $k$ is large.  

\myparatight{Limitations of existing certified defenses}Existing certified defenses~\cite{levine2020deep, jia2020intrinsic, jia2022certified} are mainly designed for traditional supervised learning. Our experimental results show they achieve sub-optimal robustness guarantees when extended to few-shot classification with foundation models. The reason is that, by design, they don't consider the high-quality feature vectors produced by foundation models for inputs. We note that some studies~\cite{wang2020certifying,weber2023rab} extend randomized smoothing~\cite{lecuyer2019certified,cohen2019certified} to provide certified robustness guarantees for data poisoning attacks. However, they can only provide guarantees when an attacker adds bounded perturbations (with respect to $\ell_2$ norm) to training inputs. In general, the attacker could arbitrarily perturb an input to craft a poisoned input in data poisoning attacks. 
\section{Problem Formulation}

\subsection{Threat Model}
We characterize the threat model with respect to the attacker's goals, background knowledge, and capabilities. Our threat model follows the previous work on certified defenses~\cite{jia2020intrinsic,levine2020deep,jia2022certified} against data poisoning attacks.

\myparatight{Attacker's goals}We consider that an attacker aims to poison the support set such that a classifier built upon the poisoned support set makes an attacker-desired, arbitrary prediction for a testing input, e.g., the testing input is misclassified by the classifier.

\myparatight{Attacker's background knowledge and capabilities}As we focus on the certified defense, we consider the strongest attacker in our threat model. In particular, we assume the attacker knows everything about the few-shot classification, including  1) all clean support samples in a clean support set, 2) the few-shot classification algorithm as well as its hyper-parameters, and 3) the parameters of the foundation model used by the few-shot classification method. 

We consider that an attacker could poison the support samples in a clean support set $\mathcal{D}$. In particular, we consider two attack scenarios. In the first scenario, we consider that the attacker could \emph{arbitrarily} poison $T$ (called \emph{poisoning size}) support samples in $\mathcal{D}$, i.e., the attacker could arbitrarily manipulate the inputs and labels of $T$ support samples in $\mathcal{D}$. 
We call this scenario \emph{Group Attack}. In the second scenario, we consider a stronger attacker, where the attacker could poison $T$ support samples from each class. We call this scenario \emph{Individual Attack}.
Specifically, suppose $\mathcal{D}^c \subseteq \mathcal{D}$ is a subset of $K$ support samples in $\mathcal{D}$ whose label is $c$, where $c=1,2,\cdots, C$. We consider that an attacker could \emph{arbitrarily} manipulate up to $T$ clean support samples in each $\mathcal{D}^c$, where $c=1,2,\cdots, C$. In other words, the attacker could \emph{arbitrarily} select $T$ support samples from each $\mathcal{D}_c$ and then \emph{arbitrarily} change the inputs and labels of those $T$ support samples.  We note that the attacker could manipulate up to $C\cdot T$ support samples in $\mathcal{D}$. 

Note that, in $C$-way-$K$-shot classification~\cite{snell2017prototypical}, the number of support samples for each class in a support set is $K$, i.e., the number of support samples is the same for different classes. Thus, to be consistent with existing studies~\cite{snell2017prototypical,pautov2022smoothed,sung2018learning,yoon2019tapnet} on few-shot classification, we consider that the poisoned support set also contains the same number of support samples for each class. This is a realistic threat model since the support samples in the support set are chosen by the user. Thus, the user could choose the same number of support samples for each class. 

We consider an attacker could poison support samples in few-shot classification. We acknowledge that an attacker may not always be able to do this in broad settings, e.g., when the number of support samples is small and they are collected from a trusted source. We note that, beyond adversarially perturbed support samples (i.e., worst-case scenario), our defense may also have the potential to enhance robustness for few-shot classification when certain support samples are out-of-distribution ones~\cite{jeong2020ood} (i.e., their feature vectors deviate from those of in-distribution support samples of the same class.)

\subsection{Certifiably Robust Few-Shot Classification}
Suppose we have a clean support set $\mathcal{D}$. Moreover, we use $\mathcal{B}(\mathcal{D}, T)$ to denote a set of all possible poisoned support sets when an attacker conducts the Group Attack (or Individual Attack) with a poisoning size $T$.
Given a testing input $\mathbf{x}_{test}$, a support set $\mathcal{D}$, a foundation model $g$, and a few-shot classification method $\mathcal{M}$, we use $\mathcal{M}(\mathbf{x}_{test}; \mathcal{D}, g)$ to denote the predicted label of a classifier built by $\mathcal{M}$ on the support set $\mathcal{D}$ with the foundation model $g$. Note that we also write $\mathcal{M}(\mathbf{x}_{test}; \mathcal{D}, g)$ as $\mathcal{M}(\mathbf{x}_{test}; \mathcal{D})$ for simplicity reasons. Given a testing input $\mathbf{x}_{test}$ and a few-shot classification method $\mathcal{M}$, we could find a maximum poisoning size $T^*$ (called \emph{certified poisoning size}) such that the predicted label of the classifier built by $\mathcal{M}$ on an arbitrary poisoned support set in $\mathcal{B}(\mathcal{D}, T)$ for $\mathbf{x}_{test}$ does not change. Formally, we could compute $T^*$ as follows:
\begin{align}
\label{definition-of-crfc}
&T^* = \argmax_{T} T \nonumber \\
    s.t.,  \text{ }& \mathcal{M}(\mathbf{x}_{test}; \mathcal{D}) = \mathcal{M}(\mathbf{x}_{test}; \mathcal{D}_p), \forall \mathcal{D}_p \in \mathcal{B}(\mathcal{D}, T).
\end{align}
Note that $T^*$ could be different for different testing inputs as the constraint in the optimization problem involves $\mathbf{x}_{test}$.

\myparatight{Design goals}
We aim to design a few-shot classification method that is accurate, efficient, and certifiably robust.

\section{Our {\name}}
\label{method}
\begin{figure*}[t]
    \centering
    \includegraphics[width=1.0\linewidth]{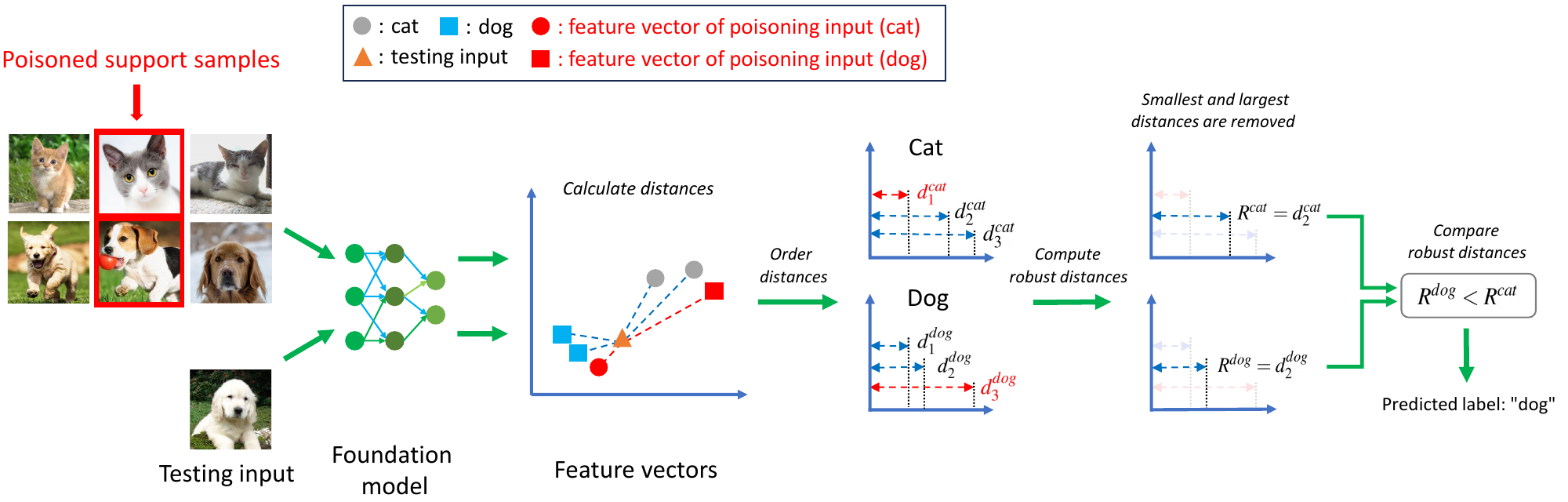}
   \vspace{-5mm}
    \caption{Overview of {\name} under Individual Attack. We have three support samples for each of the two classes (i.e., $2$-way-$3$-shot classification). An attacker could poison one support sample for each class, where the feature vectors with {\color{red}red} color are for poisoned support samples. ${\color{red}d_1^{cat}}, d_2^{cat}, d_3^{cat}$ (or $d_1^{dog}, d_2^{dog}, {\color{red}d_3^{dog}}$)  are distances between the feature vectors of three support samples whose labels are ``cat'' (or ``dog'') and the testing input, which are used to compute two robust distances $R^{cat}$ and $R^{dog}$.  Our {\name} still predicts the correct label ``dog'' for the testing input under two poisoned support samples.  }
    \label{fig-Fscert-overview}
    \vspace{-5mm}
\end{figure*}
\subsection{Overview of {\name}}
We aim to build a certifiably robust classifier against data poisoning attacks to few-shot classification. Suppose we have a testing input $\mathbf{x}_{test}$, and a poisoned support set $\mathcal{D}_p$ where $\mathcal{D}^c_p \subseteq \mathcal{D}_p$ is a subset of support samples in $\mathcal{D}_p$ whose labels are $c$ ($c=1,2,\cdots, C$). {\name} is based on two key observations. First, the feature vector produced by a foundation model for the testing input $\mathbf{x}_{test}$ would be close (or not close) to those of the support samples whose labels are the same as (or different from) the testing input $\mathbf{x}_{test}$. Second, under data poisoning attacks with a poisoning size $T$, at least $(K-T)$ support samples in each $\mathcal{D}^c_p$ would be clean since an attacker could manipulate at most $T$ support samples in $\mathcal{D}^c_p$ and the total number of support samples in  $\mathcal{D}^c_p$ is $K$. 

Based on the above two observations, we know that the feature vector of $\mathbf{x}_{test}$ would be close (or not close) to most of the support samples (i.e., $(K-T)$ clean ones) from $\mathcal{D}^c_p$ if $c$ is the same (or different from) the ground truth label of $\mathbf{x}_{test}$ when $T$ is small (e.g., less than or equal to $\lfloor \frac{K-1}{2} \rfloor$). Based on this finding, we compute a robust distance $R^c$ between $\mathbf{x}_{test}$ and each class's support set $\mathcal{D}^c_p$. In particular, we compute the distance (e.g., $\ell_2$-distance) between the feature vector of  $\mathbf{x}_{test}$ and the feature vector of each support sample in $\mathcal{D}^c_p$. Then, we remove the $K'$ largest and smallest distances and take an average over the remaining $(K-2\cdot K')$ ones as the robust distance $R^c$. Then, we predict the label (denoted as $\hat{y}$) whose robust distance is the smallest for the testing input $\mathbf{x}_{test}$, i.e., $\hat{y}=\argmin_{c=1,2,\cdots, C}R^c$. Figure~\ref{fig-Fscert-overview} shows an overview of {\name} under Individual Attack, where $K=3$, $K'=1$, and $T=1$.

We prove that {\name} consistently predicts $\hat{y}$ for a testing input under arbitrary data poisoning attacks when the poison size $T$ is no larger than a threshold (called \emph{certified poisoning size}, denoted as $T^*$). Our key idea in deriving $T^*$ is to derive an upper (or lower) bound for the robust distance of the label $\hat{y}$ (or $c\neq \hat{y}$). We formulate the derivation of the upper or lower bound as an optimization problem and derive an analytic solution to it. {\name} still predicts the label $\hat{y}$ for $\mathbf{x}_{test}$ when the upper bound of the robust distance for $\hat{y}$ is smaller than the lower bound of the robust distance for $\forall c \neq \hat{y}$, enabling us to compute the certified poisoning size $T^*$ for the testing input $\mathbf{x}_{test}$.

Finally, we show our derived certified poisoning size is tight. In particular, we prove that we could construct an empirical data poisoning attack with a poisoning size $(T^*+1)$ such that the testing input $\mathbf{x}_{test}$ could be misclassified under such an attack. In other words, it is impossible to derive a larger certified poisoning size $T^*$ than ours. 

\subsection{Our {\name}}

Suppose we have a support set $\mathcal{D}$ with $C\cdot K$ support samples, where $\mathcal{D}^c \subseteq \mathcal{D}$ is a subset of $K$ support samples in $\mathcal{D}$ whose label is $c$ ($c=1,2,\cdots, C$). We use $\mathbf{x}_i^c$ to denote the $i$-th support input in $\mathcal{D}^c$.
Given a testing input $\mathbf{x}_{test}$, a set of $K$ support samples $\mathcal{D}^c$, and a foundation model $g$, we compute a distance between the features vectors produced by the foundation model $g$ for each support input in $\mathcal{D}^c$ and the testing input $\mathbf{x}_{test}$. For simplicity, we denote $\Tilde{d}_i^c= Dist(g(\mathbf{x}_{test}), g(\mathbf{x}_i^c))$, where $Dist$ is a distance metric such as $\ell_2$-distance and $i=1,2,\cdots, K$.
We sort $\Tilde{d}_1^c, \Tilde{d}_2^c, \cdots, \Tilde{d}_{K}^c$ in ascending order to get an ordered distance sequence, i.e., $d_1^c \leq d_2^c \leq \cdots \leq d_{K}^c$.

Given $d_1^c, d_2^c, \cdots, d_{K}^c$, our key insight is that at most $T$ of them could be changed when an attacker manipulates up to $T$ samples in $\mathcal{D}^c$. This inspires us to utilize robust statistics techniques~\cite{huber2004robust} to estimate a robust distance $R^c$ between the feature vectors of the testing input $\mathbf{x}_{test}$ and support samples in $\mathcal{D}^c$. 
Specifically, we remove the largest and smallest $K'$ distances among $d_1^c, d_2^c, \cdots, d_{K}^c$ and then take an average over the remaining $K-2K'$ distances, where $K' \leq \lfloor (K-1)/2 \rfloor$ is a hyper-parameter. Formally, we compute $R^c$ as follows:
\begin{align}
\label{eqn-compute-score}
    R^c = \frac{\sum_{i=K'+1}^{K - K'}d_i^c}{K - 2 K'},
\end{align}
where $c=1,2,\cdots, C$. 
Roughly speaking, a small $R^c$ means the feature vector of the testing input $\mathbf{x}_{test}$ is close to those of support samples whose label is $c$.
Thus, given $R^1, R^2, \cdots, R^C$, we predict the following label for $\mathbf{x}_{test}$:
\begin{align}
    \mathcal{M}(\mathbf{x}_{test}; \mathcal{D}) = \argmin_{c=1,2,\cdots, C} R^c.
\end{align}

Algorithm~\ref{alg:predict} shows how our {\name} predicts a label for a testing input, where the function \textsc{SortInAscend} sorts distances in ascending order. 

\begin{algorithm}[tb]
   \caption{\emph{{\name}}}
   \label{alg:predict}
\begin{algorithmic}
   \STATE {\bfseries Input:} Support set $\mathcal{D}$, a foundation model $g$, a testing input $\mathbf{x}_{test}$, hyper-parameter $K'$, a distance metric $Dist$
   \STATE {\bfseries Output:} Predicted label for $\mathbf{x}_{test}$ \\
   \STATE $\mathcal{D}^c = \{(\mathbf{x}_i, y_i) \in \mathcal{D}| y_i =c \}, c=1,2,\cdots, C$ \\
   $K = |\mathcal{D}^c|$ \\
   \FOR{$c=1,2,\cdots, C$}
    \FOR{$(\mathbf{x}_{i}^c, c)\in \mathcal{D}^c$} 
    \STATE $\Tilde{d}_i^c = Dist(g(\mathbf{x}_{test}), g(\mathbf{x}_i^c))$ \\
    \ENDFOR
    \STATE $d_1^c, d_2^c,\cdots, d_K^c = \textsc{SortInAscend}(\Tilde{d}_1^c, \Tilde{d}_2^c, \cdots, \Tilde{d}_K^c)$
    \STATE $R^c = (\sum_{i=K'+1}^{K - K'}d_i^c)/(K - 2 K')$ \\
    \ENDFOR
   \STATE \textbf{return} $\argmin_{c=1,2,\cdots, C} R^c$
\end{algorithmic}
\end{algorithm}

\subsection{Certified Robustness Guarantee of {\name}}
Our {\name} provably predicts the same label for a testing input when the number of poisoned support samples is bounded. Suppose $\mathcal{B}(\mathcal{D}, T)$ is a set of all possible poisoned support sets when an attacker can poison $T$ support samples for each class for Individual Attack \JY{(or $T$ support samples in total across all classes for Group Attack)}. Formally, we aim to compute a maximum $T$ such that we have the following:
\begin{align}
    \mathcal{M}(\mathbf{x}_{test}; \mathcal{D}) = \mathcal{M}(\mathbf{x}_{test}; \mathcal{D}_p), \forall \mathcal{D}_p \in \mathcal{B}(\mathcal{D}, T).
\end{align}
For simplicity, we use $\hat{y} = \mathcal{M}(\mathbf{x}_{test}; \mathcal{D}) = \argmin_{c=1,2,\cdots, C} R^c$ to denote the predicted label of our {\name} for the testing input with the clean support set $\mathcal{D}$.
Given an arbitrary poisoned support set $\mathcal{D}_p \in \mathcal{B}(\mathcal{D}, T)$, we use $\mathcal{D}^1_p, \mathcal{D}^2_p, \cdots, \mathcal{D}^C_p$ to denote the $C$ subsets of $\mathcal{D}_p$, where the subset $\mathcal{D}^c_p$ contains all support samples in $\mathcal{D}_p$ whose labels are $c$ for $c=1,2,\cdots, C$. \JY{We use $T^c$ to represent the number of poisoned support samples in $\mathcal{D}^c_p$, where $T^c=T$ ($c=1,2,\cdots, C$) for Individual Attack and $\sum^{C}_{c=1}T^c = T$ for Group Attack.} We denote $\Tilde{e}_i^c = Dist(\mathbf{x}_{test}, \Tilde{\mathbf{x}}_i^c)$ for $(\Tilde{\mathbf{x}}_i^c, c) \in \mathcal{D}^c_{p}$, where $Dist$ is a distance metric, $i=1,2,\cdots, K$, and $c=1,2,\cdots, C$. We sort $\Tilde{e}_1^c, \Tilde{e}_2^c, \cdots, \Tilde{e}_K^c$ in ascending order to obtain an ordered distance sequence, i.e., $e_1^c \leq e_2^c \leq \cdots \leq e_K^c$.  Given  $e_1^c, e_2^c, \cdots, e_K^c$, we denote \JY{$R^c_p(T^c) = \frac{\sum_{i=K'+1}^{K-K'}e_i^c}{K-2K'}$, which is the robust distance computed by {\name} for a testing input $\mathbf{x}_{test}$ under the poisoned subset $\mathcal{D}^c_p$ with $T^c$ poisoned support samples.} 

Our goal is to find the maximum $T$ such that our {\name} provably predicts the label $\hat{y}$ with an arbitrary poisoned support set $\mathcal{D}_p$, i.e., $\hat{y}= \argmin_{c=1,2,\cdots, C} R^c_p(T^c)$, where \JY{$T^c = T$ for Individual Attack and $\sum^{T}_{c=1}T^c = T$ for Group Attack}. To reach the goal, our key idea is to derive an upper bound for $R^{\hat{y}}_p(T^{\hat{y}})$ and a lower bound of $\min_{c\neq \hat{y}}R^{c}_p(T^c)$. Our {\name} still predicts the label $\hat{y}$ when the upper bound of $R^{\hat{y}}_p(T^{\hat{y}})$ is smaller than the lower bound of $\min_{c\neq \hat{y}}R^{c}_p(T^c)$.

\myparatight{Deriving an upper bound of $R^{\hat{y}}_p(T^{\hat{y}})$} We denote by  $\overline{R}_p^{c}(T^c)$ the upper bound of $R^{c}_p(T^c)$ for $c=1,2,\cdots, C$ (including $\hat{y}$) when there are $T^c$ poisoned support samples in $\mathcal{D}^c_p$. With at most $T^c$ poisoned support samples in $\mathcal{D}^{c}_p$, we know that at most $T^c$ support samples in $\mathcal{D}^{c}_p$ are different from those in $\mathcal{D}^c$, i.e., $\sum_{i=1}^K \mathbb{I}(\mathbf{x}_i^c \neq \Tilde{\mathbf{x}}_i^c) \leq T^c$, where $\mathbb{I}$ is the indicator function whose output is 1 if the condition is satisfied and 0 otherwise. Based on the definition of $\Tilde{d}_i^c$ and $\Tilde{e}_i^c$, we know $\sum_{i=1}^{K}\mathbb{I}(
\Tilde{e}_i^c \neq \Tilde{d}_i^c) \leq T^c$. Formally, the derivation for the upper bound $\overline{R}_p^{c}(T^c)$ can be formulated as the following optimization problem:
\begin{align}
 \JY{\overline{R}_p^{c}(T^c) } =& \max_{e_1^c,e_2^c,\cdots, e_{K}^c} \frac{\sum_{i=K'+1}^{K-K'}e_i^c}{K-2K'}   \\
 \label{reorder-constraint}
 & s.t., \text{ } e_1^c \leq e_2^c \leq \cdots \leq e_{K}^c, \\
 &\sum_{i=1}^{K}\mathbb{I}(\Tilde{e}_i^c \neq \Tilde{d}_i^c) \leq T^c,
\end{align}
where we have Equation~\ref{reorder-constraint} because our {\name} reorders the distances in ascending order before estimating a robust distance. 
Next, we derive an analytical solution to the above optimization problem. We consider that $T^c \leq K'$ (note that the upper bound \JY{$\overline{R}_p^{c}(T^c)$} could be arbitrarily large when $T^c> K'$). \JY{The constraint $\sum_{i=1}^{K}\mathbb{I}(\Tilde{e}_i^c \neq \Tilde{d}_i^c) \leq T^c$} means that the attacker could change at most $T^c$ distances among $d_1^c, d_2^c,\cdots, d_K^c$ to maximize $\overline{R}_p^{c}(T^c)$. The optimal strategy for the attacker is to change the $T^c$ smallest distances among $d^c_{1}, d^c_{2}, \cdots, d^c_{K}$ to transform them into the 
$T^c$ largest distances (we defer the formal proof to Appendix~\ref{proof-of-theorem-1}). Thus, the upper bound $ \overline{R}_p^{c}(T^c)$ would be as follows:
\begin{align}
\label{equation-upper-bound}
\overline{R}_p^{c}(T^c) =  \frac{\sum_{i=K'+1+T^c}^{K-K'+T^c}d_i^c}{K-2\cdot K'},
\end{align}
where $T^c \leq K'$. Note that $\overline{R}_p^{c}(T^c)$ increases as $T^c$ increases.  By letting $c= \hat{y}$ in Equation~\ref{equation-upper-bound}, we obtain the upper bound $\overline{R}_p^{\hat{y}}(T^{\hat{y}})$ of $R^{\hat{y}}_p(T^{\hat{y}})$. Next, we use an intuitive example to illustrate our derived upper bound.
\begin{example}
    Suppose we have $K=5$ and $d_1^c=1, d_2^c=2, d_3^c=3, d_4^c=4, d_5^c=5$. Moreover, we set $K'=1$ and assume $T^c=1$, i.e., the attacker could change at most one distance among $d_1^c, d_2^c, \cdots, d_5^c$. Then, the upper bound $\overline{R}_p^{c}(T^c=1)$ would be $4$, i.e., $\overline{R}_p^{c}(T^c=1)=\frac{d_3^c+d_4^c+d_5^c}{3}=4$.  The optimal strategy for the attacker is to change $d_1^c$  to a value that is no smaller than $d_5^c=5$.
\end{example}

\begin{algorithm}[!tb]
   \caption{Computing Certified Poisoning Size}
   \label{alg:binary-search-IA}
\begin{algorithmic}
   \STATE {\bfseries Input:}  $d_1^c, d_2^c, \cdots, d_K^c$ for $c=1,2,\cdots, C$,  predicted label $\hat{y}$, $K'$
   \STATE {\bfseries Output:} Certified poisoning size $T^*$\\
 \STATE {${low} = 0$}
 \STATE {${high} = K'$}
 \WHILE{${low} \neq {high}$}
 \STATE{$mid = \lceil{low + high}\rceil/2$}
  \FOR{$c=1,2,\cdots, C$}
       \IF{$c = \hat{y}$}
    \STATE $\overline{R}_p^{c}(T) = (\sum_{i=K'+1+mid}^{K - K'+mid}d_i^c)/(K - 2 K')$ \\
    \ELSE
        \STATE $\underline{R}_p^{c}(T)= (\sum_{i=K'+1-mid}^{K - K'-mid}d_i^c)/(K - 2 K')$ \\
        \ENDIF
    \ENDFOR
   \IF{$\overline{R}_p^{\hat{y}}(T) < \min_{c \neq \hat{y}} \underline{R}_p^{c}(T)$}
   \STATE{${low} = {mid}$}
    \ELSE
    \STATE{${high} = {mid}-1$}
   \ENDIF
\ENDWHILE
   \STATE \textbf{return} $low$
\end{algorithmic}
\end{algorithm}

\myparatight{Deriving a lower bound of $\min_{c\neq \hat{y}}R^c_p(T^c)$} We denote by $\underline{R}_p^{c}(T^c)$ the lower bound of $R^c_p(T^c)$. As there are $T^c$ poisoned support samples in $\mathcal{D}^c_p$, we know that at most $T^c$ support samples in $\mathcal{D}^{c}_p$ are different from those in $\mathcal{D}^c$, i.e., $\sum_{i=1}^K \mathbb{I}(\mathbf{x}_i^c \neq \Tilde{\mathbf{x}}_i^c) \leq T^c$. Based on the definition of $\Tilde{d}_i^c$ and $\Tilde{e}_i^c$, we have 
$\sum_{i=1}^{K}\mathbb{I}(\Tilde{e}_i^c \neq \Tilde{d}_i^c) \leq T^c$. Formally, we can formulate finding the lower bound $\underline{R}_p^{c}(T^c)$ as the following optimization problem:

\begin{align}
\underline{R}_p^{c}(T^c) &= \min_{e_1^c,e_2^c,\cdots, e_{K}^c} \frac{\sum_{i=K'+1}^{n-K'}e_i^c}{n-2K'}  \\
 &s.t. \text{ } e_1^c \leq e_2^c \leq \cdots \leq e_{K}^c, \\
 & \sum_{i=1}^{K}\mathbb{I}(\Tilde{e}_i^c \neq \Tilde{d}_i^c) \leq T^c.
\end{align}
We also derive an analytical solution to the above optimization problem.
Similar to the derivation of $\overline{R}_p^{c}(T^c)$, we know 
that the optimal strategy for an attacker is to change the $T^c$ largest distances among $d_{1}^c, d_{2}^c, \cdots, d_{K}^c$ to transform them into the $T^c$ smallest distances. 
As a result, we have the following lower bound:
\begin{align}
\label{equation-lower-bound}
    \underline{R}_p^{c}(T^c) = \frac{\sum_{i=K'+1-T^c}^{K-K'-T^c}d_i^c}{K-2\cdot K'}.
\end{align}
where $T^c \leq K'$. Note that $\underline{R}_p^{c}(T^c)$ decreases as $T^c$ increases. The lower bound of $\min_{c\neq \hat{y}}R^c_p(T^c)$ would be $\min_{c\neq \hat{y}} \underline{R}_p^{c}(T^c)$, where $\underline{R}_p^{c}(T^c)$ is computed in Equation~\ref{equation-lower-bound}.

\myparatight{Optimization problem for Individual Attack}\JY{For Individual Attack, we have $T^c = T$ for $c=1,2,\cdots, C$. Therefore,} {\name} still predicts the label $\hat{y}$ for the testing input $\mathbf{x}_{test}$ when the upper bound of $R^{\hat{y}}_p(T)$ is smaller than the lower bound of $\min_{c\neq \hat{y}} R^c_p(T)$, i.e.,  $\overline{R}_p^{\hat{y}}(T) < \min_{c\neq \hat{y}} \underline{R}_p^{c}(T)$. We aim to find a maximum $T$ (denoted as $T^*$) such that $\overline{R}_p^{\hat{y}}(T) < \min_{c\neq \hat{y}} \underline{R}_p^{c}(T)$. Formally, we could compute $T^*$ by solving the following optimization problem:
\begin{align}
\label{final-opt-problem}
    T^* = \argmax_{T=1,\cdots, K'} T,
    \text{ }s.t.\text{ } \overline{R}_p^{\hat{y}}(T) < \min_{c\neq \hat{y}} \underline{R}_p^{c}(T),
\end{align}
where $\overline{R}_p^{\hat{y}}(T)$ and $\min_{c\neq \hat{y}} \underline{R}_p^{c}(T)$ are computed based on Equation~\ref{equation-upper-bound} and~\ref{equation-lower-bound}, respectively.  We could use binary search to solve the above optimization problem as $\overline{R}_p^{\hat{y}}(T)$ (or $\min_{c\neq \hat{y}} \underline{R}_p^{c}(T)$) increases (or decreases) as $T$ increases. 
Specifically, given an arbitrary $T$, we can compute $\overline{R}_p^{\hat{y}}(T)$ and $\min_{c\neq \hat{y}} \underline{R}_p^{c}(T)$ efficiently based on Equation~\ref{equation-upper-bound} and~\ref{equation-lower-bound}. Then, we could verify whether the constraint of the optimization problem in Equation~\ref{final-opt-problem} is satisfied or not. 
Algorithm~\ref{alg:binary-search-IA} shows our binary search algorithm to compute $T^*$ for a testing input. 

\myparatight{Optimization problem for Group Attack}
\JY{For Group Attack, an attacker could arbitrarily poison at most $T$ support samples in total across all classes, i.e., $\sum^{C}_{c=1} T^c = T$. Without loss of generality, suppose $c^* \neq \hat{y}$ is the class with the smallest robust distance after the attack, i.e., $c^* = \argmin_{c\neq \hat{y}} {R}_p^{c}(T^c)$. 
 Then, the optimal attack strategy for the attacker is to only poison $\mathcal{D}^{\hat{y}} \subseteq \mathcal{D}$ and $\mathcal{D}^{c^*} \subseteq \mathcal{D}$ ($c^* \neq \hat{y}$), while keeping other subsets of support samples untouched. 
 The reason is that poisoning a support sample from $\mathcal{D}^{\hat{y}}$ (or $\mathcal{D}^{c^*}$) instead of $\mathcal{D}^{c}$ ($c\neq c^*, \hat{y}$) could increase $\overline{R}_p^{\hat{y}}$ (or decrease $\underline{R}_p^{c^*}$), resulting in a stronger attack.}
 
 \JY{We use $T^{\hat{y}}$ to denote the number of poisoned support samples in $\mathcal{D}^{\hat{y}}_p \subseteq \mathcal{D}_p$. Since the total number of poisoned samples is $T$, there are at most $T-T^{\hat{y}}$ poisoned samples in $\mathcal{D}^{c^*}_p \subseteq \mathcal{D}_p$, i.e., $T^{c^*} \leq T-T^{\hat{y}}$. To guarantee our {\name} still predict the label $\hat{y}$ for a testing input $\mathbf{x}_{test}$, we need to ensure the upper bound of $R^{\hat{y}}_p$ be smaller than the lower bound of $R^c_p$ for all possible $T^{\hat{y}}$, i.e., $\overline{R}_p^{\hat{y}}(T^{\hat{y}}) <  \underline{R}_p^{c^*}(T-T^{\hat{y}}), \forall T^{\hat{y}}: 0\leq T^{\hat{y}} \leq T$. Since $c^*$ is unknown, we find the maximum $T$ (i.e., certified poisoning size $T^*$) such that the previous equation holds for every possible $c^*\neq \hat{y}$. Formally, we formulate the computation of $T^*$ as the following optimization problem:
 \begin{align}
 \label{final-opt-problem-group}
    &T^* = \argmax_{T=1,\cdots, K'} T,  \\
   \text{ }s.t.\text{ }& \overline{R}_p^{\hat{y}}(T^{\hat{y}}) < \underline{R}_p^{c}(T-T^{\hat{y}}), \forall c \neq \hat{y}, \forall T^{\hat{y}}: 0\leq T^{\hat{y}} \leq T. \nonumber
 \end{align}
 where $\overline{R}_p^{\hat{y}}(T^{\hat{y}})$ and $\underline{R}_p^{c}(T-T^{\hat{y}})$ could be computed based on Equation~\ref{equation-upper-bound} and~\ref{equation-lower-bound}, respectively. Similar to Individual Attack, we use a binary search algorithm to compute $T^*$. The details can be found in Algorithm~\ref{alg:binary-search-GA} in Appendix.}
 
Formally, we use the following theorem to summarize our previous derivations:
\begin{theorem}[Certified Poisoning Size]
    Suppose we have a clean support set $\mathcal{D}$ for $C$-way-$K$-shot classification, where $\mathcal{D}^c \subset \mathcal{D}$ is a subset of $K$ support samples in $\mathcal{D}$ whose labels are $c$ ($c=1,2,\cdots, C$). Given a foundation model $g$ and a distance metric $Dist$, we denote by $d_i^c = Dist(g(\mathbf{x}_{test}), g(\mathbf{x}_i^c))$ for each $\mathbf{x}_i^c \in \mathcal{D}^c$, where $i=1,2,\cdots, K$. Without loss of generality, we assume $d_1^c \leq d_2^c \leq \cdots \leq d_{K}^c$. Suppose $\mathcal{B}(\mathcal{D}, T)$ is a set of all poisoned support sets when an attacker could manipulate up to $T$ support samples in each $\mathcal{D}^c$ \JY{for Individual Attack (or $T$ support samples in total from $\mathcal{D}$ for Group Attack)}. Then, we have the following:
    \begin{align}
        \mathcal{M}(\mathbf{x}_{test};\mathcal{D}) = \mathcal{M}(\mathbf{x}_{test};\mathcal{D}_p), \forall \mathcal{D}_p \in \mathcal{B}(\mathcal{D}, T^*),
    \end{align}
    where $\mathcal{M}(\mathbf{x}_{test};\mathcal{D})$ represents the predicted label of our {\name} for $\mathbf{x}_{test}$ with the support set $\mathcal{D}$. \JY{For Individual Attack,} $T^*$ is the solution to following optimization problem:
    \begin{align}
         T^* = \argmax_{T=1,\cdots, K'} T,
    \text{ }s.t.\text{ } \overline{R}_p^{\hat{y}}(T) < \min_{c\neq \hat{y}} \underline{R}_p^{c}(T).
    \end{align}
    \JY{For Group Attack, $T^*$ is calculated as follows:
       \begin{align}
    &T^* = \argmax_{T=1,\cdots, K'} T, \nonumber \\
   \text{ }s.t.\text{ }& \overline{R}_p^{\hat{y}}(T^{\hat{y}}) < \underline{R}_p^{c}(T-T^{\hat{y}}), \forall c \neq \hat{y}, \forall T^{\hat{y}}: 0\leq T^{\hat{y}} \leq T, 
 \end{align}}
    where $\overline{R}_p^{\hat{y}}(\cdot)$ and $\underline{R}_p^{c}(\cdot)$ are computed based on Equation~\ref{equation-upper-bound} and~\ref{equation-lower-bound}, respectively.
\end{theorem}
\proof{Please refer to Appendix~\ref{proof-of-theorem-1} for the details.}

In our evaluation, following~\cite{levine2020deep,jia2020intrinsic}, we evaluate certified poisoning size $T^*$ by reporting certified accuracy.

\subsection{Tightness of Certified Poisoning Size} We show that our derived certified poisoning size $T^*$ is tight, i.e., there exists an empirical attack with a poisoning size $T^*+1$ such that $\hat{y}$ is not guaranteed to be predicted for the testing input $\mathbf{x}_{test}$. In particular, $T^*$ is the maximum value such that Equation~\ref{final-opt-problem} holds for Individual Attack (or Equation~\ref{final-opt-problem-group} holds for Group Attack), which means the constraint in Equation~\ref{final-opt-problem} \JY{(or~\ref{final-opt-problem-group})} does not hold for $T^*+1$.

We first show the upper bound $\overline{R}_p^{\hat{y}}(T^{\hat{y}})$ could be reached when the attacker can manipulate $T^{\hat{y}}$ support samples from class $\hat{y}$. Without loss of generality, we assume $\mathbf{x}_K^{\hat{y}}$ is the support input where the distance  $Dist(g(\mathbf{x}_{test}), g(\mathbf{x}_K^{\hat{y}}))$ is the largest among the support inputs in $\mathcal{D}^{\hat{y}}$. Given a subset of $K$ clean support samples $\mathcal{D}^{\hat{y}}=\{\mathbf{x}_i^{\hat{y}}, \mathbf{x}_2^{\hat{y}}, \cdots, \mathbf{x}_K^{\hat{y}}\}$. We construct the following poisoned subset for an arbitrary poisoning size $T^{\hat{y}}$: $\mathcal{D}^c_p = \{\Tilde{\mathbf{x}}_i^{\hat{y}}, \Tilde{\mathbf{x}}_2^{\hat{y}}, \cdots, \Tilde{\mathbf{x}}_K^{\hat{y}}\}$, where $\Tilde{\mathbf{x}}_i^{\hat{y}} = \mathbf{x}_K^{\hat{y}}$ when $i=K'+1, K'+2, \cdots, K'+T^{\hat{y}}$, and $\Tilde{\mathbf{x}}_i^{\hat{y}} = \mathbf{x}_i^{\hat{y}}$ otherwise. We could verify that $R^{\hat{y}}_p(T^{\hat{y}}) = \overline{R}_p^{\hat{y}}(T^{\hat{y}})$ with the poisoned subset $\mathcal{D}^{\hat{y}}_p$. \JY{Similarly, when $T^c$ support samples from class $c\neq \hat{y}$ can be manipulated, we could construct poisoned subsets $\mathcal{D}^c_p$ for $c \neq \hat{y}$ such that the lower bound is reached, i.e., $R^c_p(T^c) = \underline{R}_p^{c}(T^c)$.} 

\JY{For Individual Attack,} since the constraint in Equation~\ref{final-opt-problem} does not hold for $T^*+1$ (i.e., $\overline{R}_p^{\hat{y}}(T^*+1) \geq \min_{c\neq \hat{y}} \underline{R}_p^{c}(T^*+1)$), and the lower or upper bounds in the constraint in Equation~\ref{final-opt-problem} could be reached \JY{by letting $T^{\hat{y}} = T^*+1$ and $T^{c} = T^*+1$}, we know there exists an empirical attack with a poisoning size $T^*+1$ such that the label $\hat{y}$ is not guaranteed to be predicted.

\JY{For Group Attack, we know that the constraint in Equation~\ref{final-opt-problem-group} does not hold for $T^*+1$, i.e., $\exists c\neq \hat{y}, \exists T^{\hat{y}},  \overline{R}_p^{\hat{y}}(T^{\hat{y}}) \geq  \underline{R}_p^{c}(T^*+1-T^{\hat{y}})$. From previous analysis, we know that $\overline{R}_p^{\hat{y}}(T^{\hat{y}})$ and $\underline{R}_p^{c}(T^*+1-T^{\hat{y}})$ can be reached by manipulating $T^{\hat{y}}$ support samples from class $\hat{y}$ and $T^c = T^*+1-T^{\hat{y}}$ support samples from class $c$. Therefore, we obtain a empirical attack that manipulates $T^*+1$ support samples in total and causes our defense to make an incorrect prediction.}

Our following theorem summarizes above derivation:
\begin{theorem}[Tightness of Our Certified Poisoning Size]
    There exists an empirical data poisoning attack with a poisoning size $(T^*+1)$ such that the label $\hat{y}$ is not predicted by our {\name} for the testing input $\mathbf{x}_{test}$ or there exist ties.
\end{theorem}

\section{Evaluation}
\label{evaluation}

\subsection{Experimental Setup}
\label{exp:setup}

\myparatight{Datasets} We use three benchmark datasets in our evaluation. In particular, we use CUB200-2011~\cite{triantafillou2019meta}, CIFAR-FS~\cite{bertinetto2018meta}, and \emph{tiered}ImageNet~\cite{ren2019incremental}. 

\begin{itemize}
    \item \myparatight{CUB200-2011}The dataset contains 11,745 bird images, where each image belongs to one of the 200 bird species.
    Moreover, the dataset is split into 8,204 training images from 140 classes, 1,771 validation images from 30 classes, and 1,770 testing images from 30 classes. 

    \item \myparatight{CIFAR-FS} CIFAR-FS is a benchmark dataset used for few-shot classification created from CIFAR-100. In particular, the dataset contains 38,400 training images of 64 classes, 9,600 validation images of 16 classes, and 12,000 testing images of 20 classes. 

    \item \myparatight{\emph{tiered}ImageNet}This dataset is a subset of ImageNet, which contains 779,165 images from 608 classes.
    In particular, it contains 448,695 training images from 351 classes, 124,261 validation images from 97 categories, and 206,209 testing images from 160 classes. 
\end{itemize}

\myparatight{Foundation models}We use two foundation models, i.e., CLIP~\cite{radford2021learning} and DINOv2~\cite{oquab2023dinov2}, which are released by OpenAI and Meta, respectively. In particular, CLIP is pre-trained on 400 million image-text pairs collected from the Internet, and DINOv2 is pre-trained on a curated dataset of 142 million images.  
Given an image as input, those foundation models output a feature vector for the given image. 

\myparatight{Few-shot classification setup}We follow the setting in the previous work on few-shot classification~\cite{ProtoNet_implementation,pautov2022smoothed,sung2018learning,yoon2019tapnet}. Specifically, we randomly select 20 batches of samples from the test dataset, where each batch contains $C\cdot K$ support samples ($K$ support samples for each of the $C$ classes) and $C$ testing samples (one testing sample for each of the $C$ classes).  
For each batch, we use the $C\cdot K$ support samples to build/train a few-shot learning classifier and evaluate it on $C$ testing samples. We report the average results on testing inputs in 20 batches. We note that in the testing phase, existing studies~\cite{ProtoNet_implementation,pautov2022smoothed,sung2018learning} select both support samples and testing inputs from the testing dataset because the samples in the training and validation datasets have different classes from those in the testing dataset. Unless otherwise mentioned, we set $C=5$ and $K=5$, i.e., 5-way-5-shot classification.

\myparatight{Compared methods} We compare our {\name} with the following baselines:
\begin{itemize}
    \item \myparatight{State-of-the-art few-shot classification methods} We compare our {\name} with state-of-the-art few-shot classification methods, including \emph{ProtoNet}~\cite{snell2017prototypical} and \emph{Linear Probing (LP)}~\cite{chen2019closer}. We note that ProtoNet does not have any hyperparameters. LP trains a linear classifier using the feature vectors of support samples. 
    Following CLIP~\cite{CLIP_implementation}, we use the Limited-memory BFGS solver in the scikit-learn package~\cite{sklearn} to optimize the linear classifier while the maximum number of iterations is set to 1,000. Additionally, instead of training a linear classifier, we also consider training a fully connected neural network (we call this method \emph{FCN}). Due to space reasons, we put the comparison of our {\name} with FCN in Figure~\ref{fig-finetune-all} in Appendix. 

    \item \myparatight{State-of-the-art provable defenses} We compare {\name} with other provable robust methods for few shot learning, including Bagging~\cite{jia2020intrinsic}, DPA~\cite{levine2020deep}, and $k$-NN~\cite{jia2022certified}. Given a support set, Bagging first builds $N_{B}$ base classifiers, where each base classifier is built upon a set of $M_B$ support samples selected uniformly at random from the support set. Given a testing input, Bagging takes a majority vote over the labels predicted by those base classifiers for the given testing input. DPA divides a training dataset into $N_D$ disjoint sub-datasets and uses each of them to build a base classifier. Similar to Bagging, given a testing input, DPA takes a majority vote over the labels predicted by those $N_D$ base classifiers for the given testing input to make a prediction. Given a testing input and a support set, $k$-NN finds the $k$ nearest neighbors of the testing input in the training dataset using a distance metric and takes a majority vote over the labels of those $k$ nearest neighbors as the given testing input.

    We note that both Bagging and DPA need to build multiple base classifiers, which could be very inefficient in practice. As the set of support samples could be different for different testing inputs, we need to repeat this process multiple times. To address the challenge, we use ProtoNet to build each base classifier for Bagging and DPA as it achieves state-of-the-art performance. For $C$-way and $K$-shot few-shot learning task, we set $N_B = 1000$ and $M_B=C$ for Bagging and we set $N_D=K$ for DPA. Note that Bagging (or DPA) is more robust when $M_B$ is smaller (or $N_D$ is larger). 
    We set those hyper-parameters for Bagging and DPA such that the set of support samples used to train each base classifier contains one support sample from each class on average. We set $k=K$ for $k$-NN since each class has at most $K$ support samples. 
\end{itemize}

\myparatight{Evaluation metrics} Following previous certified defenses~\cite{jia2020intrinsic,levine2020deep,jia2022certified}, we use \emph{certified accuracy} as the evaluation metric. Specifically, certified accuracy measures the classification accuracy of a few-shot learning method under arbitrary data poisoning attacks.
Formally, the certified accuracy is defined as the fraction of testing inputs whose labels are correctly predicted as well as whose certified poisoning sizes are at least $T$. Given a set of testing inputs $\mathcal{D}_{test}$, we can compute certified accuracy as follows:
    \begin{align}
        &\text{Certified accuracy} \nonumber \\
        =& \frac{\sum_{(\mathbf{x}_{test}, y_{test}) \in \mathcal{D}_{test}}\mathbb{I}(\hat{y}_{test}=y_{test})\cdot \mathbb{I}(T^*(\mathbf{x}_{test})\geq T)}{|\mathcal{D}_{test}|},
    \end{align}
    where $\hat{y}_{test}$ and $y_{test}$ are the predicted label and the ground truth label of the testing input $\mathbf{x}_{test}$,  $T^*(\mathbf{x}_{test})$ represents the certified poisoning size of $\mathbf{x}_{test}$, $\mathbb{I}$ is the indicator function, and $|\mathcal{D}_{test}|$ is the total number of testing inputs in $\mathcal{D}_{test}$. Certified accuracy is a lower bound of the classification accuracy that a few-shot classification method achieves on the testing dataset $\mathcal{D}_{test}$ under \emph{arbitrary} data poisoning attacks with a poisoning size $T$.

    We note that existing state-of-the-art few-shot classification methods such as ProtoNet~\cite{snell2017prototypical} and Linear Probing (LP)~\cite{chen2019closer} cannot provide certified robustness guarantee against data poisoning attacks. In other words, we cannot calculate the certified accuracy for them. To address the challenge, we calculate the \emph{empirical accuracy} of those methods on a testing dataset under an empirical data poisoning attack, which is an upper bound of their certified accuracy under arbitrary attacks. 
    In particular, we extend feature collision attack~\cite{shafahi2018poison} to compute the empirical accuracy for 
    ProtoNet~\cite{snell2017prototypical} and Linear Probing~\cite{chen2019closer}. Given a test sample $\mathbf{x}_{test}$ whose ground truth label is ${y}$, an attacker aims to craft 
    $T$ poisoned support samples for each of the $C$ classes for Individual Attack (or $T$ poisoned support samples in total across the $C$ classes for Group Attack) such that the testing input $\mathbf{x}_{test}$ is misclassified by the few-shot classifier built upon the poisoned support samples.  We reuse the notations used in Section~\ref{method}. 
    
    \emph{Individual Attack:} For support samples in $\mathcal{D}^{c}$ for $c\neq y$ (i.e., the set of $K$ support samples whose ground truth label is not $y$), we randomly select $T$ samples and then craft poisoned samples via performing feature-collision attacks. In particular, we solve the optimization problem in Equation~\ref{eqn-collision} to craft each poisoned support sample. Note that we set $\lambda=0$ when solving the optimization problem to consider a strong empirical attack. For support samples in $\mathcal{D}^{y}$, our key idea is to craft $T$ poisoned support samples such that their feature vectors are far away from that of the testing input $\mathbf{x}_{test}$. In particular, for LP, we directly modify $T$ support samples (without loss of generality, denoted by $(\mathbf{x}^y_1,y),\cdots,(\mathbf{x}^y_T,y)$) in $\mathcal{D}^{y}$ to $(\mathbf{x}^c_1,y),\cdots,(\mathbf{x}^c_T,y)$, where $c\neq y$ is an arbitrary class. As for ProtoNet,
     we use the Projected Gradient Descent (PGD)~\cite{madry2017towards} algorithm to find a poisoned image $\mathbf{x}^*$ such that $Dist(g(\mathbf{x}_{test}),g(\mathbf{x}^*))$ is maximized, where $g$ is the foundation model. Then, we poison each of the $T$ support samples in $\mathcal{D}^{y}$ by replacing it with $(\mathbf{x}^*,y)$. Note that the label of each poisoned support sample remains unchanged. So the number of support samples in each class is still $K$ after the attack. 
    
   { \color{black} {\emph{Group Attack:} We first find the easiest target label $c^* \neq y$ for the feature collision attack. In particular, we select $c^*$ as the label whose prototype is closest to the feature vector of a testing input $\mathbf{x}_{test}$, i.e., $c^* = \argmin_{c\neq y} Dist (g(\mathbf{x}_{test}),\frac{1} {K}\sum_{i=1}^{K}g(\mathbf{x}_i^c))$, where $g(\mathbf{x}_i^c)$ is the feature vector produced by the foundation model $g$ for the support input $\mathbf{x}_i^c$.
    Then, for support samples in $\mathcal{D}^{c^*}$, we randomly select $T$ samples and craft poisoned samples by solving Equation~\ref{eqn-collision} with $\lambda=0$.}}
    
\myparatight{Parameter settings} Our {\name} has the following hyper-parameters: $K'$ and the distance metric $Dist$. Unless otherwise mentioned, we set $K'=\lfloor (K-1)/2 \rfloor$ and use squared $\ell_2$-distance as $Dist$. 

\begin{figure*}
\vspace{-10mm}
\centering
\begin{minipage}[b]{0.29\textwidth}
{\includegraphics[width=1\textwidth]{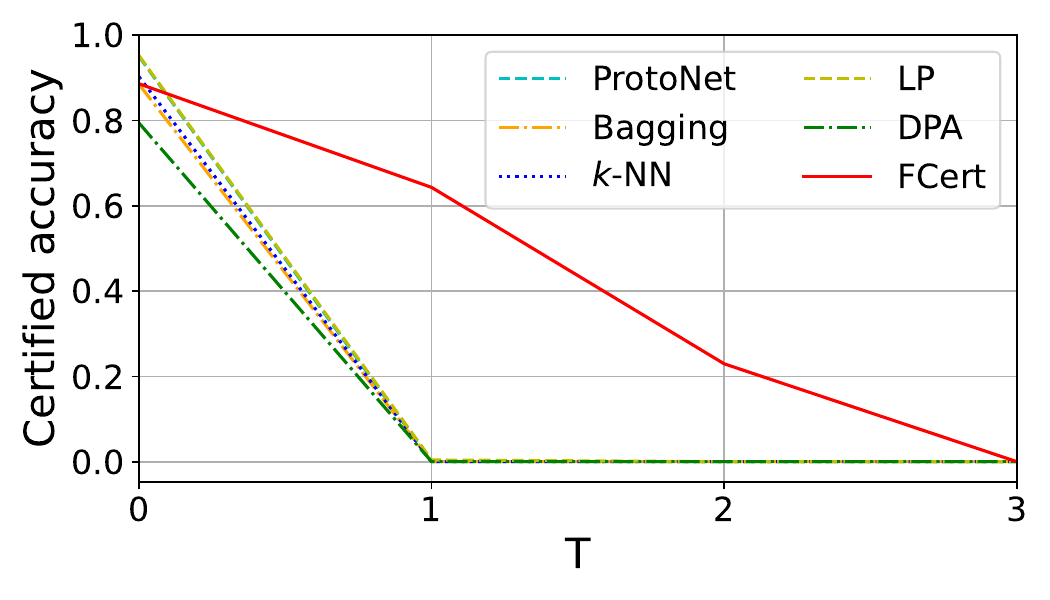}}
\end{minipage}
\vspace{0.3mm}
{\includegraphics[width=0.29\textwidth]{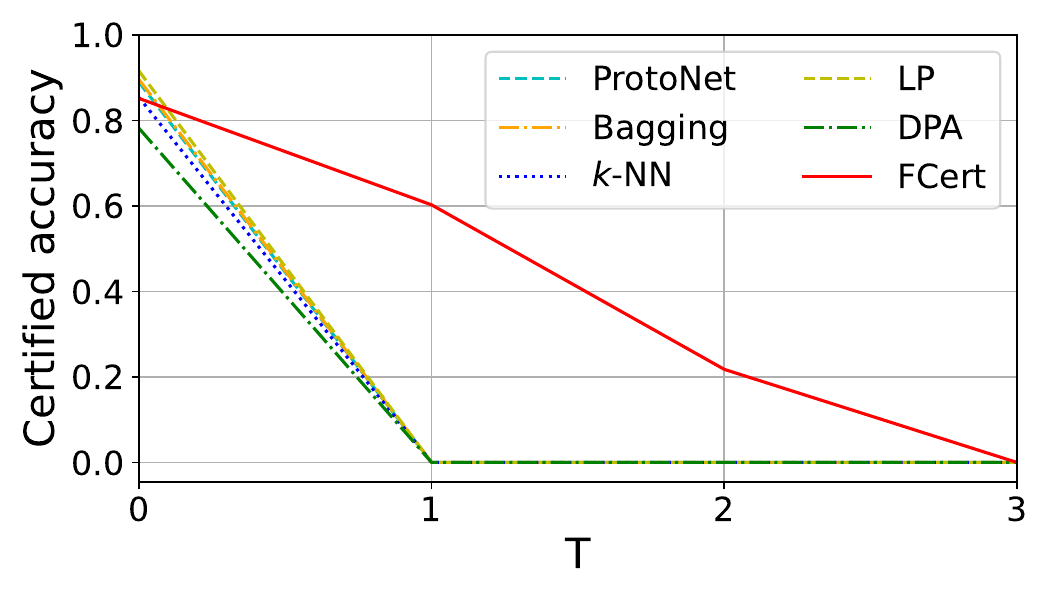}}\vspace{-0.3mm}
{\includegraphics[width=0.29\textwidth]{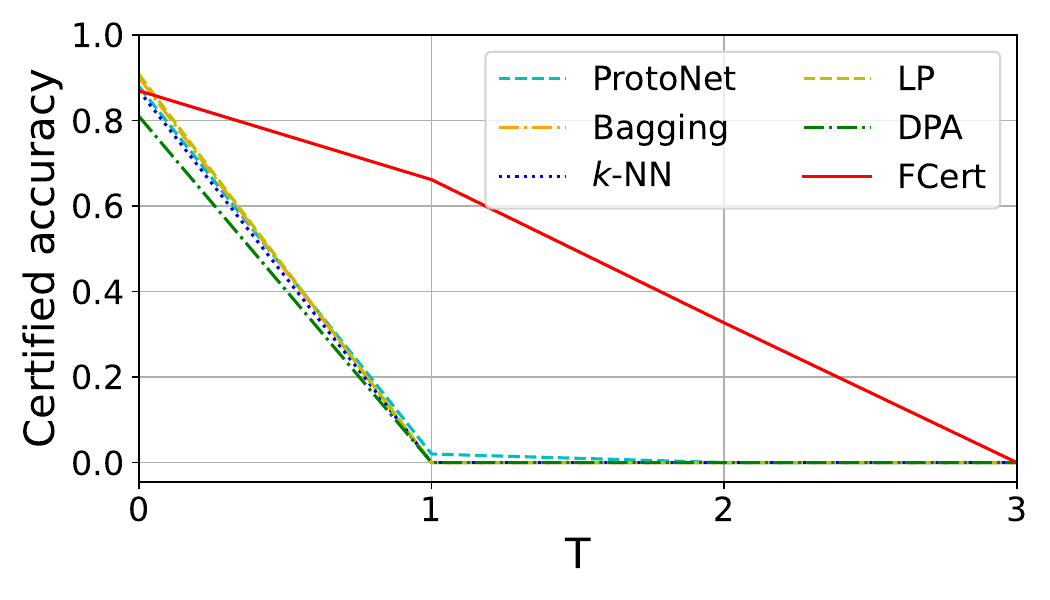}}\vspace{-0.3mm}
{\includegraphics[width=0.29\textwidth]{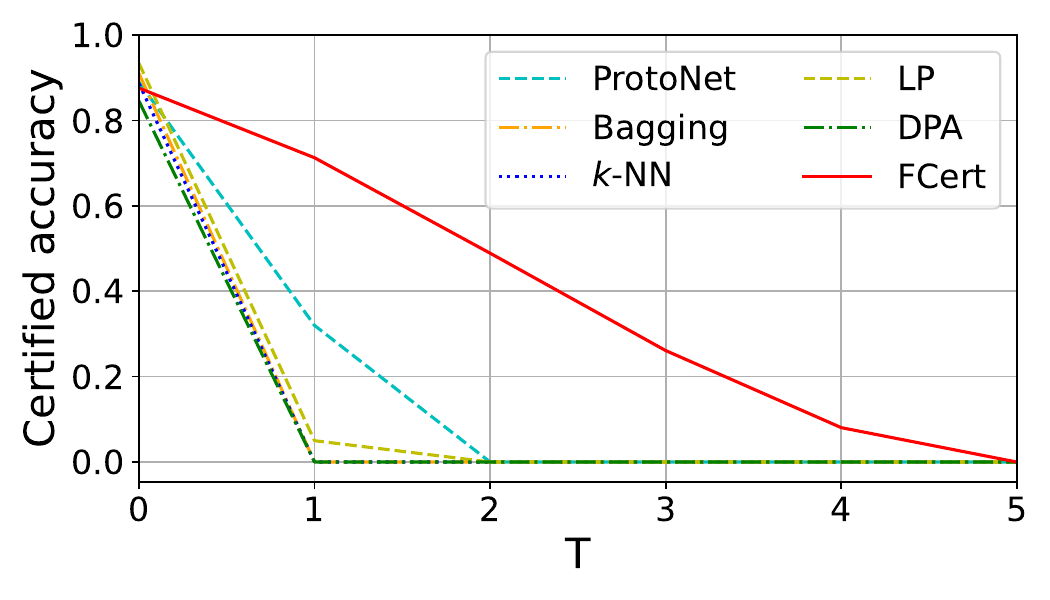}}\vspace{0.4mm}
{\includegraphics[width=0.29\textwidth]{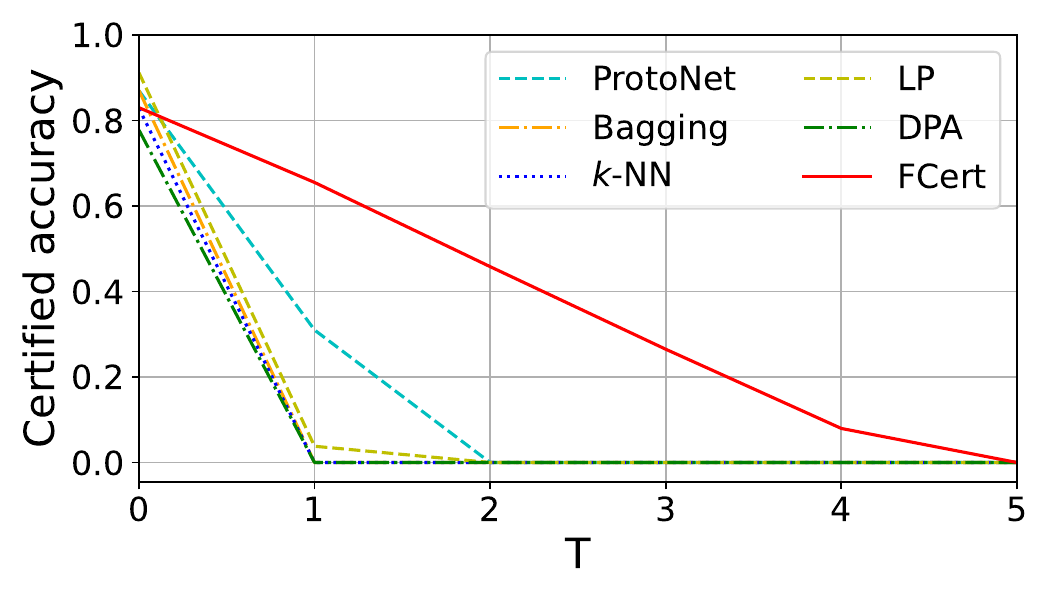}}\vspace{0.4mm}
{\includegraphics[width=0.29\textwidth]{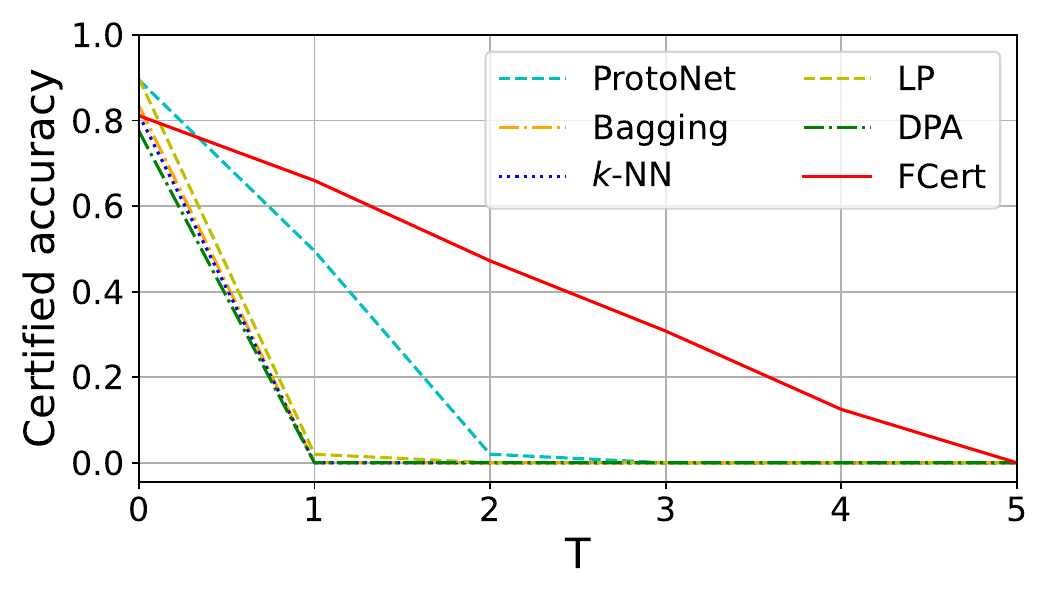}}\vspace{0.4mm}
\hspace*{-0.0005\textwidth}
\subfloat[CUB200-2011]{\includegraphics[width=0.29\textwidth]{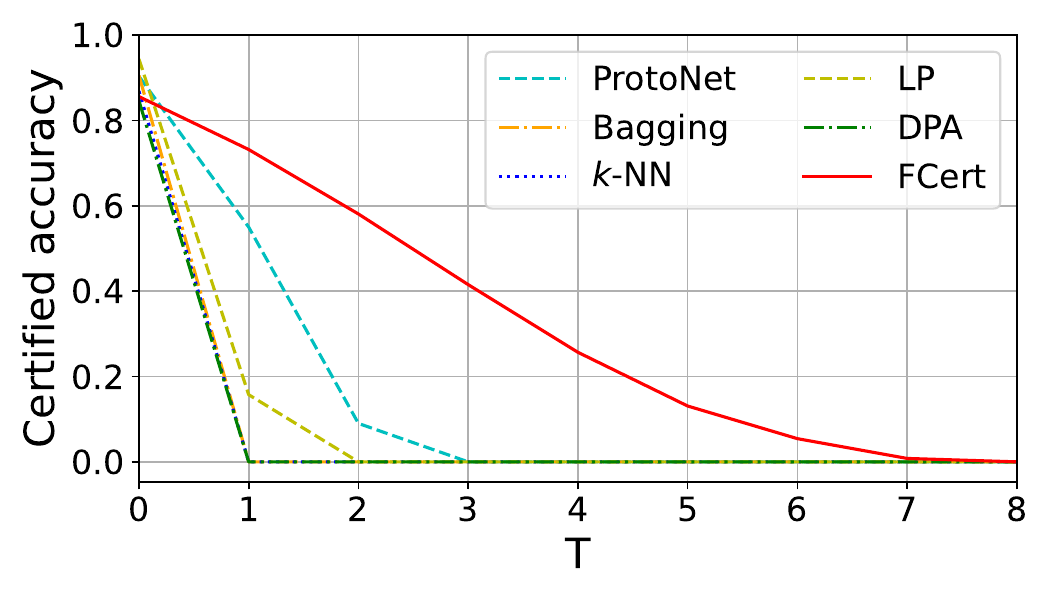}}\vspace{1mm}
\subfloat[CIFAR-FS]{\includegraphics[width=0.29\textwidth]{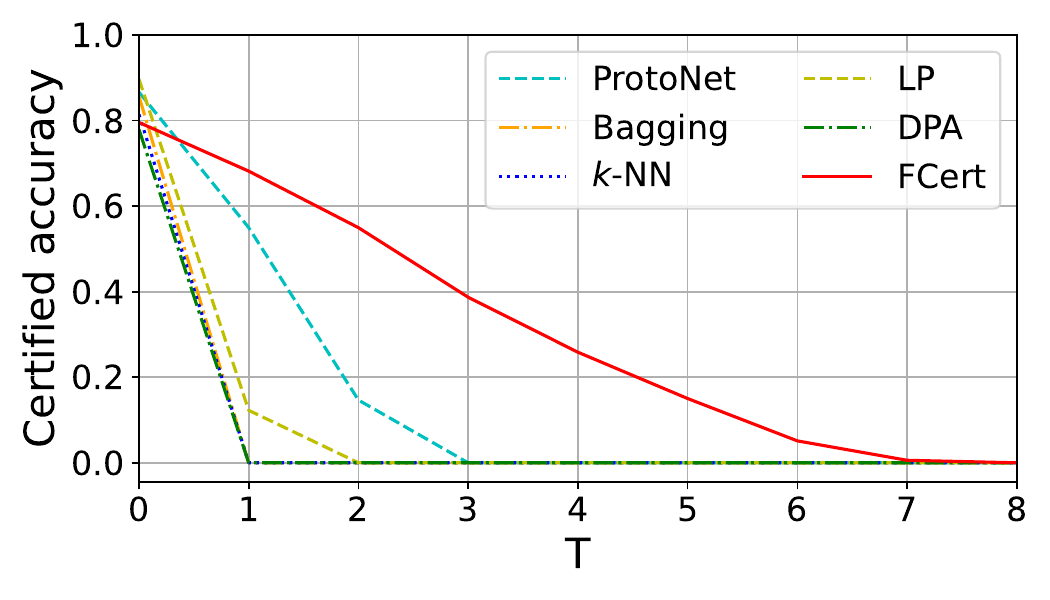}}\vspace{1mm}
\subfloat[\emph{tiered}ImageNet]{\includegraphics[width=0.29\textwidth]{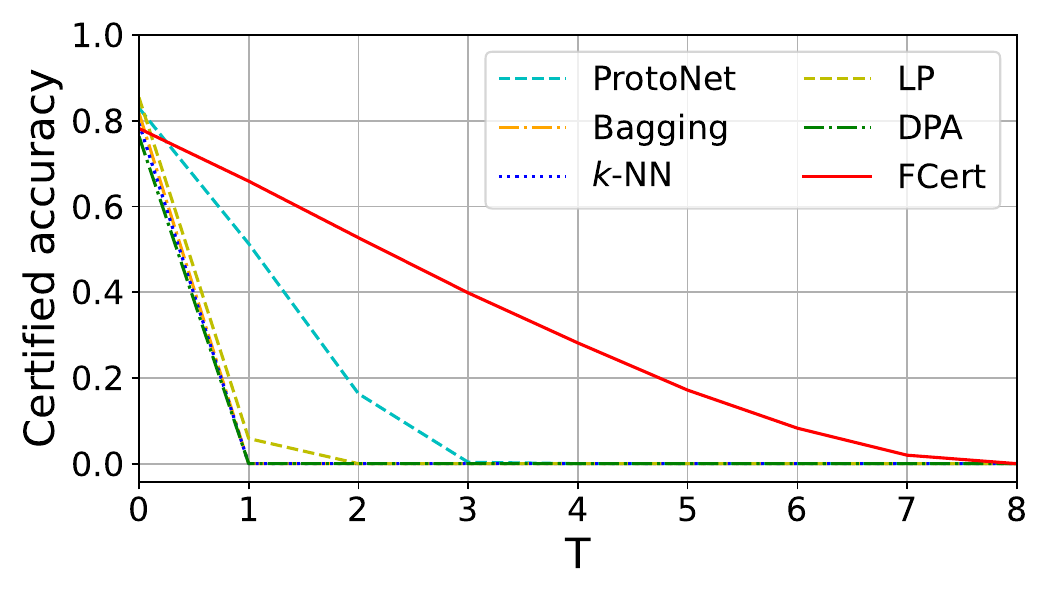}}\vspace{1mm}
\vspace{-4mm}
\caption{Comparing the certified accuracy of {\name} with existing provable defenses (or empirical accuracy of existing few-shot learning methods) for $C$-way-$K$-shot few-shot classification with CLIP. The attack type is individual attack. $K$ = 5, $C$ = 5 (first row); $K$ = 10, $C$ = 10 (second row); $K$ = 15, $C$ = 15 (third row). $T$ is poisoning size. 
}
\label{compare-certified-accuracy-clip-ind}
\vspace{0mm}
\end{figure*}

\begin{figure*}
\vspace{0mm}
\centering
{\includegraphics[width=0.29\textwidth]{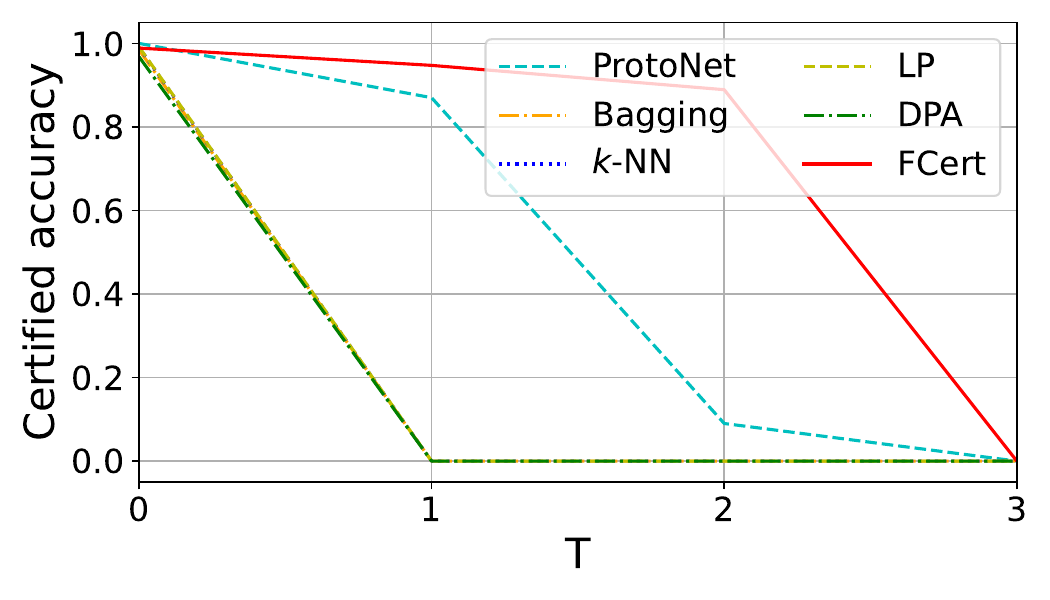}}\vspace{-0.15mm}
{\includegraphics[width=0.29\textwidth]{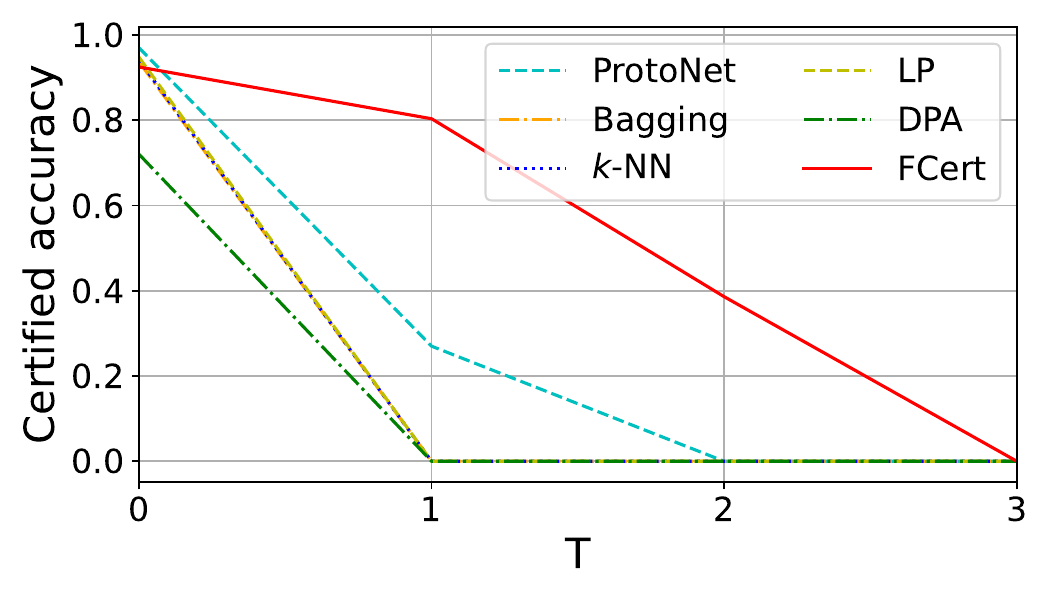}}\vspace{-0.15mm}
{\includegraphics[width=0.29\textwidth]{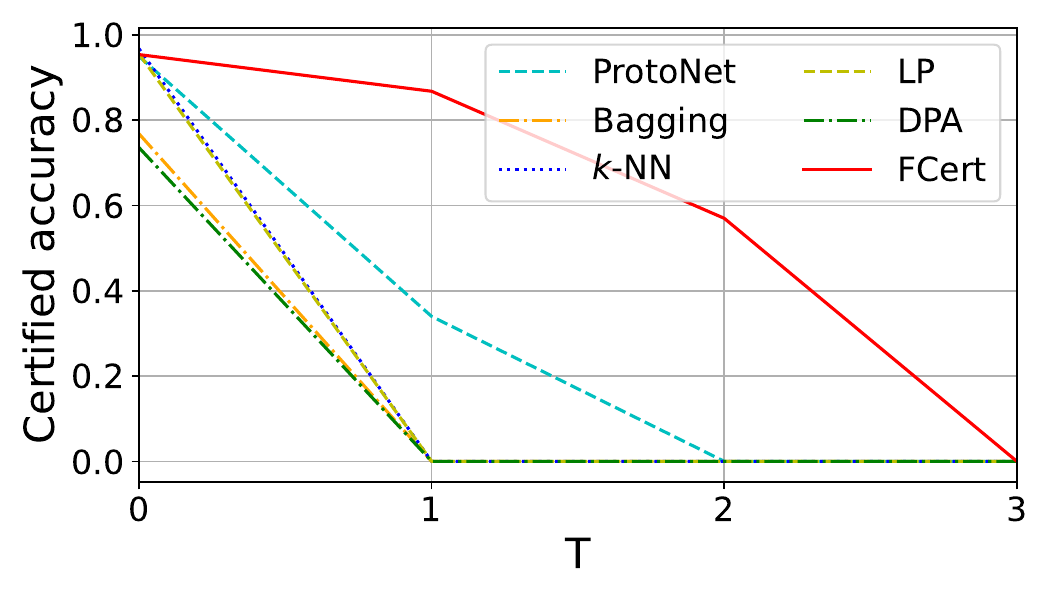}}\vspace{-0.15mm}

{\includegraphics[width=0.29\textwidth]{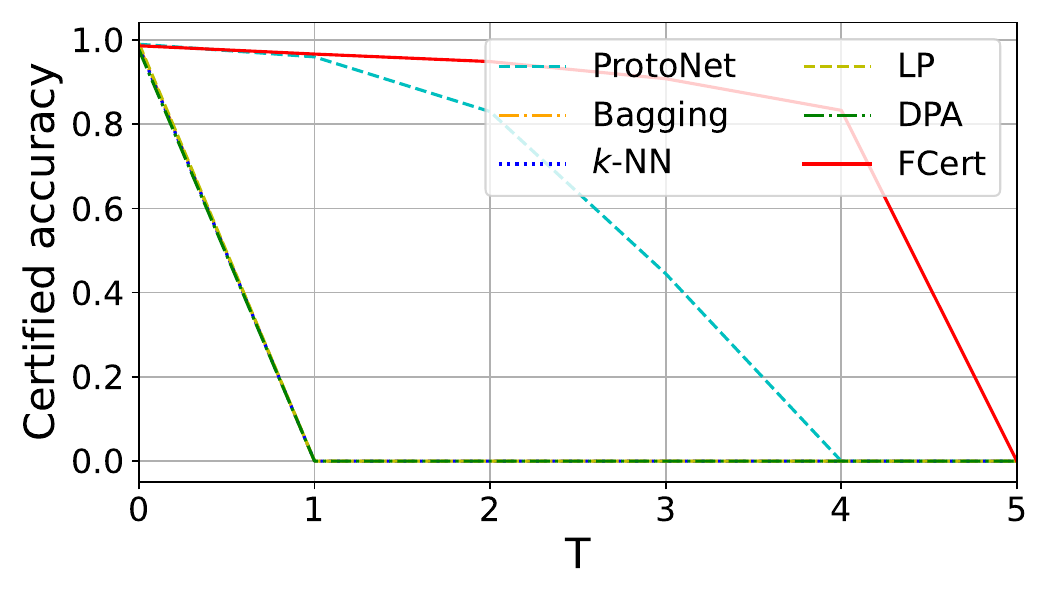}}\vspace{0.4mm}
{\includegraphics[width=0.29\textwidth]{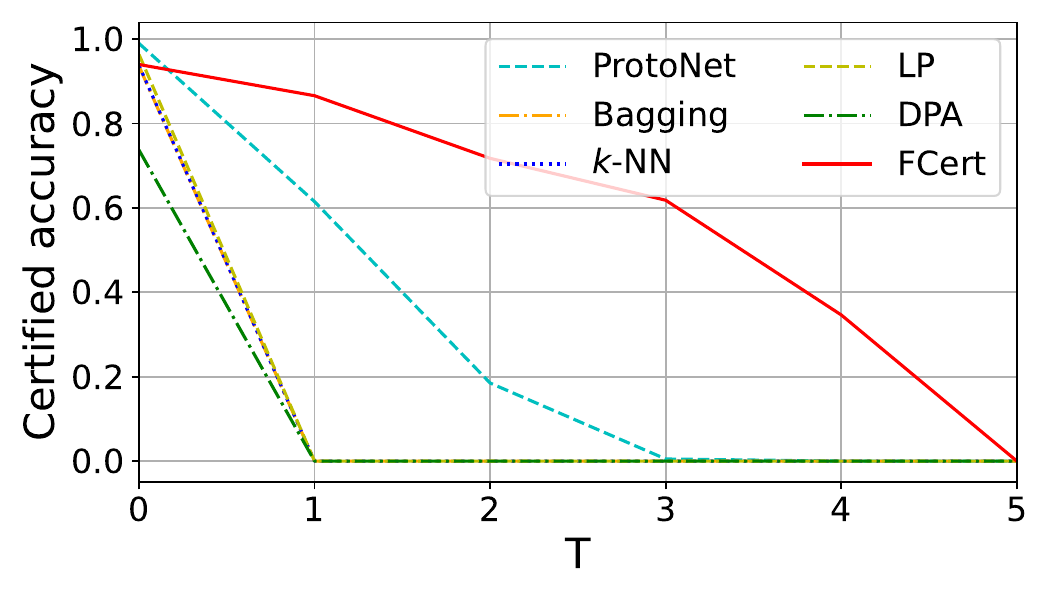}}\vspace{0.4mm}
{\includegraphics[width=0.29\textwidth]{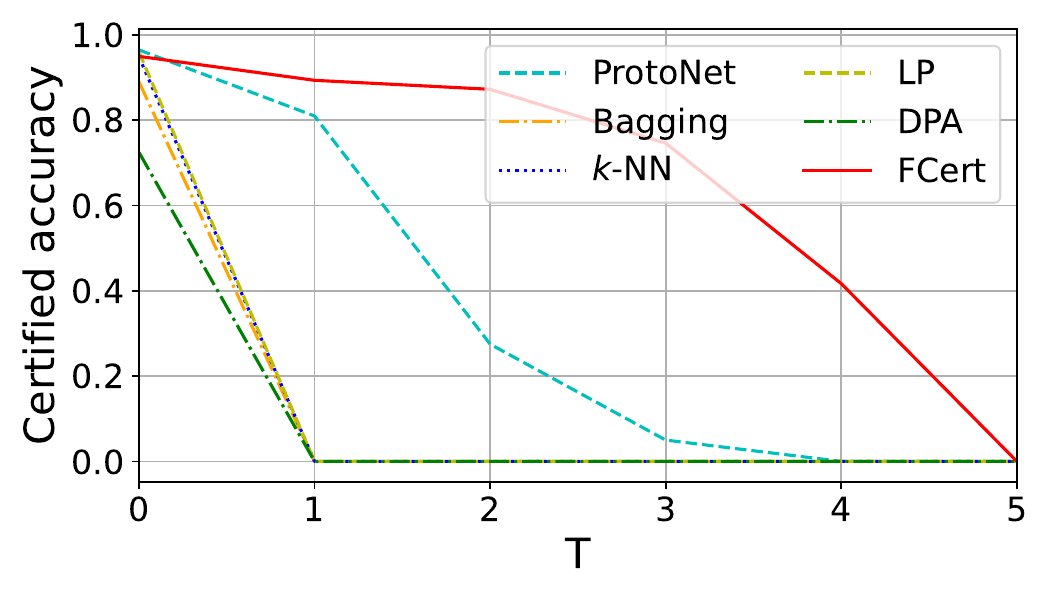}}\vspace{0.4mm}

\hspace*{0.0005\textwidth}
\subfloat[CUB200-2011]{\includegraphics[width=0.29\textwidth]{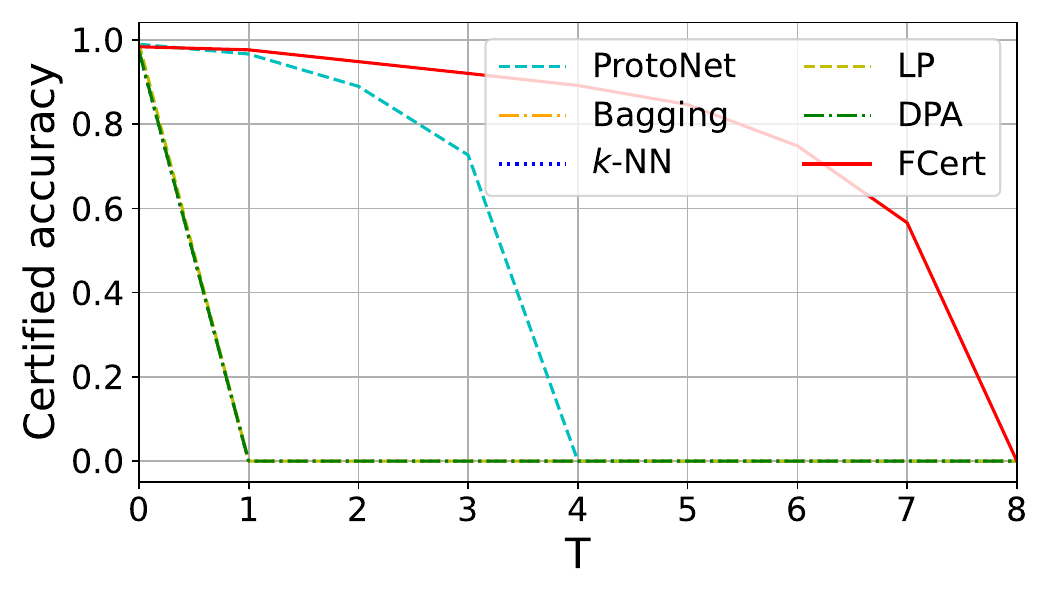}}\vspace{1mm}
\subfloat[CIFAR-FS]{\includegraphics[width=0.29\textwidth]{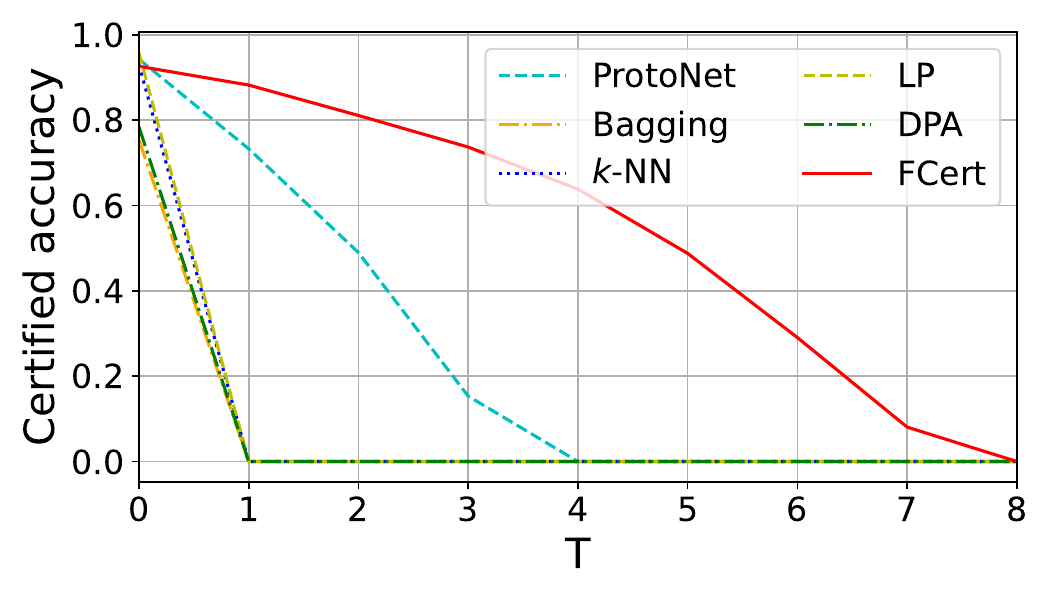}}\vspace{1mm}
\subfloat[\emph{tiered}ImageNet]{\includegraphics[width=0.29\textwidth]{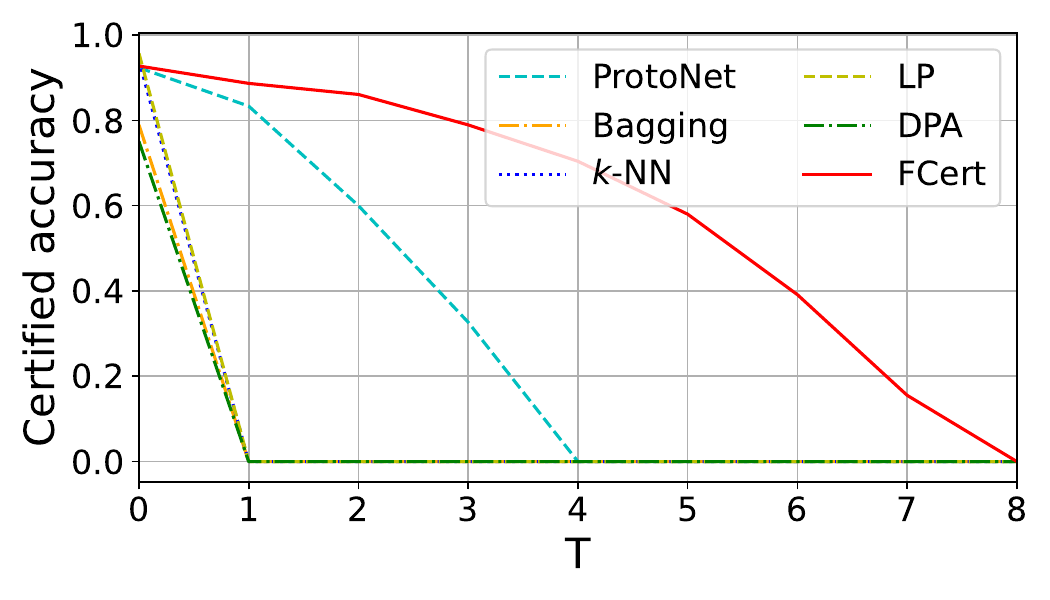}}\vspace{1mm}
\vspace{-4mm}
\caption{Comparing the certified accuracy of {\name} with existing provable defenses (or empirical accuracy of existing few-shot learning methods) for $C$-way-$K$-shot few-shot classification with DINOv2. The attack type is individual attack. $K$ = 5, $C$ = 5 (first row); $K$ = 10, $C$ = 10 (second row); $K$ = 15, $C$ = 15 (third row). $T$ is poisoning size.  
}
\label{compare-certified-accuracy-dinov2-ind}
\vspace{0mm}
\end{figure*}

\begin{figure*}
\vspace{-10mm}
\centering
\begin{minipage}[b]{0.29\textwidth}
{\includegraphics[width=1\textwidth]{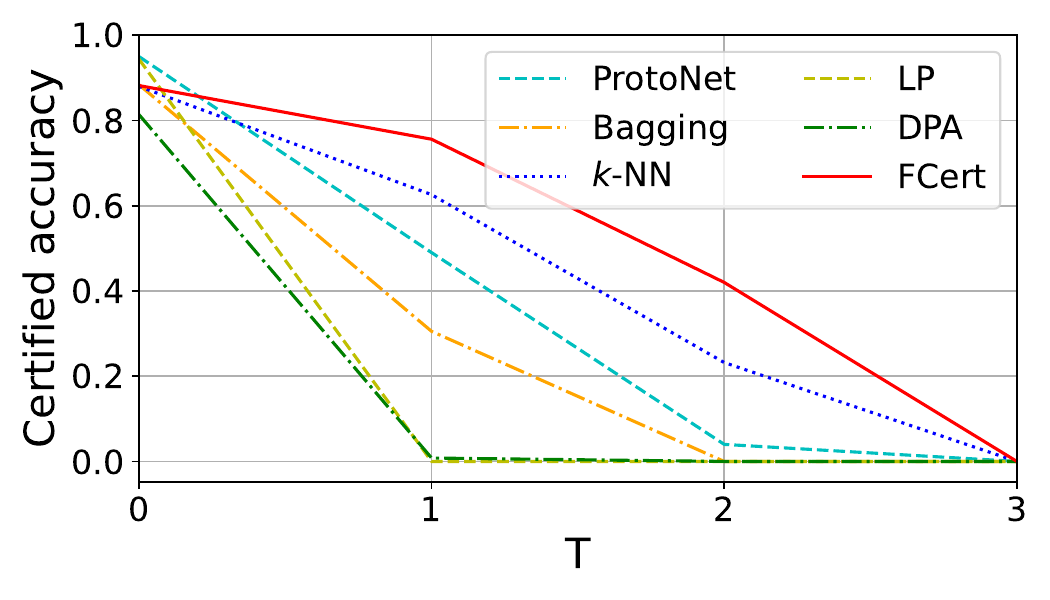}}
\end{minipage}
\vspace{0.3mm}
{\includegraphics[width=0.29\textwidth]{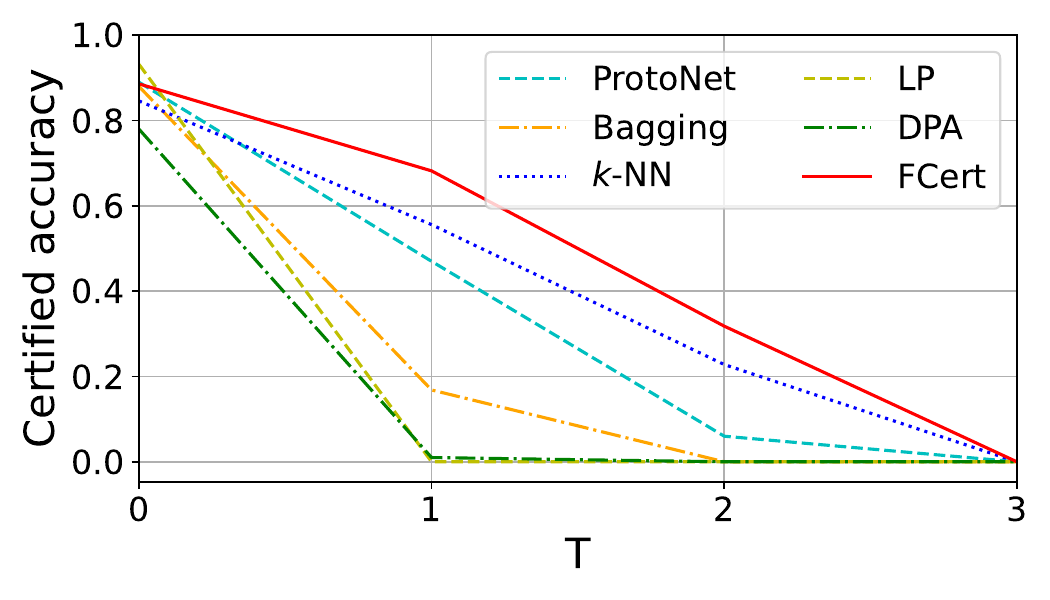}}\vspace{-0.3mm}
{\includegraphics[width=0.29\textwidth]{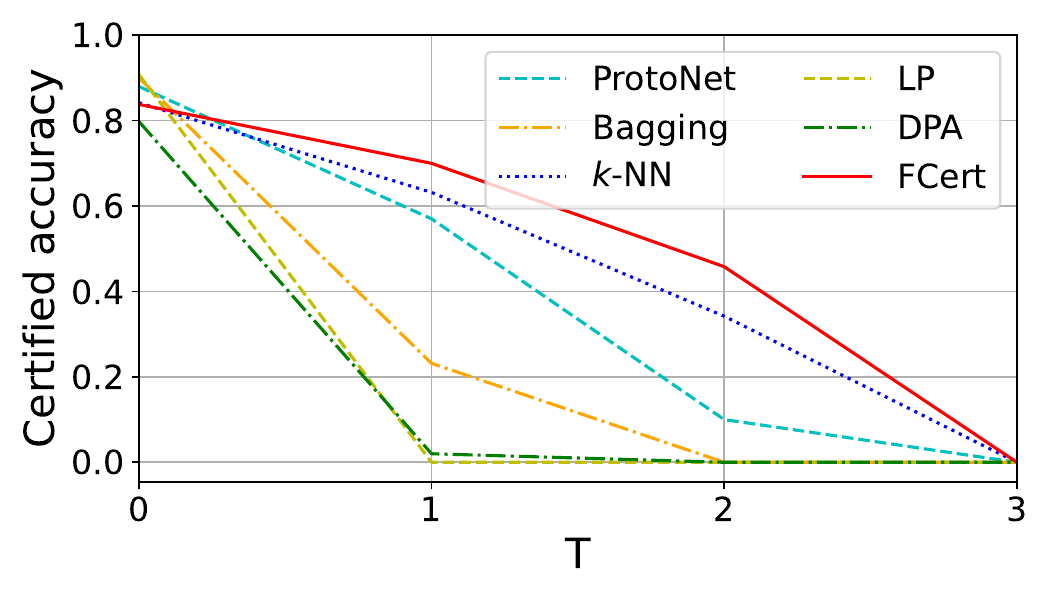}}\vspace{-0.3mm}
{\includegraphics[width=0.29\textwidth]{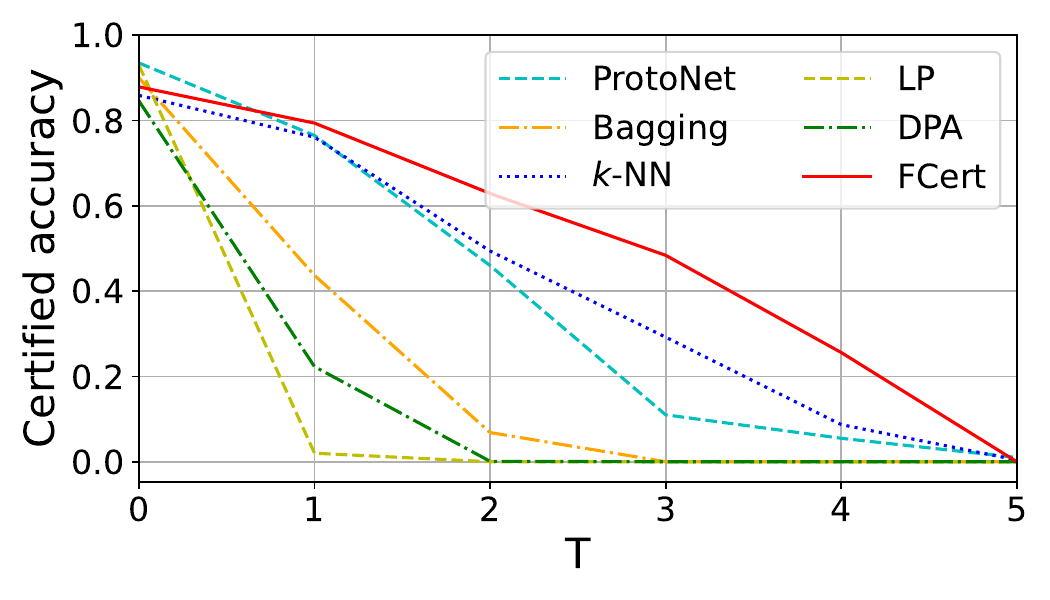}}\vspace{0.4mm}
{\includegraphics[width=0.29\textwidth]{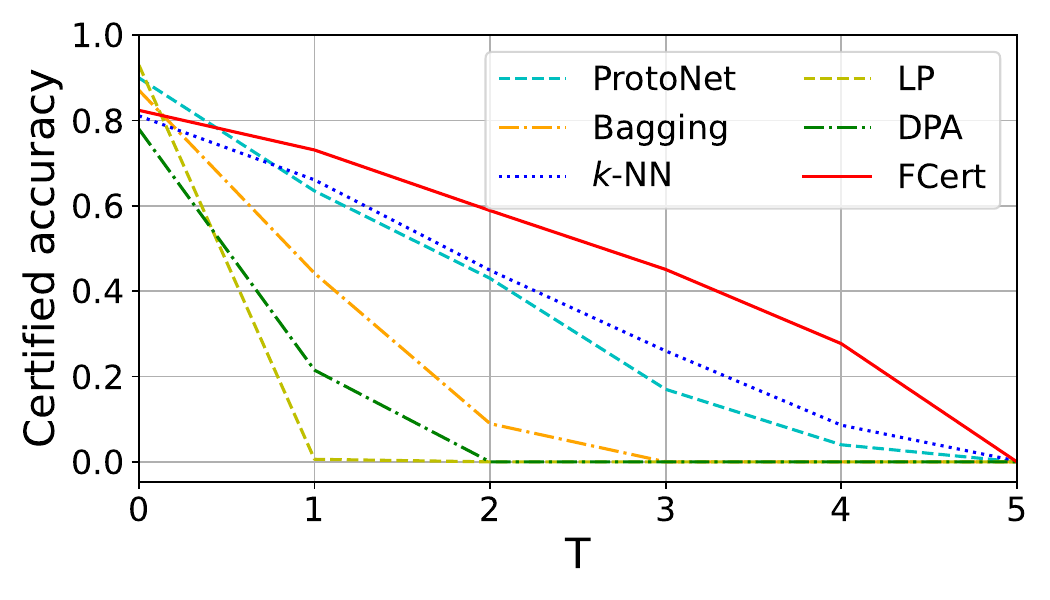}}\vspace{0.4mm}
{\includegraphics[width=0.29\textwidth]{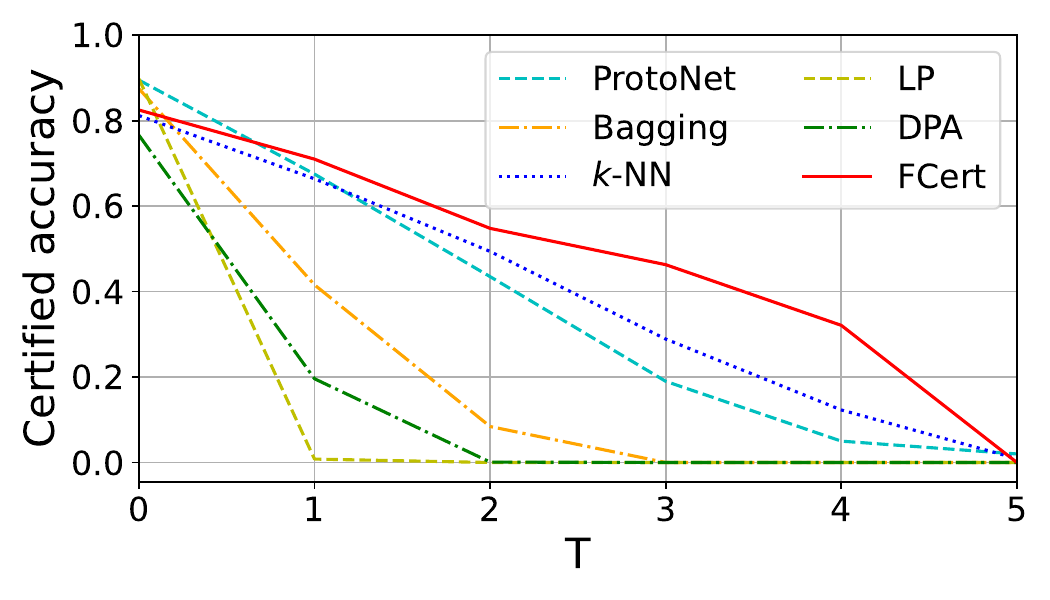}}\vspace{0.4mm}
\hspace*{-0.0005\textwidth}
\subfloat[CUB200-2011]{\includegraphics[width=0.29\textwidth]{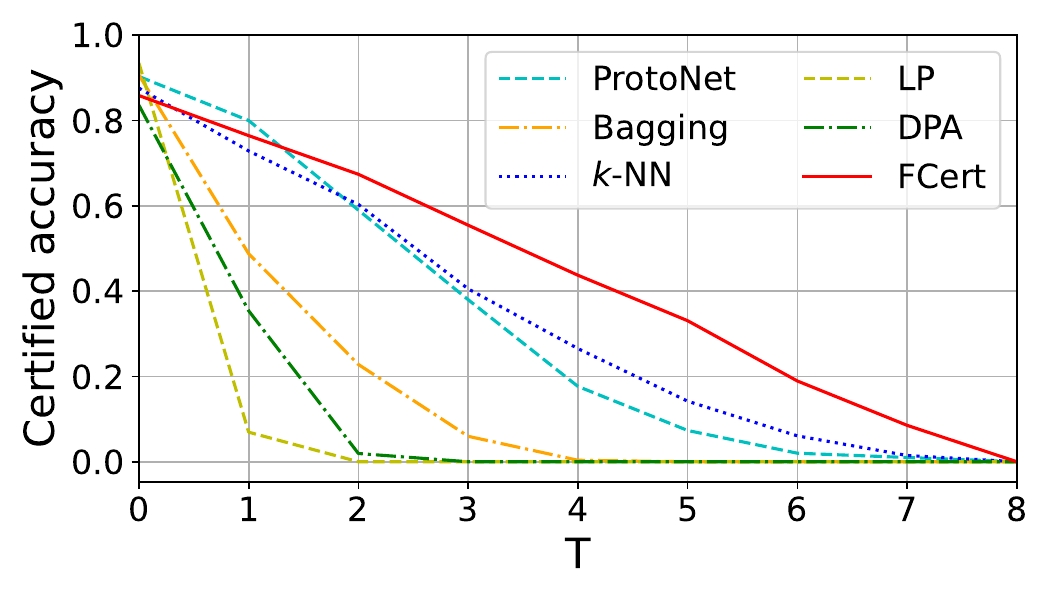}}\vspace{1mm}
\subfloat[CIFAR-FS]{\includegraphics[width=0.29\textwidth]{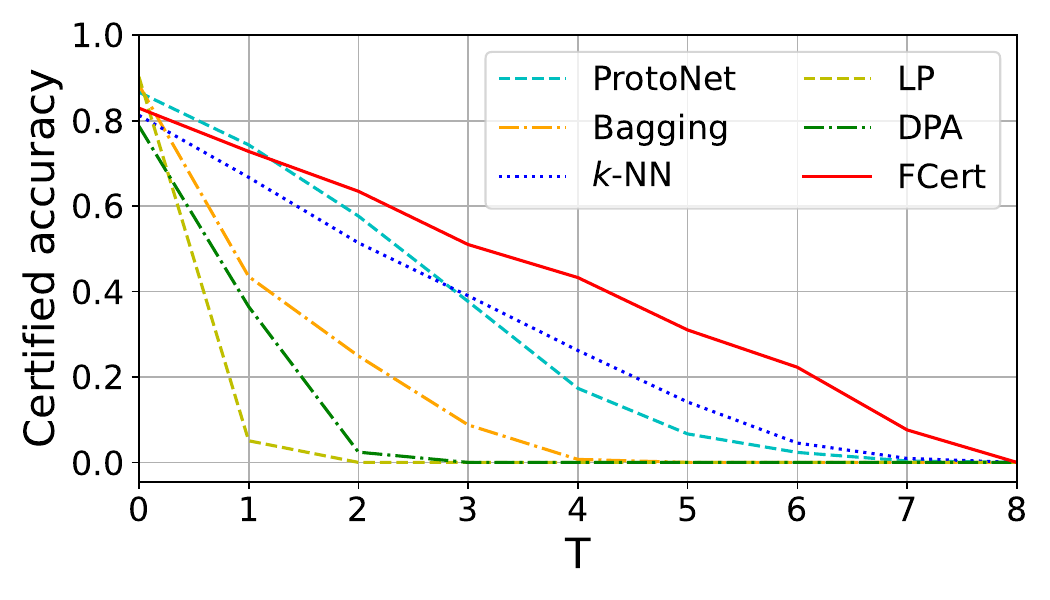}}\vspace{1mm}
\subfloat[\emph{tiered}ImageNet]{\includegraphics[width=0.29\textwidth]{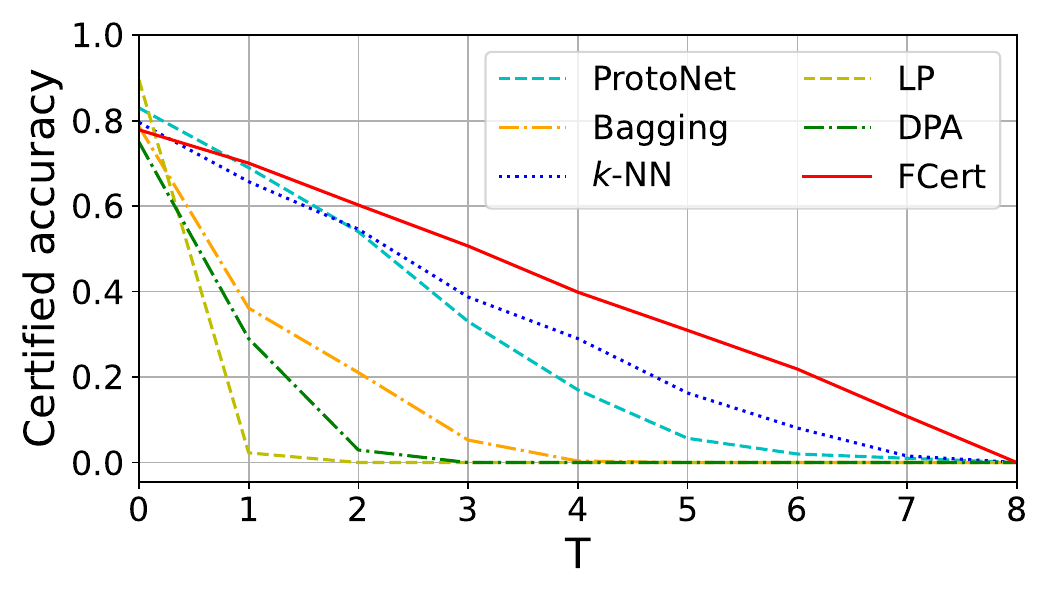}}\vspace{1mm}
\vspace{-4mm}
\caption{Comparing the certified accuracy of {\name} with  existing provable defenses (or empirical accuracy of exisiting few-shot learning methods) for $C$-way-$K$-shot few-shot classification with CLIP. The attack type is group attack. $K$ = 5, $C$ = 5 (first row); $K$ = 10, $C$ = 10 (second row); $K$ = 15, $C$ = 15 (third row). $T$ is poisoning size. 
}
\label{compare-certified-accuracy-clip-group}
\vspace{0mm}
\end{figure*}

\begin{figure*}
\vspace{0mm}
\centering
\begin{minipage}[b]{0.29\textwidth}
{\includegraphics[width=1\textwidth]{figs/empirical_cubirds200_k=5_c=5_encoder=CLIP_attack-type=joint.pdf}}
\end{minipage}
\vspace{0.3mm}
{\includegraphics[width=0.29\textwidth]{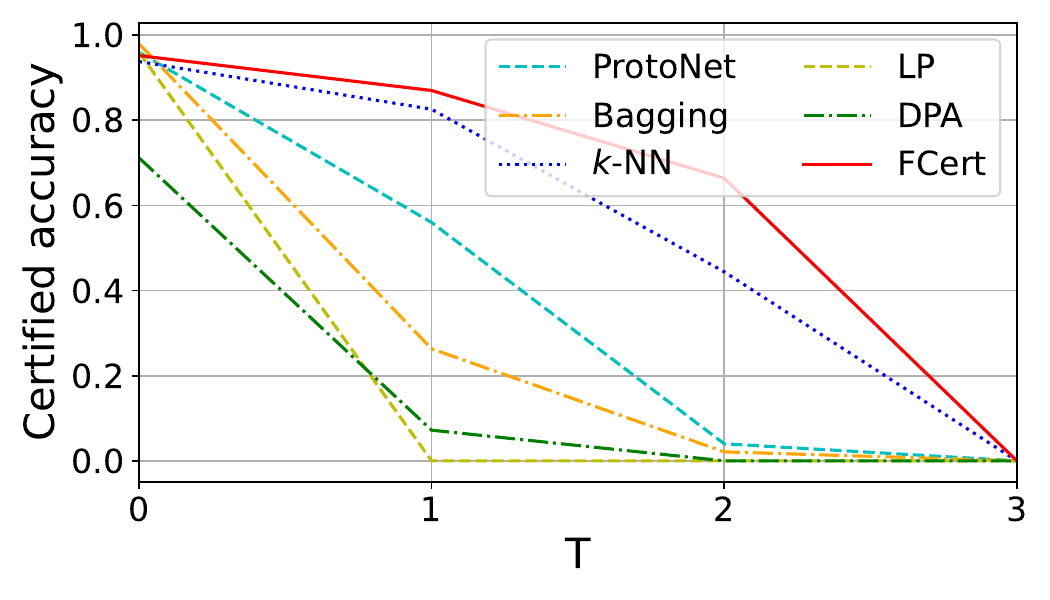}}\vspace{-0.3mm}
{\includegraphics[width=0.29\textwidth]{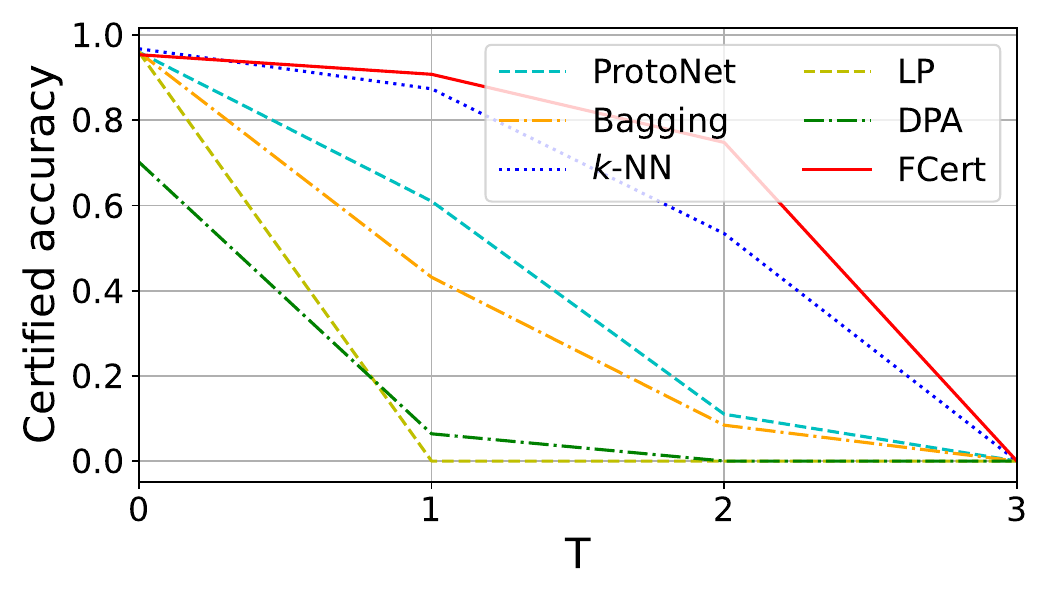}}\vspace{-0.3mm}
{\includegraphics[width=0.29\textwidth]{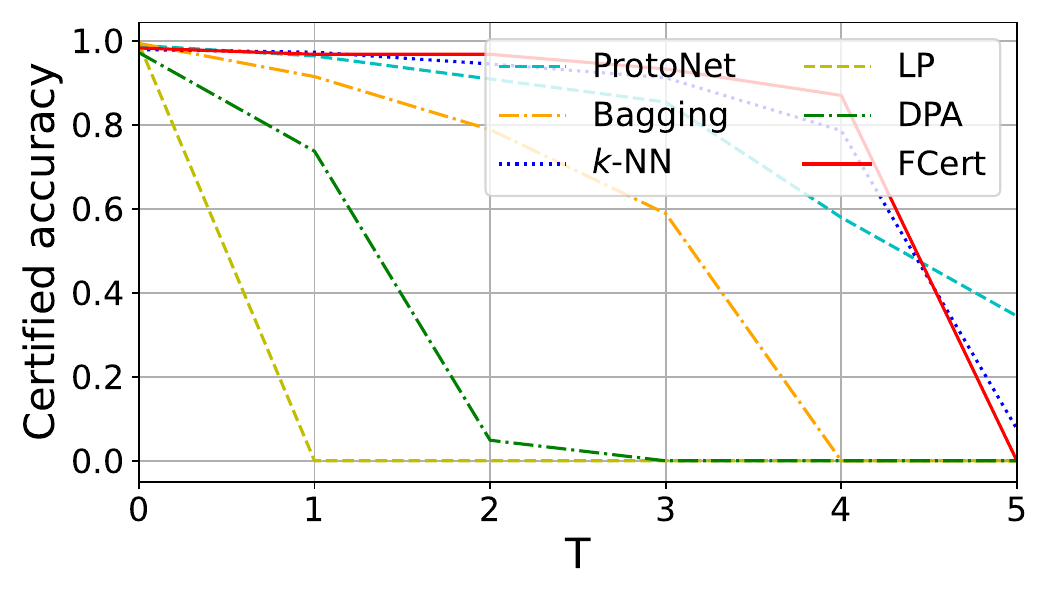}}\vspace{0.4mm}
{\includegraphics[width=0.29\textwidth]{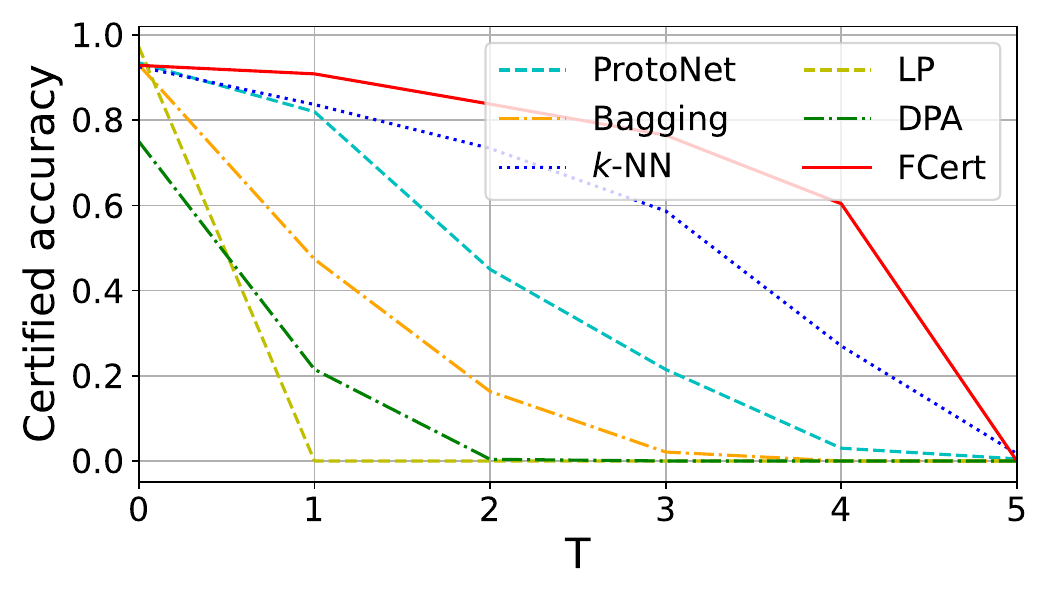}}\vspace{0.4mm}
{\includegraphics[width=0.29\textwidth]{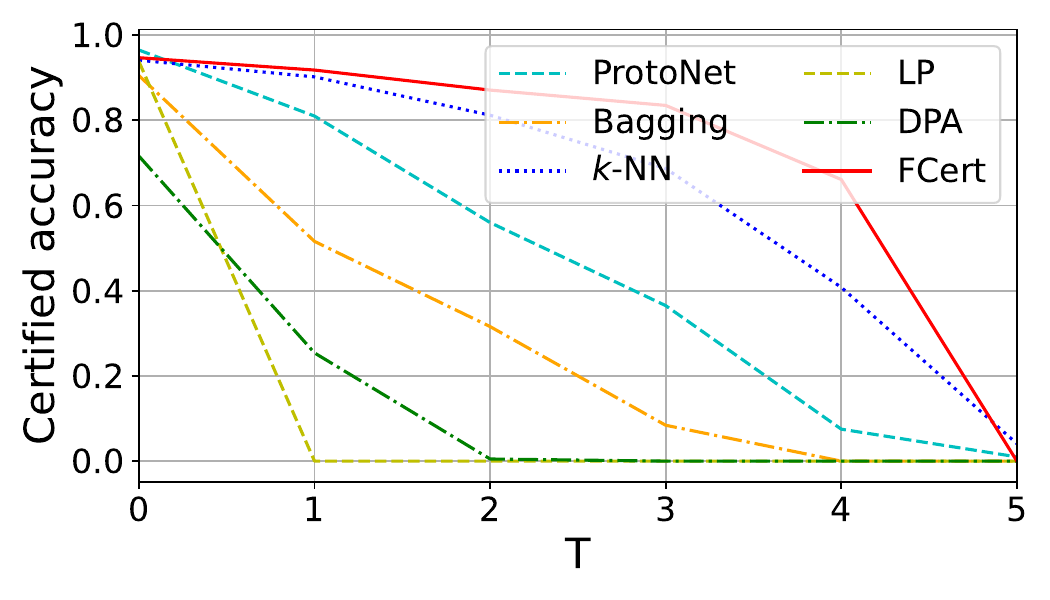}}\vspace{0.4mm}
\hspace*{-0.0005\textwidth}
\subfloat[CUB200-2011]{\includegraphics[width=0.29\textwidth]{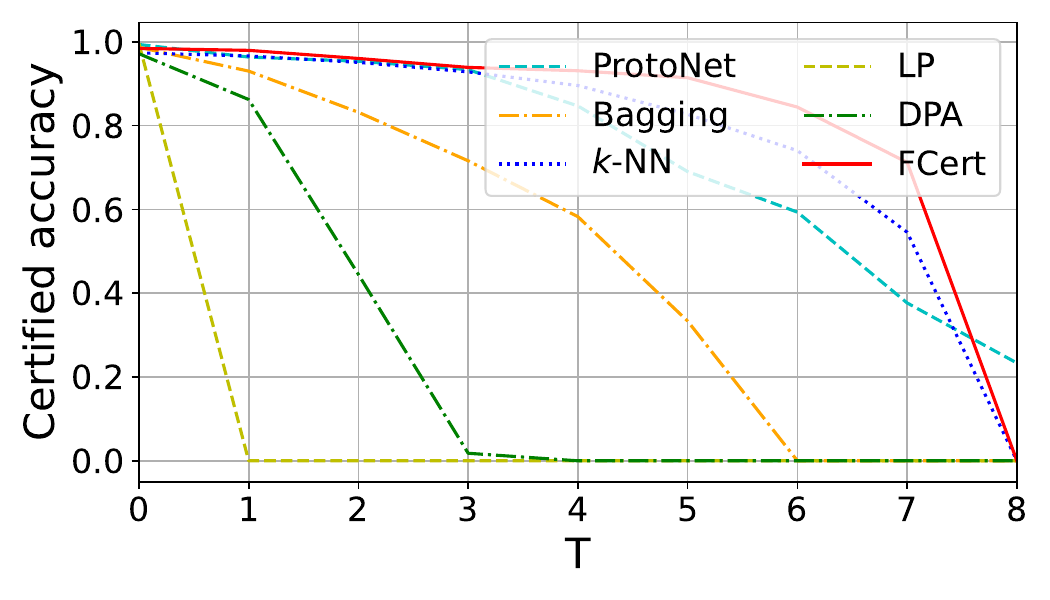}}\vspace{1mm}
\subfloat[CIFAR-FS]{\includegraphics[width=0.29\textwidth]{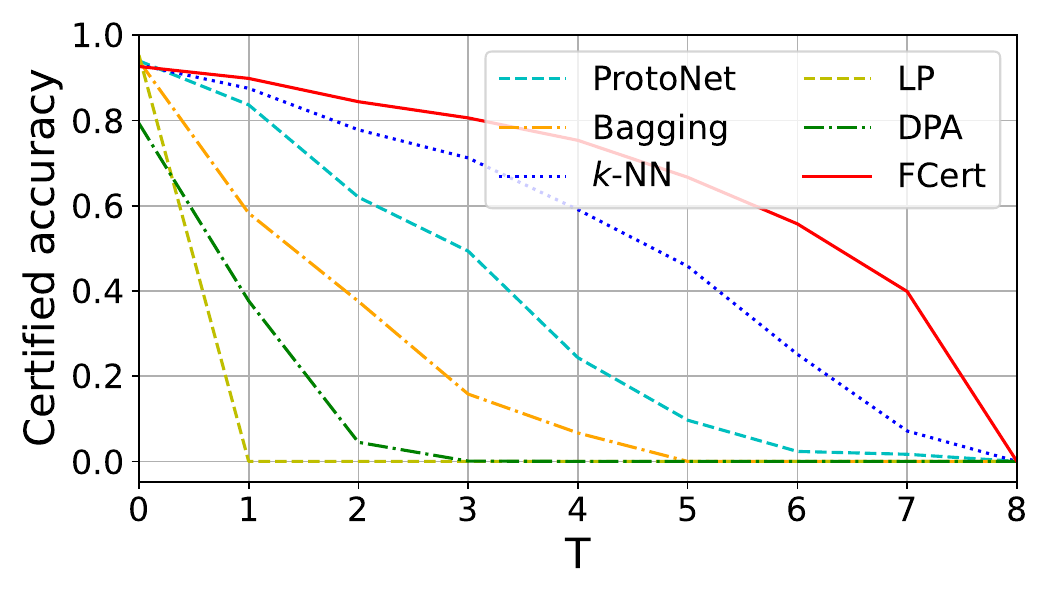}}\vspace{1mm}
\subfloat[\emph{tiered}ImageNet]{\includegraphics[width=0.29\textwidth]{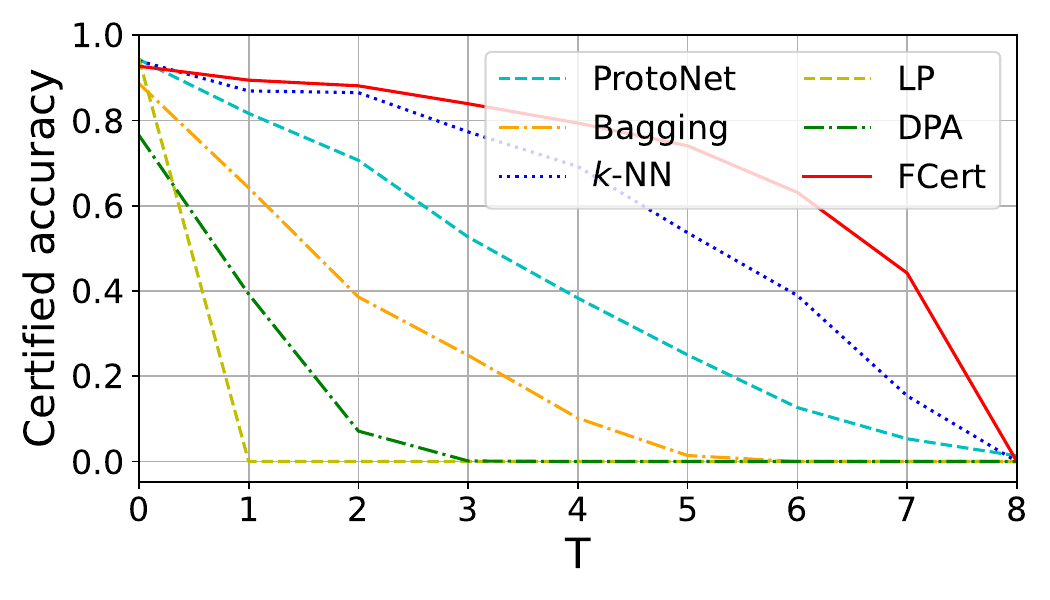}}\vspace{1mm}
\vspace{-2mm}
\caption{Comparing the certified accuracy of {\name} with existing provable defenses (or empirical accuracy of existing few-shot learning methods) for $C$-way-$K$-shot few-shot classification with DINOv2. The attack type is group attack. $K$ = 5, $C$ = 5 (first row); $K$ = 10, $C$ = 10 (second row); $K$ = 15, $C$ = 15 (third row).  $T$ is poisoning size.
}
\label{compare-certified-accuracy-dinov2-group}
\vspace{-2mm}
\end{figure*}

\begin{figure*}[!t]
\centering
\subfloat[CUB200-2011]
{\includegraphics[width=0.29\textwidth]{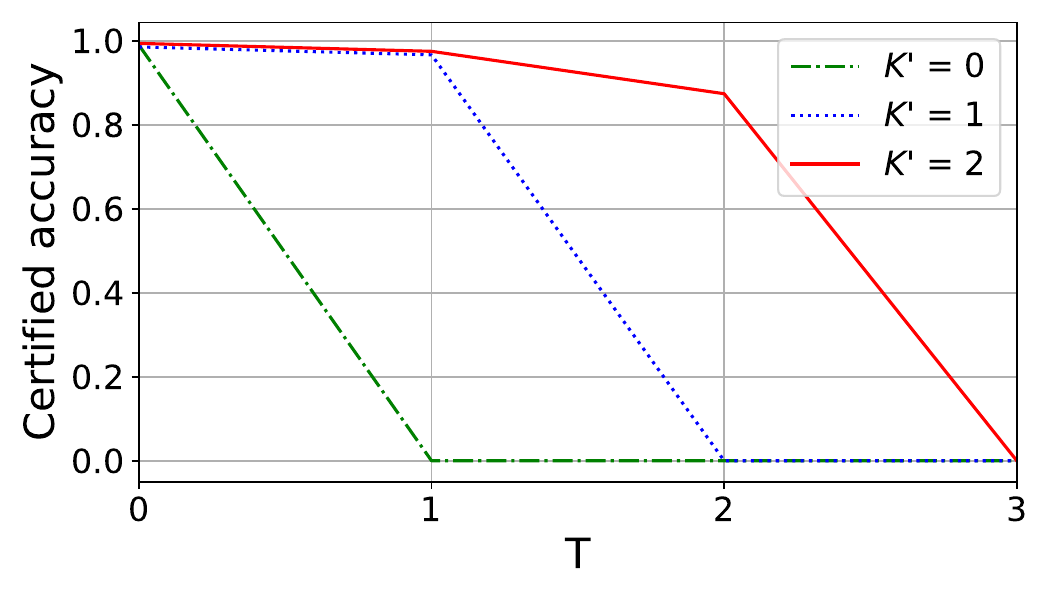}}\vspace{-0.5mm}
\subfloat[CIFAR-FS]
{\includegraphics[width=0.29\textwidth]{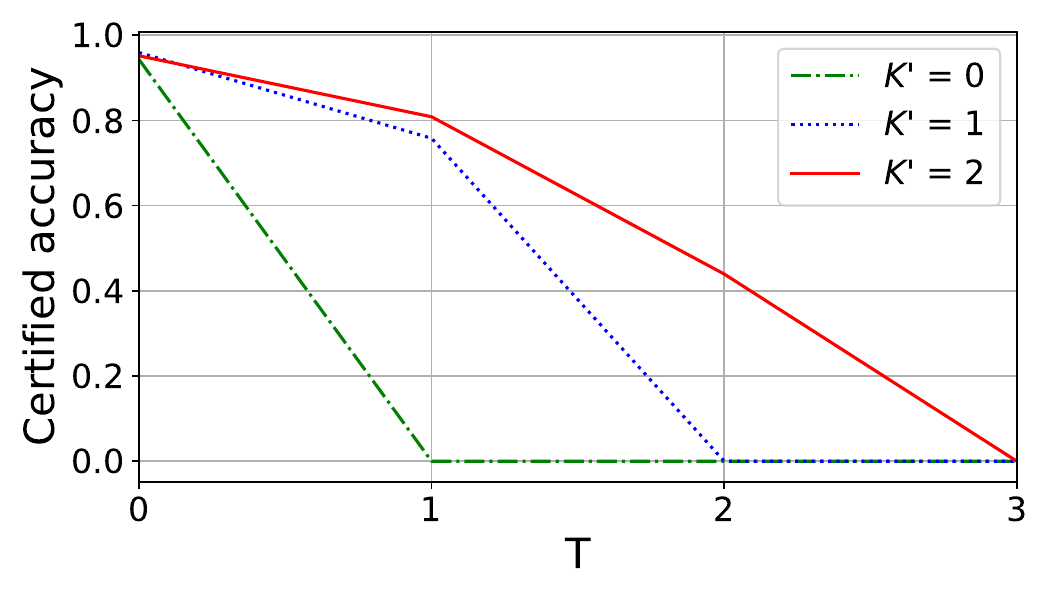}}\vspace{-0.5mm}
\subfloat[\emph{tiered}ImageNet]
{\includegraphics[width=0.29\textwidth]{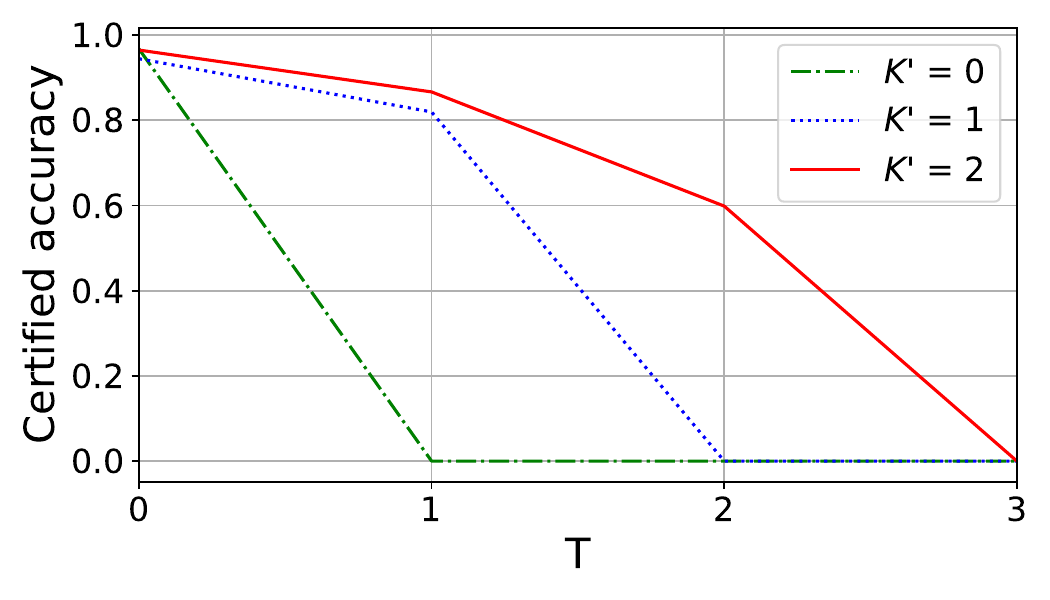}}\vspace{-0.5mm}
\caption{Impact of $K'$ on our {\name} for few-shot classification (5-way-5-shot) with DINOv2.
}
\label{fig-impact-of-Kprime}
\end{figure*}

\begin{figure*}[!t]
\vspace{-2mm}
\centering

\hspace*{0.005\textwidth}\subfloat[CUB200-2011]{\includegraphics[width=0.29\textwidth]{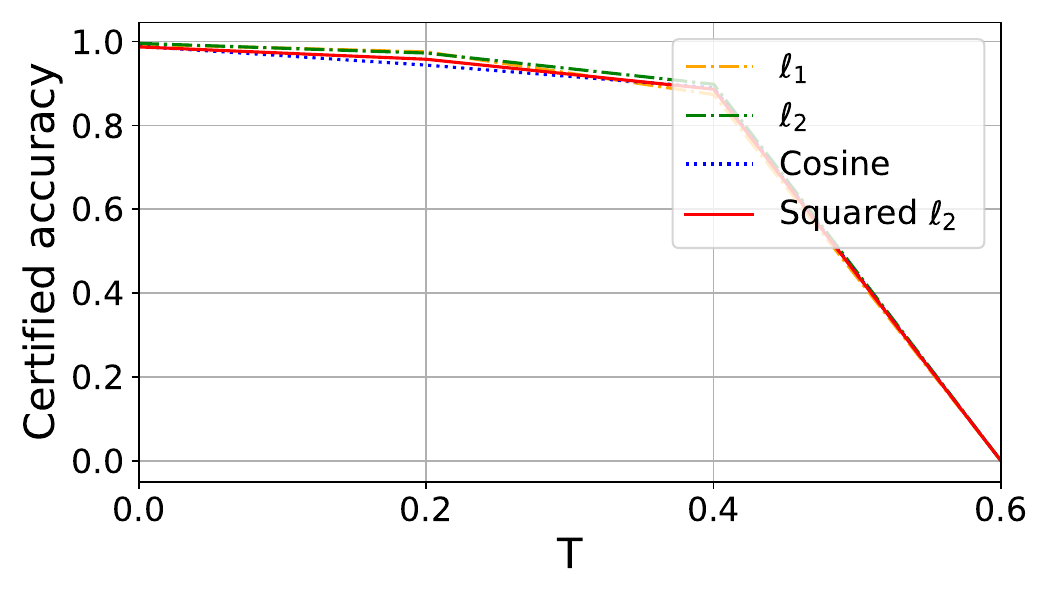}}\vspace{1mm}
\subfloat[CIFAR-FS]{\includegraphics[width=0.29\textwidth]{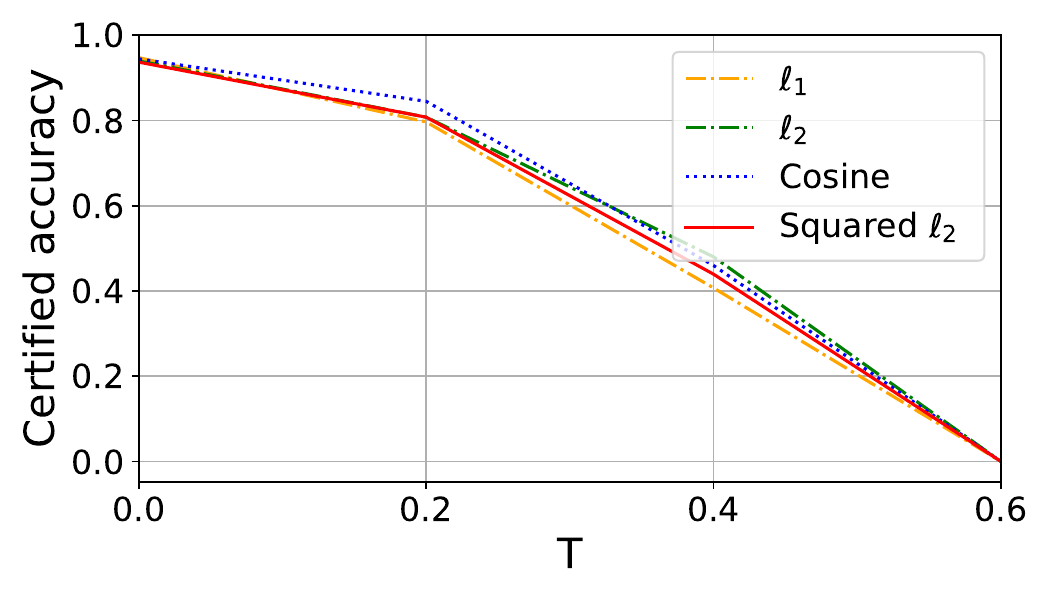}}\vspace{1mm}
\subfloat[\emph{tiered}ImageNet]{\includegraphics[width=0.29\textwidth]{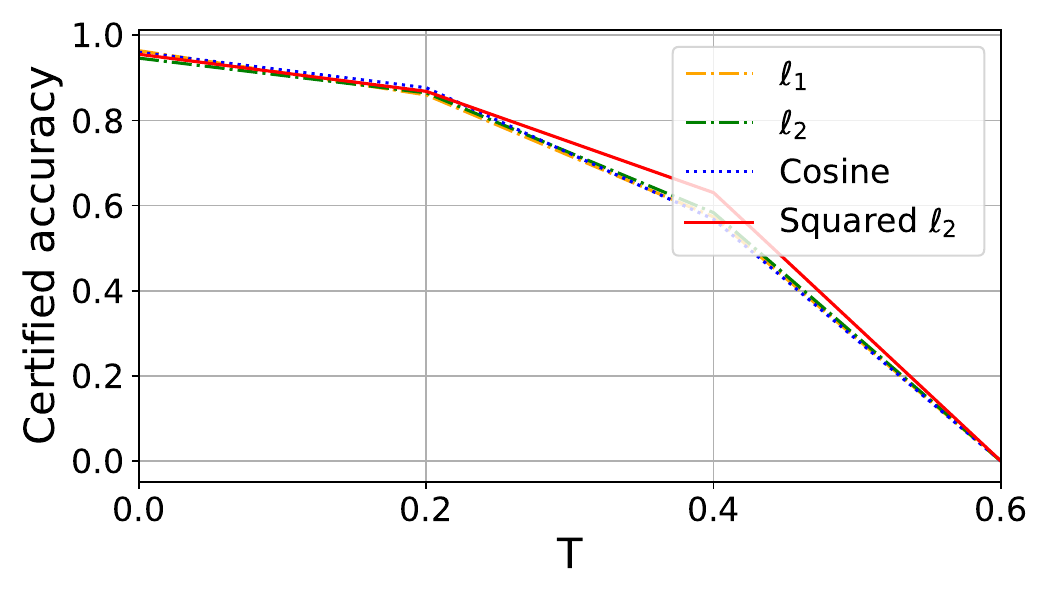}}\vspace{1mm}
\vspace{-4mm}
\caption{Impact of the distance metric $Dist$ on our {\name} for few-shot classification (5-way-5-shot) with DINOv2.
}
\label{fig-imapct-of-distance-metric}
\vspace{-3mm}
\end{figure*}

\subsection{Experimental Results}

\myparatight{Comparing our {\name} with existing provable defenses and few-shot learning methods}
Figure~\ref{compare-certified-accuracy-clip-ind},~\ref{compare-certified-accuracy-dinov2-ind},~\ref{compare-certified-accuracy-clip-group}, and~\ref{compare-certified-accuracy-dinov2-group} compare our {\name} with state-of-the-art provable defenses (i.e., Bagging, DPA, and $k$-NN) and few-shot learning methods (i.e., ProtoNet and linear probing (LP)) for data poisoning attacks under different few-shot classification settings (i.e., 5-way-5-shot, 10-way-10-shot, and 15-way-15-shot), different attack types (i.e., Individual Attack and Group Attack), and different foundation models (i.e., CLIP~\cite{radford2021learning} and DINOv2~\cite{oquab2023dinov2}). 
We have the following observations from the experimental results. First, our {\name} achieves similar classification accuracy with state-of-the-art few-shot learning methods when there is no attack (i.e., the number of poisoned support samples is 0), which means our {\name} maintains utility without attacks. In other words, our {\name} achieves the accuracy goal. Second, our {\name} consistently outperforms all existing methods under attacks. In particular, our {\name} could tolerate a large number of poisoned support samples. By contrast, existing methods can tolerate a small number of poisoned samples. 

Our {\name} outperforms existing certified defenses because our {\name} could better utilize the high-quality feature vectors produced by the foundation model when building a few-shot classifier. Recall that existing state-of-the-art few-shot learning methods cannot provide certified robustness guarantees. Thus, we compare the certified accuracy of our {\name} with their empirical accuracies. Our experimental results show that the certified accuracy of our {\name} is better than the empirical accuracy of these methods. This means the classification accuracy of {\name} under arbitrary data poisoning attacks is better than the classification accuracy of those methods under a particular attack, demonstrating that {\name} is strictly more robust than those methods.

\myparatight{Comparing the computation cost of our {\name} with existing few-shot classification methods and provabe defenses}Table~\ref{compare-computation-cost} compares the computation cost of {\name} with existing few-shot classification methods and provable defenses. We find that {\name} is as efficient as state-of-the-art methods for both the training and testing computation costs. The reason is that {\name} aggregates feature vectors in a very efficient way. 

\begin{table}[!t]\renewcommand{\arraystretch}{1.3}
\centering

\caption{Comparing the average training cost (ms) and testing cost (ms) per testing input for different methods. The dataset is CUB200-2011.\label{finetuning surrogate}}
\resizebox{\linewidth}{!}{
\begin{tabular}{|c|cc|cc|}
\hline
\multirow{2}{*}{\makecell{Compared \\methods}} & \multicolumn{2}{c|}{CLIP} & \multicolumn{2}{c|}{DINOv2}  \\ \cline{2-5} 
        & \multicolumn{1}{c|}{Training (ms)}    & Testing (ms) & \multicolumn{1}{c|}{Training (ms)}    & Testing (ms)  \\ \hline
ProtoNet   & \multicolumn{1}{c|}{8.66} & 0.34 & \multicolumn{1}{c|}{7.47} & 0.32 \\ \hline
LP    & \multicolumn{1}{c|}{112.30} & 0.52   & \multicolumn{1}{c|}{99.39} & 0.49    \\ \hline
Bagging & \multicolumn{1}{c|}{283.52} & 13.92 & \multicolumn{1}{c|}{264.35} & 9.69  \\ \hline
DPA & \multicolumn{1}{c|}{10.13} & 0.41 & \multicolumn{1}{c|}{8.07} & 0.34  \\ \hline
$k$-NN & \multicolumn{1}{c|}{8.89} & 0.39 & \multicolumn{1}{c|}{7.29} & 0.41  \\ \hline
{\name} & \multicolumn{1}{c|}{7.63} & 0.36 & \multicolumn{1}{c|}{6.82} & 0.31  \\ \hline
\end{tabular}

}
\vspace{-3mm}
\label{compare-computation-cost}
\end{table}

\subsection{Impact of Hyper-parameters}

We study the impact of hyper-parameters on {\name}. In particular, our {\name} has the following hyper-parameters: $K'$ and distance metric $Dist$. {\color{black} {We focus on Individual Attack, which is considered stronger than Group Attack. }} 

\myparatight{Impact of $K'$} Figure~\ref{fig-impact-of-Kprime} and~\ref{fig-impact-of-Kprime-Appendix} (in Appendix) show the impact of $K'$ on our {\name} under our default setting. We have following observations. First, our {\name} achieves similar classification accuracy when there is no attack (i.e., the percentage of poisoned support samples is 0). Second, our {\name} achieves a larger certified accuracy when $K'$ is larger. Our two observations imply that, in practice, we could set $K'$ to be a large value to achieve good classification accuracy without attacks and robustness under attacks.

\myparatight{Impact of distance metric $Dist$} Figure~\ref{fig-imapct-of-distance-metric} and~\ref{fig-imapct-of-distance-metric-appendix} (in Appendix) shows the impact of distance metric $Dist$ under our default setting. Our experimental results show our {\name} achieves similar performance for different metrics, meaning our {\name} is insensitive to the distance metric. In other words, our {\name} is effective with different distance metrics.

\subsection{Applying our {\name} to NLP}
\begin{figure}

 \begin{minipage}{0.45\columnwidth} 
    \centering

  \subfloat[PaLM-2 API]{ \includegraphics[width=1.1\linewidth]{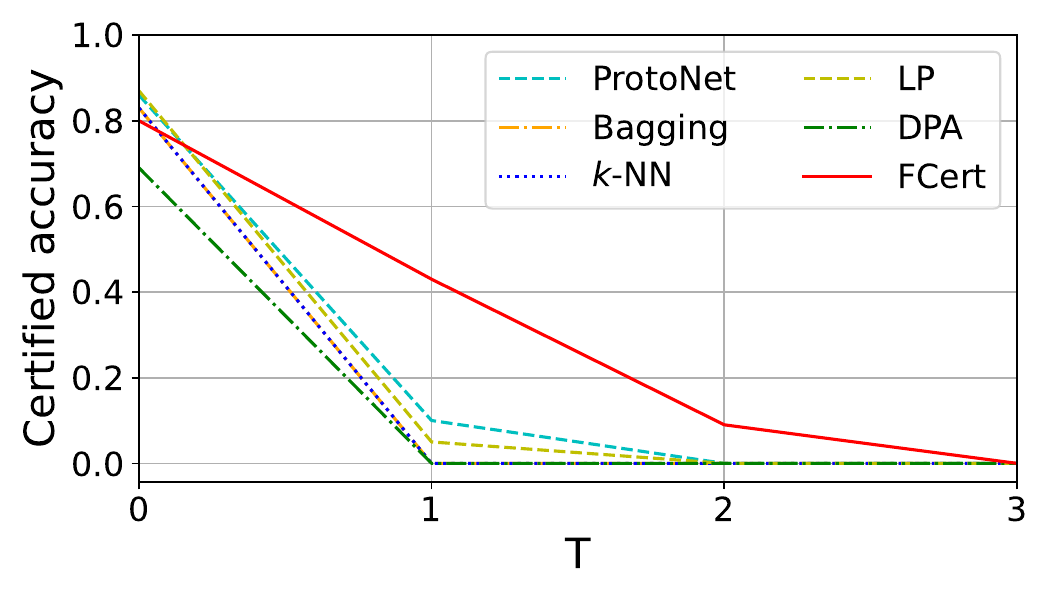}} 
 \end{minipage}%
 \vspace{-0.005\textwidth}
  \hspace{0.05\columnwidth} 
\begin{minipage}{0.45\columnwidth} 
   \centering
  \subfloat[OpenAI API]{\includegraphics[width=1.1\linewidth]{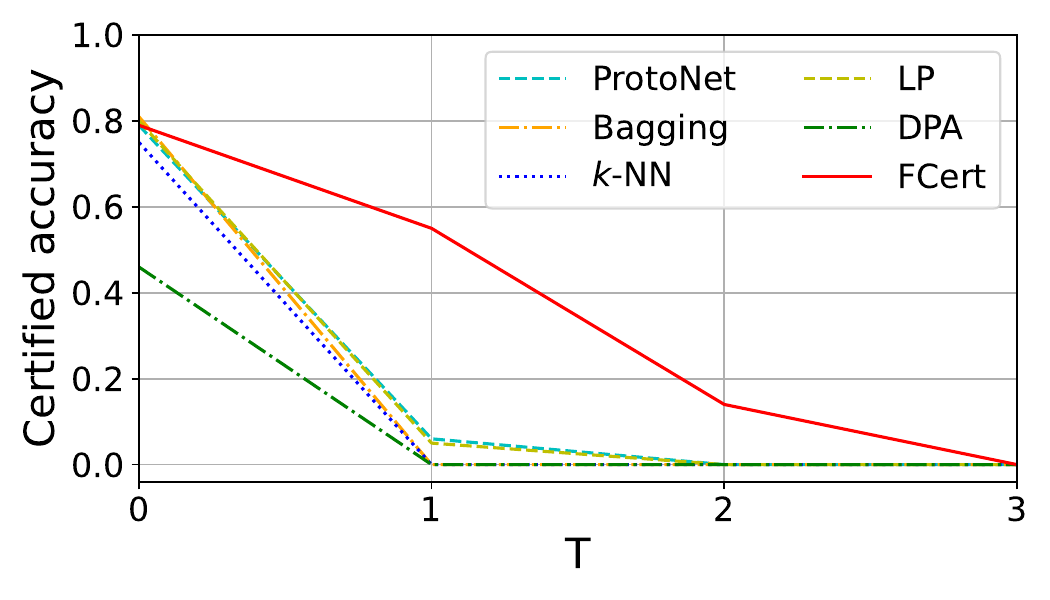}} 
  
\end{minipage}
\vspace{-0.005\textwidth}
      \caption{Comparing the certified accuracy of our {\name} with those of existing provable defenses (or empirical accuracy of existing few-shot learning methods) under Individual Attack in the text domain. }
      \label{fig-comparing-certified-accuracy-nlp-ind}

\end{figure}

\begin{figure}

 \begin{minipage}{0.45\columnwidth} 
    \centering
 \subfloat[PaLM-2 API]{ \includegraphics[width=1.1\linewidth]{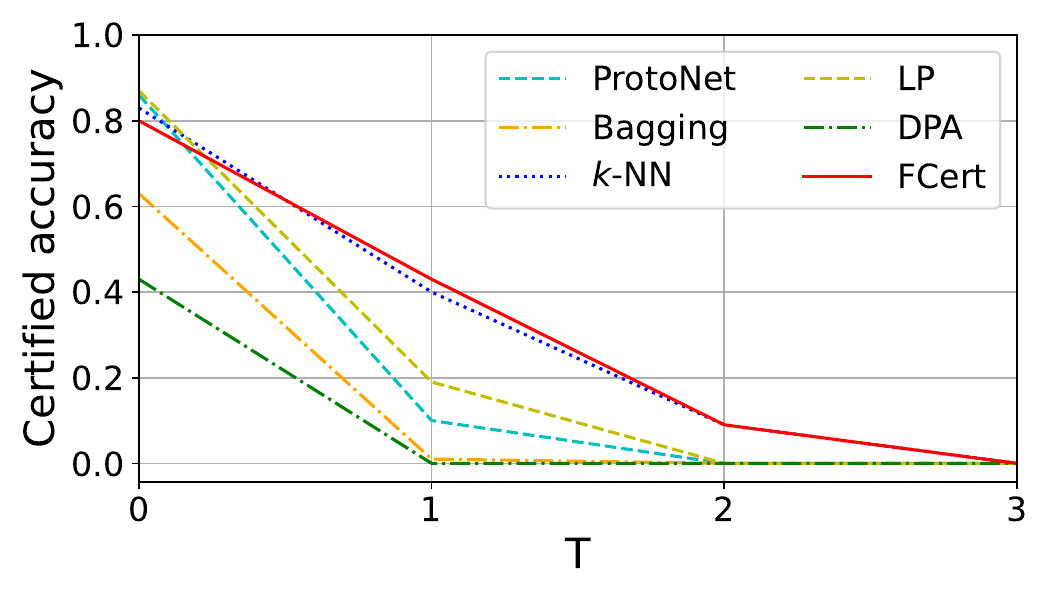}} 
 \vspace{-0.05\textwidth}
 \end{minipage}%
  \hspace{0.05\columnwidth} 
\begin{minipage}{0.45\columnwidth} 
   \centering
  \subfloat[OpenAI API]{\includegraphics[width=1.1\linewidth]{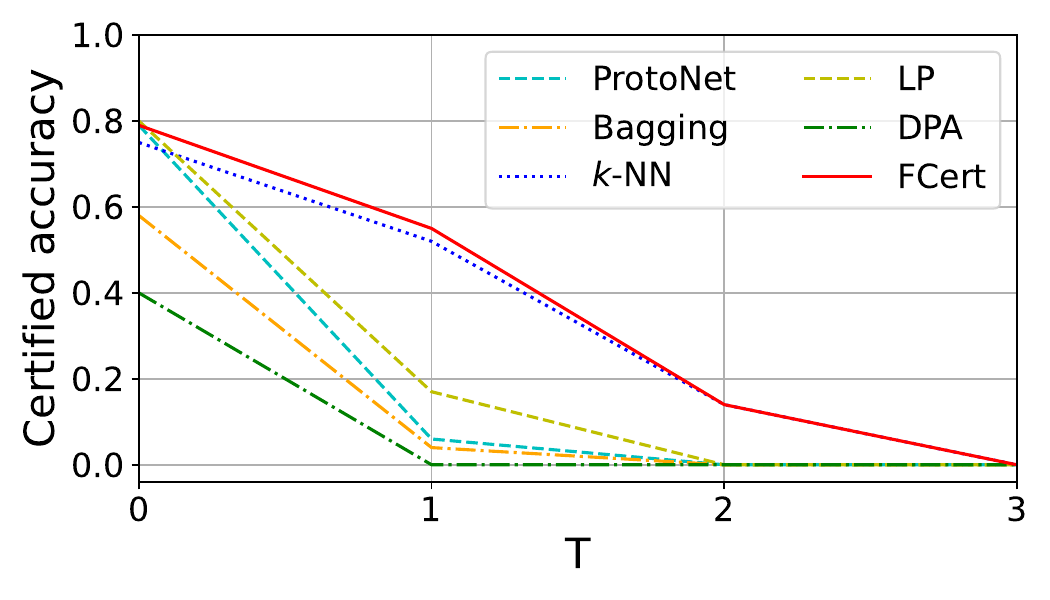}} 
  \vspace{-0.05\textwidth}
\end{minipage}
      \caption{Comparing the certified accuracy of our {\name} with those of existing provable defenses (or empirical accuracy of existing few-shot learning methods) under Group Attack in the text domain.}
      \label{fig-comparing-certified-accuracy-nlp-group}
      \vspace{-5mm}
\end{figure}

In previous experiments, we conducted experiments in the image domain. {\name} could also be applied to few-shot classification in the natural language processing (NLP) domain. In particular, we use PaLM-2 API~\cite{palm_api,anil2023palm} 
deployed by Google and OpenAI API~\cite{openai_api} deployed by OpenAI as foundation models. These APIs could return the feature vector for a text. We use 20 Newsgroups Text Dataset~\cite{20newsgroups} for the few-shot text classification. 20 Newsgroups Text Dataset contains 18,000 newsgroups posts, each of which belongs to one of the 20 topics. We adopt the same setup for few-shot classification as the image domain (please refer to Section~\ref{exp:setup} for details). Additionally, we also adapt the feature collision attack to NLP domains to evaluate the empirical accuracy of existing few-shot classification methods and provable defenses (details in Appendix~\ref{nlp-empirical-attack}). {\color{black}{Figure~\ref{fig-comparing-certified-accuracy-nlp-ind} and~\ref{fig-comparing-certified-accuracy-nlp-group} present our experimental results on Individual Attack and Group Attack, respectively.}} The results indicate that {\name} consistently surpasses existing few-shot classification methods and provable defenses in the text domain.

\section{\CR{Discussion and Limitations}}
\label{discussion_limitation}

\myparatight{Security of foundation models} In our work, we consider that the foundation model is pre-trained and shared by a trusted party (e.g., OpenAI, Meta, and Google), i.e., the parameters of foundation model are not manipulated by a malicious third party~\cite{shen2021backdoor,jia2022badencoder}.  We note that many existing studies~\cite{bommasani2021opportunities,carlini2021poisoning, liu2022poisonedencoder, he2022indiscriminate,li2023embarrassingly,carlini2023poisoning,zhang2022corruptencoder,jia202310} showed that an attacker could compromise a foundation model by manipulating its pre-training data. For instance, Carlini et al.~\cite{carlini2021poisoning} and Liu et al.~\cite{liu2022poisonedencoder} proposed targeted poisoning attacks to foundation models. However, they can only manipulate the feature vectors produced by a foundation model for a few target samples. By contrast, we aim to derive a lower bound of the classification accuracy of our {\name} for a large number of testing inputs in few-shot classification. He et al.~\cite{he2022indiscriminate} proposed untargeted poisoning attacks to foundation models. However, they need to compromise the entire pre-training dataset, which is less realistic in the real world. We leave the development of the certifiably robust few-shot classification with a compromised foundation model as the future work. 

\myparatight{Other few-shot classification methods} By default, we consider LP and ProtoNet in our experiments. We note that there are also other methods. For instance, instead of fine-tuning the last linear layer in LP, we could also fine-tune both the foundation model and the appended linear layer (called \emph{LP-Finetune-all}). We could also append two fully connected layers (called \emph{FCN}) instead of a linear layer. Figure~\ref{fig-finetune-all} in the Appendix shows the comparison of our {\name} with LP, LP-Finetune-all, and FCN under Indivisual Attack. We find that LP, LP-Finetune-all, and FCN achieve similar empirical accuracy. Moreover, our {\name} consistently outperforms them. We note that LP-Finetune-all achieves a small empirical accuracy in some scenarios due to the over-fitting of the foundation model to the limited number of support samples. 

\myparatight{Other attacks} In our work, we focus on data poisoning attacks. We note that there are many other attacks such as adversarial examples~\cite{szegedy2014intriguing,goodfellow2014explaining} and backdoor attacks~\cite{gu2017badnets,liu2018trojaning} to machine learning models. 
For instance, in backdoor attacks, an attacker could manipulate both the training data and testing data, while our {\name} is designed to resist attacks on training data.
We leave it as a future work to extend our framework to defend against those attacks, e.g., we could combine our {\name} with certified defenses~\cite{lecuyer2019certified,cohen2019certified} against adversarial examples to resist backdoor attacks. 

\myparatight{Robust statistics~\cite{huber2004robust}} Our {\name} utilize robust statistics techniques~\cite{huber2004robust,yin2018byzantine} to aggregate feature vectors of support samples. Our key technique contribution is to derive the certified robustness guarantees and prove its tightness. We believe it is an interesting future work to explore other robust statistics techniques~\cite{huber2004robust} under our {\name} framework.
\section{Conclusion and Future Work}
\label{conclusion}
We propose {\name}, the first certified defense against data poisoning attacks to few-shot classification with foundation models. Our {\name} provably predicts the same label for a testing input when the percentage of the poisoned support samples is bounded. We conduct extensive evaluations on both image and text domains. Moreover, we compare our {\name} with state-of-the-art certified defenses for data poisoning attacks and state-of-the-art few-shot classification methods. Potential future directions involve 1) expanding our defense to tasks like image segmentation, and 2) extending our defense to defend against other types of attacks, e.g., backdoor attacks.
\section{Acknowledgments}

We thank the anonymous reviewers and shepherd for constructive feedback on our paper. Their valuable feedback significantly enhanced the quality of our work.

\bibliographystyle{ieeetr}
\bibliography{refs}

\appendices
\section{Detailed Proof for Theorem 1}
\label{proof-of-theorem-1}

\proof{We only need to prove that Equation~\ref{equation-upper-bound} and~\ref{equation-lower-bound} provide the correct upper and lower bounds for ${R}_p^{c}(T^c)$ for any class $c = 1,2, \cdots, C$. That is, we show that the upper bound of $R_p^{c}(T^c)$ is given by $\overline{R}_p^{c}(T^c) =  \frac{\sum_{i=K'+1+T^c}^{K-K'+T^c}d_i^c}{K-2\cdot K'}$, and 
    the lower bound of ${R}_p^{c}(T^c)$ is given by $\underline{R}_p^{c}(T^c) = \frac{\sum_{i=K'+1-T^c}^{K-K'-T^c}d_i^c}{K-2\cdot K'}$, where $d_1^c \leq d_2^c \leq \cdots \leq d_{K}^c$ represents the distances of feature vectors between each support sample in class $c$ and $\mathbf{x}_{test}$. 
    
    Since we focus on a single class, we abuse the notation and use $T$ instead of $T^c$ to represent the number of modified support samples in class $c$. We use $e_1^c \leq e_2^c \leq \cdots \leq {e}_{K}^c$ to represent all distances sorted in ascending order after the poisoning attack. Note that $d_i^c$ and $e_i^c$, where $i = 1,2,\cdots, K$, could be different due to the reordering of distances after data poisoning attacks.
    
    Here we prove the correctness of the upper bound, and the lower bound can be proven in a similar way. We find that if we change the $T$ smallest distances among $d_1^c, d_2^c, \cdots, d_{K}^c$ (which are $d_1^c, d_2^c, \cdots, d_{T}^c$) to the largest distance $d_{K}^c$, then the upper bound $\overline{R}_p^{c}(T)=  \frac{\sum_{i=K'+1+T}^{K-K'+T}d_i^c}{K-2\cdot K'}$ is achieved. For simplicity, we call this strategy \emph{our poisoning strategy} in the rest of the proof. We show by exchange arguments~\cite{exchange_argument} that our poisoning strategy is optimal, i.e., there does not exist a poisoning strategy that results in a ${R}_p^{c}(T)$ that is larger than $\overline{R}_p^{c}(T)=  \frac{\sum_{i=K'+1+T}^{K-K'+T}d_i^c}{K-2\cdot K'}$ (the upper bound obtained by our poisoning strategy). In the following, we follow the standard procedure of exchange arguments~\cite{exchange_argument} to show our proof.
    Specifically, we show that if exists an optimal poisoning strategy that is different from ours (the proof is done if there does not exist such a strategy), then the upper bound obtained by the assumed optimal poisoning strategy is no better than ours. Our idea is to prove that we could transform the assumed optimal poisoning strategy to our poisoning strategy without decreasing ${R}_p^{c}(T)$ in the transformation process. In particular, we consider two steps. In Step I, we show that changing $T$ smallest distances (i.e., $d_1^c, d_2^c, \cdots, d_{T}^c$) among $d_1^c, d_2^c, \cdots, d_{K}^c$ could be the optimal attack strategy. In Step II, we show changing $d_1^c, d_2^c, \cdots, d_{T}^c$ to $d_{K}^c$ could result in the largest upper bound $\overline{R}_p^{c}(T)$, which complete our proof. Next, we discuss details of the two steps. 

    \myparatight{Step I}
    In this step, we show that if the assumed optimal poisoning strategy changes at most $(T-1)$ distances among $d_1^c, d_2^c, \cdots, d_{T}^c$, then this assumed poisoning strategy can be transformed into a new poisoning strategy that changes all $T$ distances among $d_1^c, d_2^c, \cdots, d_{T}^c$ without decreasing ${R}_p^{c}(T)$. 
    
    Without loss of generality, suppose the assumed optimal poisoning strategy changes $d_{s_1}^c, d_{s_2}^c, \cdots, d_{s_T}^c$ (in ascending order, $1 \leq s_i \leq K$ for $1 \leq i \leq T$) to $\Tilde{e}_{s_1}^c, \Tilde{e}_{s_2}^c, \cdots, \Tilde{e}_{s_T}^c$ correspondingly, where $s_1, s_2, \cdots, s_T$ are the $T$ indices and $\Tilde{e}_{s_1}^c, \Tilde{e}_{s_2}^c, \cdots, \Tilde{e}_{s_T}^c$ are the $T$ arbitrary distances.   Suppose $d_{s_i}^c \neq d_{i}^c$ for some $1\leq i\leq T$ (since the assumed optimal poisoning strategy is different from the new poisoning strategy). We denote by $i^*$ the smallest $i$ such that $d_{s_i}^c \neq d_{i}^c$. Based on the definition of $i^*$, we have $d_{s_{i^*}}^c > d_{i^*}^c$. We show that if the attacker changes $d_{i^*}^c$ instead of $d_{s_{i^*}}^c$ to $\Tilde{e}_{s_{i^*}}^c$ (the changes to all other distances are the same), then the attacker gets a new poisoning strategy that is either better or equivalent to the assumed optimal poisoning strategy. 
    We use $e_1^c \leq e_2^c \leq \cdots \leq {e}_{K}^c$ and $\hat{e}_1^c \leq \hat{e}_2^c \leq \cdots \leq \hat{e}_{K}^c$ to represent all distances (sorted in ascending order) of (poisoned) support inputs (whose ground truth labels are $c$) with a testing input under the assumed optimal poisoning attack strategy and the new poisoning attack strategy, respectively.
    
    We note that the only difference between the new poisoning attack strategy and the assumed optimal poisoning attack strategy is that $d_{i^*}^c$ instead of $d_{s_{i^*}}^c$ is changed to another value. Thus, only a small number of distances between $e_1^c, e_{2}^c, \cdots, e_{K}^c$ and $\hat{e}_1^c, \hat{e}_2^c, \cdots, \hat{e}_{K}^c$ are different. 
    Formally, 
    we have $e_i^c = \hat{e}_i^c$ for $l\leq i\leq K$, where $l$ is the smallest index for which $e_l^c > d_{s_{i^*}}^c$. 
    Similarly, we have $e_i^c = \hat{e}_i^c$ for $1\leq i\leq l'$, where $l'$ is the largest index for which $e_{l'}^c < d_{{i^*}}^c$. 
    When $l'+1\leq i \leq l-1$, $e_{{l'+1}}^c, e^c_{l'+2}, \ldots,e^c_{l-1}$ for the assumed poisoning strategy would be $d_{{i^*}}^c, e^c_{l'+2}, \ldots,e^c_{l-1}$ (in ascending order). By contrast, $\hat{e}_{l'+1}^c, \hat{e}_{l'+2}^c, \cdots, \hat{e}_{l-1}^c$ for the new poisoning strategy would be $e^c_{l'+2}, \ldots,e^c_{l-1}, d_{s_{i^*}}^c$ (in ascending order). Therefore, we have $\hat{e}^c_{i}\geq e^c_{i}$ for every $1\leq i \leq K$, which means the new poisoning strategy does not decrease  ${R}_p^{c}(T)$. We repeat the above process until the assumed optimal poisoning strategy is transformed to changing $T$ smallest distances among $d_1^c, d_2^c, \cdots, d_{K}^c$. 

    \myparatight{Step II}
    In this step, we show that if the assumed optimal poisoning strategy changes $d_1^c, d_2^c, \cdots, d_{T}^c$ to values other than $d_{K}^c$, then this poisoning strategy is no better than our poisoning strategy. Without loss of generality, suppose the assumed optimal poisoning strategy changes $d_{1}^c, d_{2}^c, \cdots, d_{T}^c$ to $\Tilde{e}_{s_1}^c, \Tilde{e}_{s_2}^c, \cdots, \Tilde{e}_{s_T}^c$ correspondingly. Since the assumed optimal poisoning strategy is different from ours, there exists some $\Tilde{e}_{s_i}^c \neq d_{K}^c$, where $1 \leq i\leq T$.  We use $e_1^c \leq e_2^c \leq \cdots \leq {e}_{K}^c$ to represent all distances sorted in ascending order under the assumed optimal poisoning attack. We show that changing $d_{i }^c$ to $d_{K}^c$ instead of $\Tilde{e}_{s_i}^c$ results in a new poisoning strategy with equal or larger ${R}_p^{c}(T)$. We consider two cases.
    \begin{itemize}
    \item In this case, we consider that
    $\Tilde{e}_{s_i}^c < d_{K}^c$. It is obvious that  $e^c_{i}$ would not be decreased by changing $\Tilde{e}_{s_i}^c$ to $d_{K}^c$ for all $1\leq i \leq K$. Thus, ${R}_p^{c}(T) =  \frac{\sum_{i=K'+1}^{K-K'}e_i^c}{K-2\cdot K'}$ would not be decreased by this shift of poisoning strategy. 

    \item In this case, we consider that
    $\Tilde{e}_{s_i}^c > d_{K}^c$. Given that all of $\Tilde{e}_{s_1}^c, \Tilde{e}_{s_2}^c, \cdots, \Tilde{e}_{s_T}^c$ are no smaller than $d_{K}^c$, they are the $T$ largest distances among ${e}_{1}^c, {e}_{2}^c, \cdots, {e}_{K}^c$ (note that the attacker has to modify more than $T$ distances if this is not the case). Since $T<K'$, $\Tilde{e}_{s_1}^c, \Tilde{e}_{s_2}^c, \cdots, \Tilde{e}_{s_T}^c$ are all removed when calculating the robust distance ${R}_p^{c}(T) =  \frac{\sum_{i=K'+1}^{K-K'}e_i^c}{K-2\cdot K'}$, which means changing some $\Tilde{e}_{s_i}^c > d_{K}^c$ to $d_{K}^c$ would not influence ${R}_p^{c}(T)$. 
    \end{itemize}

Combining these two steps, we can conclude that changing the $T$ smallest distances from $d_1^c, d_2^c, \cdots, d_{K}^c$ (which are $d_1^c, d_2^c, \cdots, d_{T}^c$) to $d_{K}^c$ is an optimal poisoning attack strategy that produces the maximum ${R}_p^{c}(T)$, i.e., our derived upper bound is correct.

\begin{table}[!t]\renewcommand{\arraystretch}{1.3}
\centering
\caption{Comparing the empirical accuracy of {\name} with weighted average variant\label{tab-variant}}
\resizebox{\linewidth}{!}{
\begin{tabular}{|c|c|c|c|c|c|}
\hline

       Compared methods & $T=0$   & $T=2$ & $T = 4$    & $T=6$ &$T=8$ \\ \hline
 Weighted average variant  & 0.982 & 0.970 & 0.935& 0.407&0\\ \hline
{\name} (default)   & 0.980 & 0.974   & 0.940 & 0.536& 0 \\ \hline
\end{tabular}

}
\label{table_compare-computation-cost-tiered_imagenet}
\vspace{-5mm}
\end{table}

\section{Empirical Attacks to Existing Defenses in the Text Domain}
\label{nlp-empirical-attack}
For the text domain, we also develop feature collision attacks~\cite{shafahi2018poison} to compute the empirical accuracy for few-shot classification methods~\cite{snell2017prototypical,chen2019closer}. For Group Attack, we use the same implementation as we used in the image domain.

For Individual Attack, given a test sample $(\mathbf{x}_{test},y)$, the goal of the attacker is to modify $T$ poisoned support samples for each of the $C$ classes to change the prediction of $\mathbf{x}_{test}$ by the few shot classifier. We reuse notations used in Section~\ref{method}. 
    
    For support samples whose ground truth labels are not $y$, we perform the same poisoning strategy as we did for the image domain. Specifically, we randomly select $T$ support samples from each class and modify them by solving the optimization problem in Equation~\ref{eqn-collision}. We set $\lambda=0$ to consider a strong attack.
    The value of $\lambda$ is set to $0$ to because we assume the attacker can arbitrarily modify these support samples.
    {\color{black}
    {For support samples whose ground truth labels are $y$, we want to create $T$ poisoned support samples such that their feature vectors are very different from that of the testing input $\mathbf{x}_{test}$. Since we cannot obtain the gradient information from the OpenAI API and PaLM-2 API, for both ProtoNet and LP, we directly modify $T$ support samples, denoted by $(\mathbf{x}^y_1,y),\cdots,(\mathbf{x}^y_T,y)$ in $\mathcal{D}^{y}$, to $(\mathbf{x}^c_1,y),\cdots,(\mathbf{x}^c_T,c)$, where $c$ is an arbitrary class different from $y$.}}

\section{Weighted Average Variant of {\name}}
In {\name}, we computed the robust distance by averaging the remaining distances after excluding $2\cdot K'$ extreme values. A variant of {\name} is to calculate the robust distance by taking a weighted average of the remaining distances, e.g., using the cosine similarity between the feature vectors of a testing input and each support sample as a weight. 
We conducted experiments on the CUB200-2011 dataset with $K = 15$, $C = 15$, and $K' = 5$. In particular, we compare the empirical accuracy of {\name} under an individual empirical attack to that of the weighted average variant. The result is shown in Table~\ref{tab-variant}. Our results reveal that {\name} achieves similar performance with its variant without attacks (i.e., $T = 0$). However, as $T$ becomes large, {\name} achieves better performance compared with its variant. 


\begin{algorithm}[!tb]
   \caption{Computing Certified Poisoning Size for Group Attack}
   \label{alg:binary-search-GA}
\begin{algorithmic}
   \STATE {\bfseries Input:}  $d_1^c, d_2^c, \cdots, d_K^c$ for $c=1,2,\cdots, C$,  predicted label $\hat{y}$, $K'$
   \STATE {\bfseries Output:} Certified poisoning size $T^*$\\
 \STATE {${low} = 0$}
 \STATE {${high} = K'$}
 \WHILE{${low} \neq {high}$}
 \STATE{$mid = \lceil{low + high}\rceil/2$}
  \FOR{$c\neq \hat{y}$}
      \FOR{$T^{\hat{y}} = 0,1,\ldots, mid$}
      \STATE $T^c = mid - T^{\hat{y}}$
    \STATE $\overline{R}_p^{\hat{y}}(T^{\hat{y}}) = (\sum_{i=K'+1+T^{\hat{y}}}^{K - K'+T^{\hat{y}}}d_i^c)/(K - 2 K')$ \\
    
        \STATE $\underline{R}_p^{c}(T^c)= (\sum_{i=K'+1-T^c}^{K - K'-T^c}d_i^c)/(K - 2 K')$ \\
        \IF{$\underline{R}_p^{c}(T^c)<\overline{R}_p^{\hat{y}}(T^{\hat{y}})$}
        \STATE{${high} = {mid}-1$}
        \STATE{\textbf{break}}
        \STATE{\textbf{break}}
        \ENDIF
       \ENDFOR
 
    \ENDFOR

   \STATE{${low} = {mid}$}

\ENDWHILE
   \STATE \textbf{return} $low$

\end{algorithmic}
\end{algorithm}

\begin{figure*}[!t]
\vspace{-5mm}
    \centering
\subfloat[]{\includegraphics[width=0.47\linewidth]{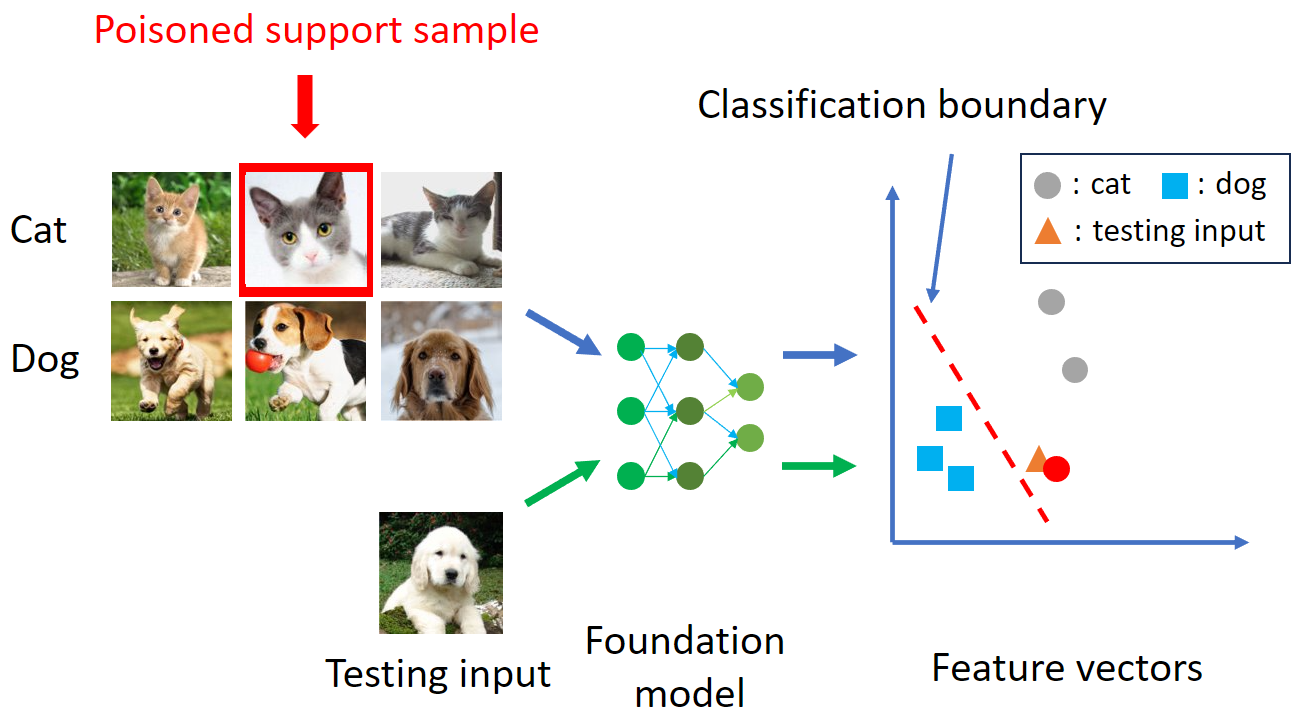}}
\hspace{8mm}
\subfloat[]{\includegraphics[width=0.47\linewidth]{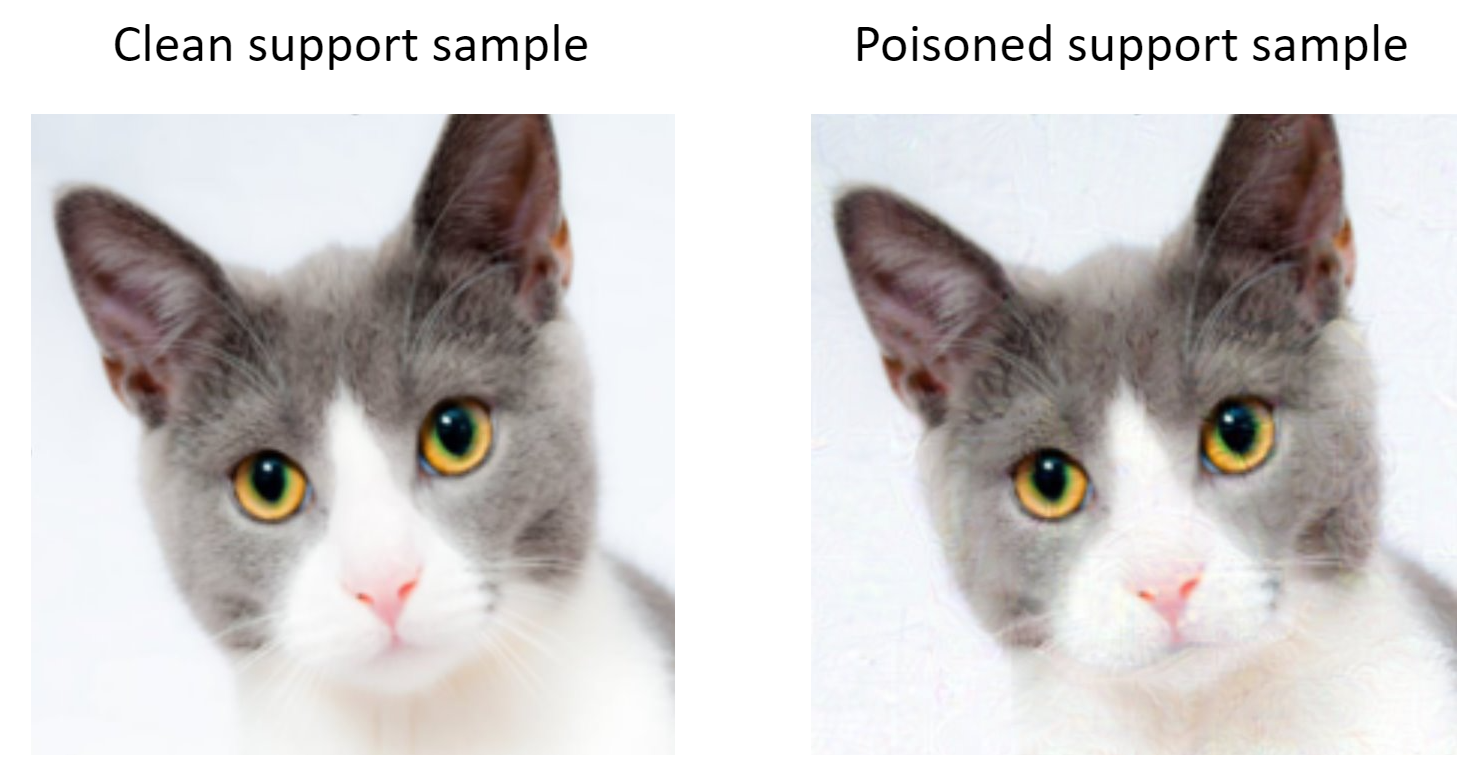}}
    \caption{An illustration of clean-label poisoning attack~\cite{shafahi2018poison} to few-shot classification. (a): The testing input is misclassified as ``cat'' under attack with a single poisoned support sample. (b): The poisoned support sample is visually indistinguishable from the clean one. \label{fig-compare-poisoned}}
    \label{clean-label-poisoning-attack}
   \vspace{-2mm}
\end{figure*}

\begin{figure*}[!t]
\centering

\subfloat[CUB200-2011]
{\includegraphics[width=0.3\textwidth]{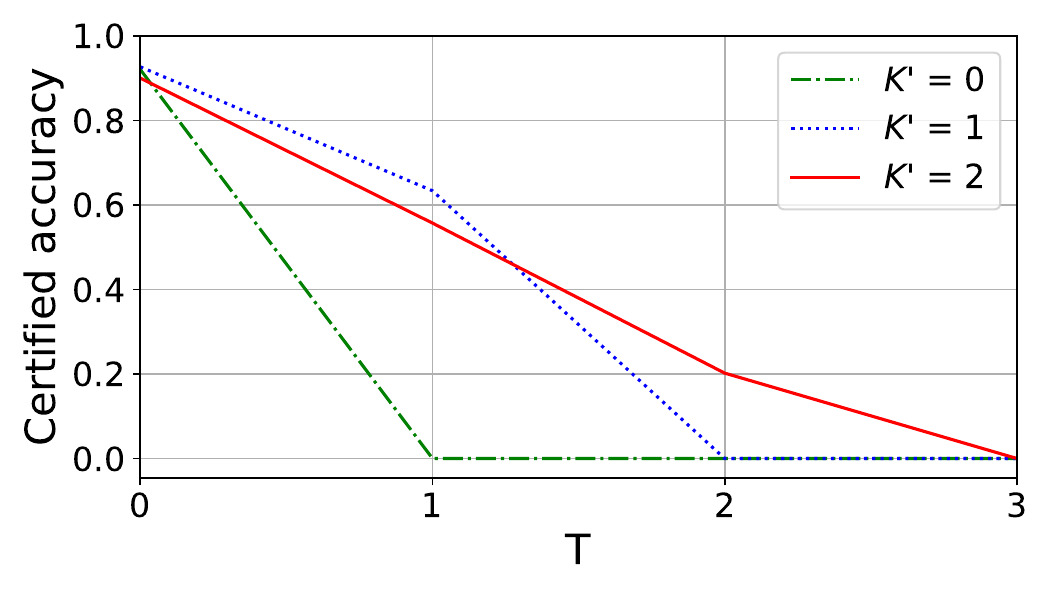}}\vspace{-0.5mm}
\subfloat[CIFAR-FS]
{\includegraphics[width=0.3\textwidth]{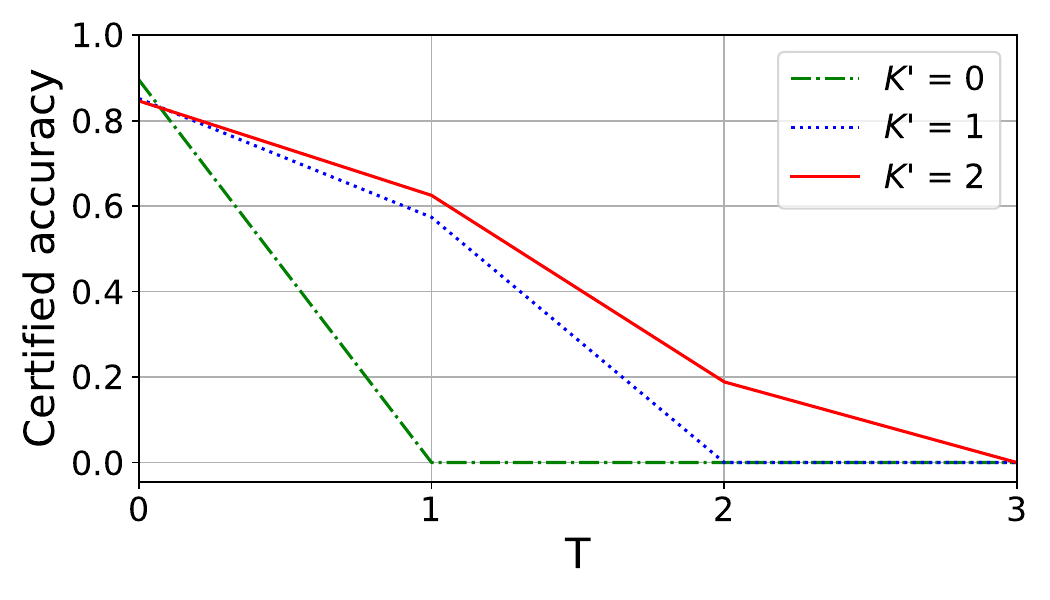}}\vspace{-0.5mm}
\subfloat[\emph{tiered}ImageNet]
{\includegraphics[width=0.3\textwidth]{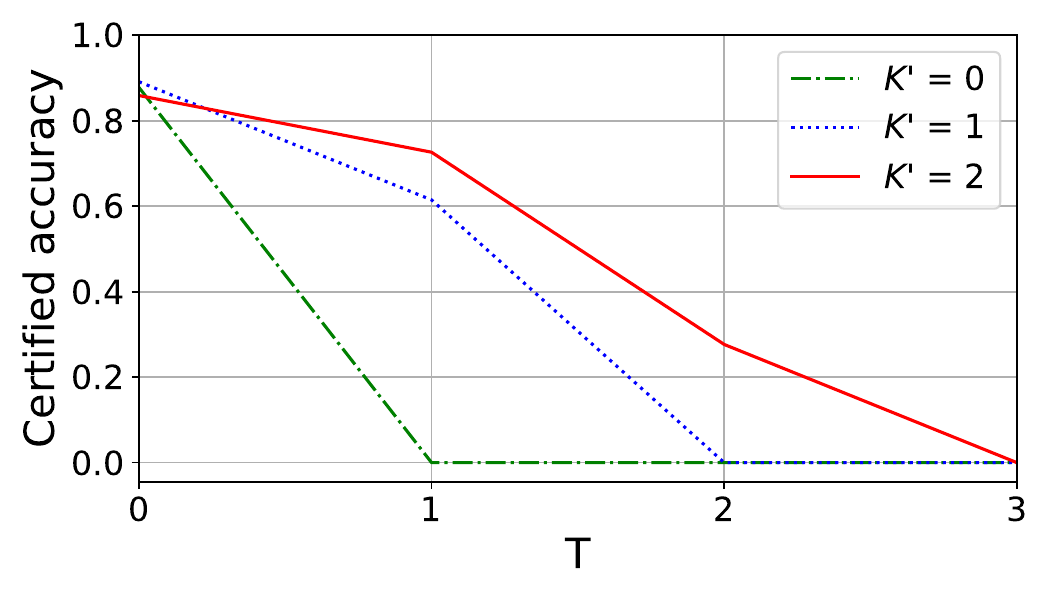}}\vspace{-0.5mm}

\caption{Impact of $K'$ on our {\name} for few-shot classification (5-way-5-shot) with CLIP. 
}
\label{fig-impact-of-Kprime-Appendix}

\end{figure*}

\begin{figure*}[!t]
\vspace{-2mm}
\centering

\subfloat[CUB200-2011]{\includegraphics[width=0.3\textwidth]{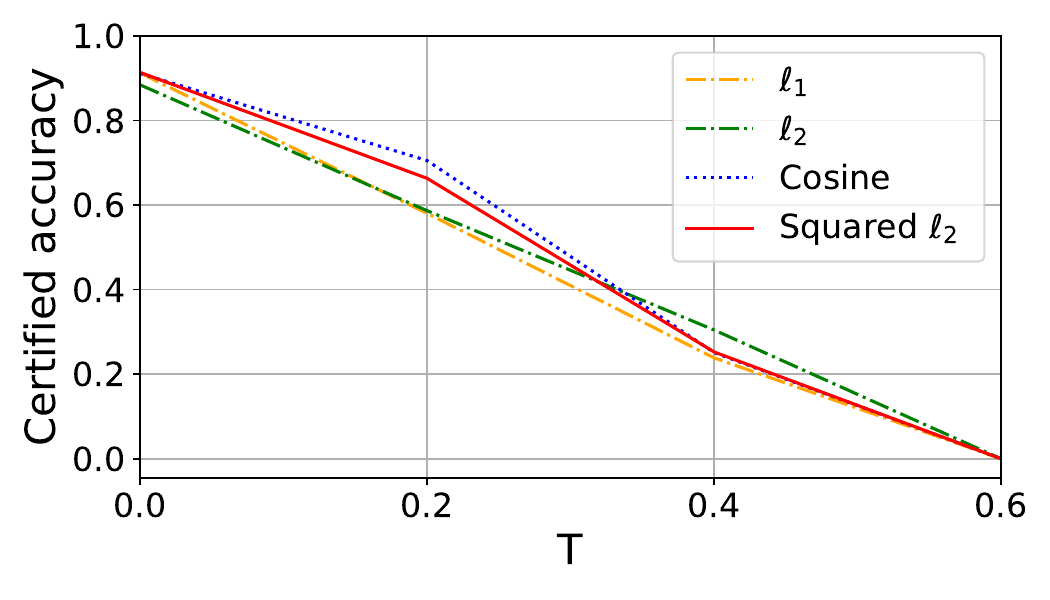}}\vspace{1mm}
\subfloat[CIFAR-FS]{\includegraphics[width=0.3\textwidth]{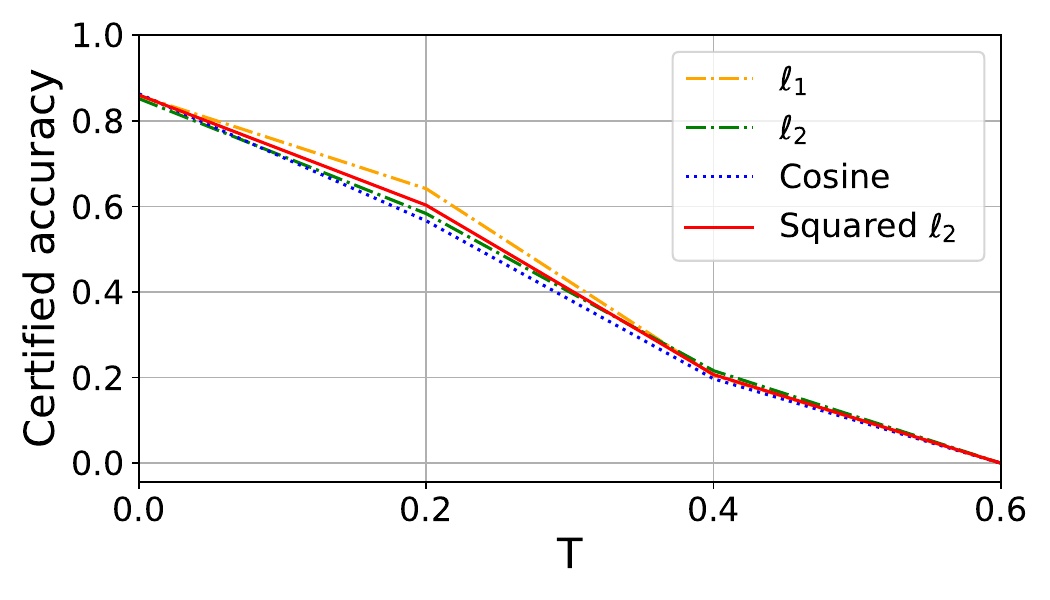}}\vspace{1mm}
\subfloat[\emph{tiered}ImageNet]{\includegraphics[width=0.3\textwidth]{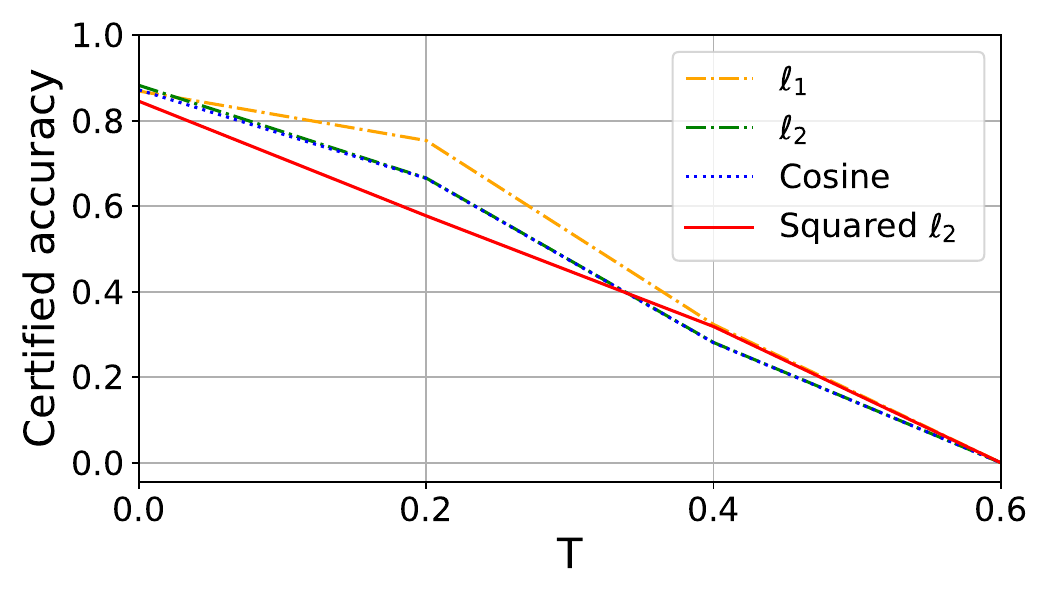}}\vspace{1mm}
\vspace{-4mm}
\caption{Impact of the distance metric $Dist$ on our {\name} for few-shot classification (5-way-5-shot) with CLIP.
}
\label{fig-imapct-of-distance-metric-appendix}
\vspace{-2mm}
\end{figure*}

\begin{figure*}[!t]
\centering
{\includegraphics[width=0.3\textwidth]{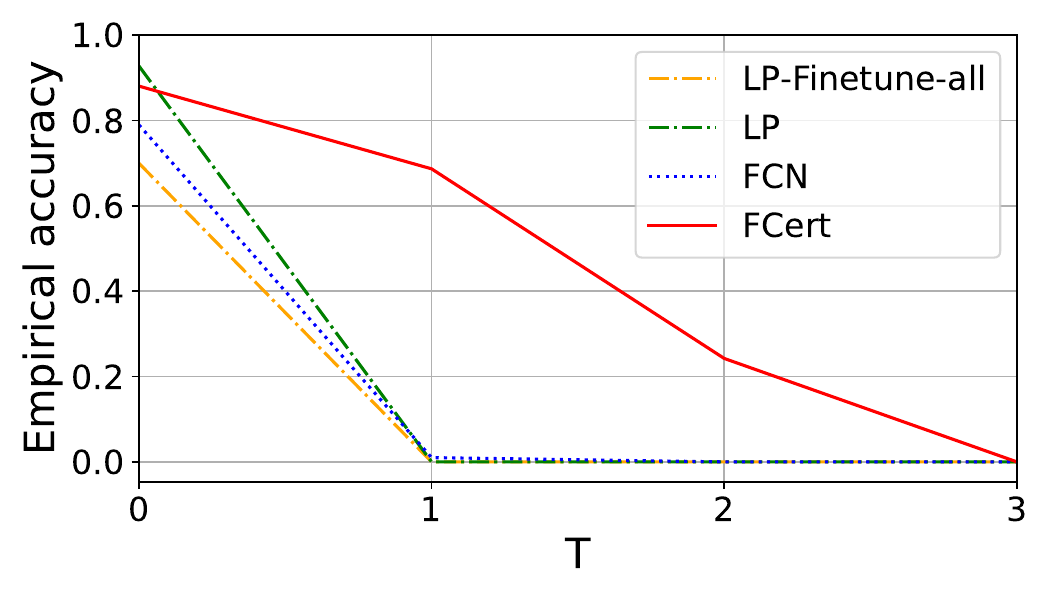}}\vspace{-0.5mm}
{\includegraphics[width=0.3\textwidth]{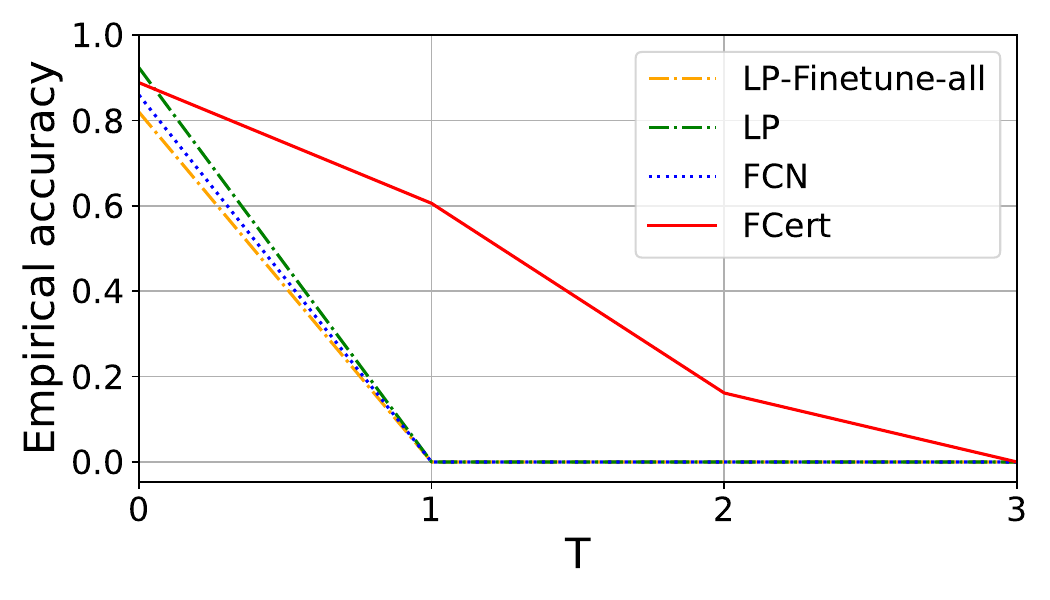}}\vspace{-0.5mm}
{\includegraphics[width=0.3\textwidth]{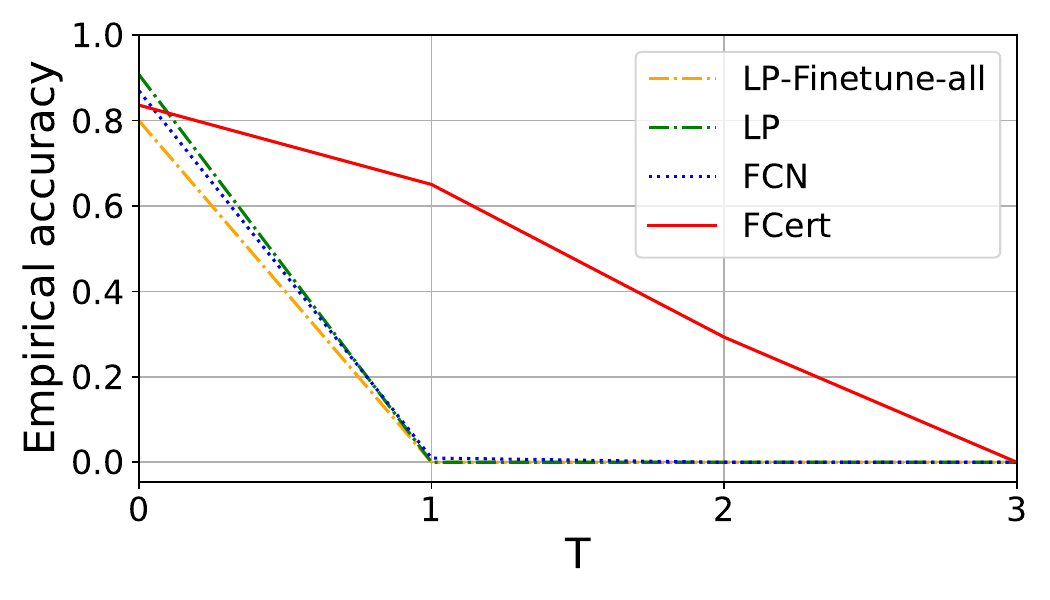}}\vspace{-0.5mm}

\subfloat[CUB200-2011]{\includegraphics[width=0.3\textwidth]{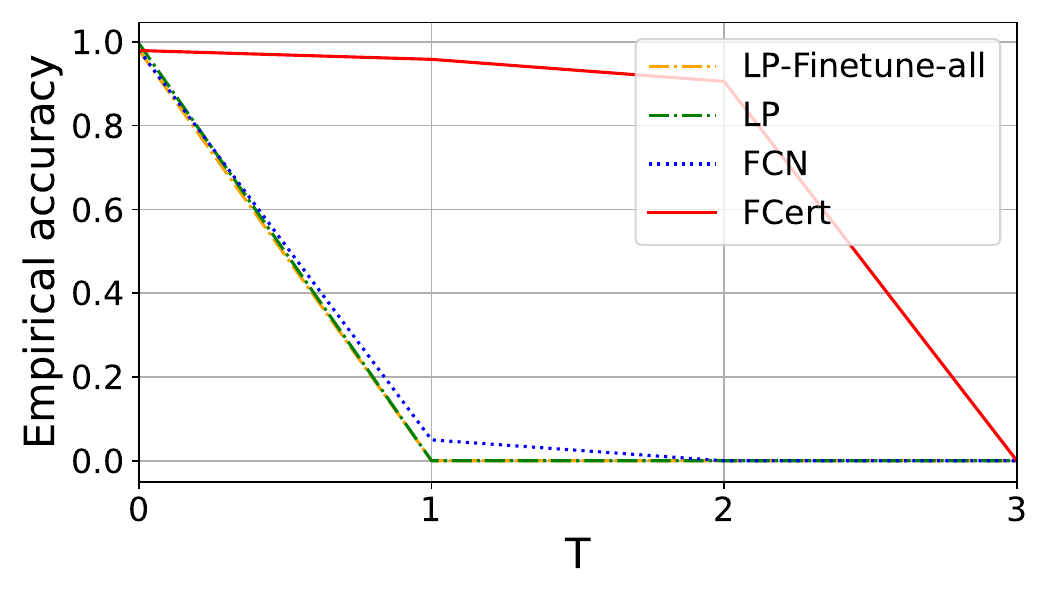}}\vspace{1mm}
\subfloat[CIFAR-FS]{\includegraphics[width=0.3\textwidth]{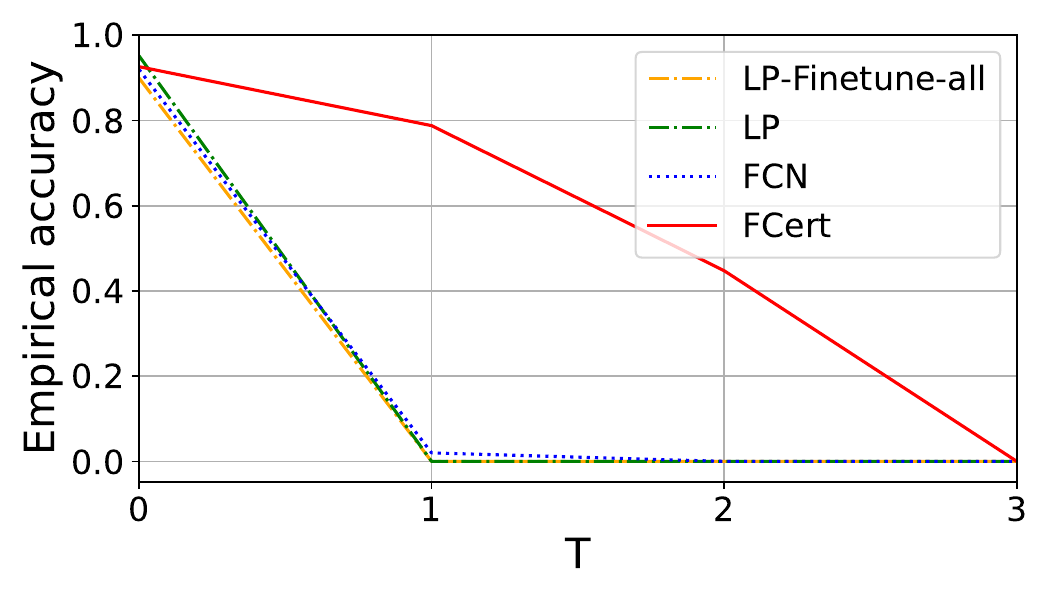}}\vspace{1mm}
\subfloat[\emph{tiered}ImageNet]{\includegraphics[width=0.3\textwidth]{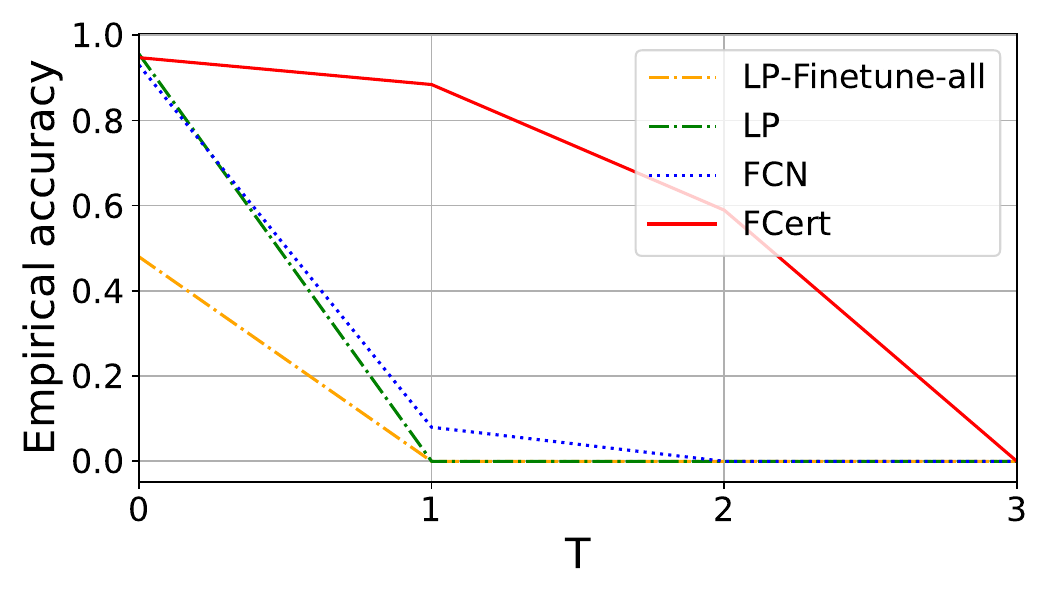}}\vspace{1mm}
\vspace{-4mm}
\caption{Comparing the emprical accuracy of our {\name} with different fine-tuning strategies and downstream classifiers for 5-way-5-shot few-shot classification with CLIP (first row) and DINOv2 (second row).  
}
\label{fig-finetune-all}
\end{figure*}

\newpage
\section{Meta-Review}
The following meta-review was prepared by the program committee for the 2024
IEEE Symposium on Security and Privacy (S\&P) as part of the review process as
detailed in the call for papers.

\subsection{Summary}
The paper introduces {\name}, a certified defense mechanism tailored for few-shot classification systems to mitigate data poisoning attacks. FCert computes robust distances between test inputs and support samples, offering theoretical guarantees of robustness while also demonstrating promising empirical results.

\subsection{Scientific Contributions}
\begin{itemize}
\item Creates a New Tool to Enable Future Science
\item Provides a Valuable Step Forward in an Established Field
\end{itemize}

\subsection{Reasons for Acceptance}
\begin{enumerate}
\item FCert addresses a significant gap in the field by providing a certified defense mechanism for few-shot classification, offering both theoretical guarantees and empirical evidence of its effectiveness.
\item The paper presents a novel approach that differentiates itself from existing defenses by focusing on few-shot learning and robust distance calculation.
\item Extensive experiments across various datasets and foundation models showcase FCert's robustness and potential applicability to real-world scenarios.

\end{enumerate}

\subsection{Noteworthy Concerns} 
\begin{enumerate} 
\item Reviewers had concerns regarding the breadth of the contribution, as the threat model associated with poisoning in few-shot learning settings may not be pervasive or realistic on broad settings.
\end{enumerate}

\end{document}